\documentclass[aps,showpacs,eqsecnum,10pt]{revtex4}
\usepackage{graphics,graphicx,graphpap,color}
\usepackage{setspace}

\newcommand\bmat{\left( \begin{array}{cc}}
\newcommand\emat{\end{array}\right)}

\def\msbar{\ifmmode{\overline{\rm MS}} \else{$\overline{\rm MS}$} \fi}
\def\drbar{\ifmmode{\overline{\rm DR}} \else{$\overline{\rm DR}$} \fi}

\def\ti              {\tilde}

\def\a               {\alpha}
\def\b               {\beta}
\def\d               {\delta}
\def\D               {\Delta}

\def\g               {\gamma}
\def\G               {\Gamma}
\def\l               {\lambda}
\def\t               {\theta}
\def\s               {\sigma}

\def\x               {\chi}

\def\sf              {{\ti f}}
\def\sfe              {{\ti f}}

\def\sfL             {{\ti f_L^{}}}
\def\sfR             {{\ti f_R^{}}}

\def\st              {{\ti t}}
\def\sb              {{\ti b}}
\def\stau            {{\ti \tau}}

\def\ch                {{\ti \chi}}

\def\chp             {\ti \x^+}
\def\chm             {\ti \x^-}
\def\nt              {\ti \x^0}

\newcommand{\msf}[1]   {m_{\ti f_{#1}}}
\newcommand{\mst}[1]   {m_{\ti t_{#1}}}
\newcommand{\msb}[1]   {m_{\ti b_{#1}}}
\newcommand{\mstau}[1] {m_{\ti\tau_{#1}}}

\def\tw              {\t_{W}}

\def\tsf              {\theta_\sfe}

\def\onehf           {{\textstyle \frac{1}{2}}}
\def\onehfb           {\frac{1}{2}}
\def\onehfbi         {{\displaystyle\frac{1}{2}}}

\def\tev             {{\rm TeV}}

\def\kslash          {\not{\!k}\,}
\newcommand{\rpi}[1]{{\hat{\Pi}_{#1}^T(s)}}

\def\non             {\nonumber}

\newcommand{\MM}{{\cal{M}}}

\renewcommand\d{\delta}

\newcommand{\vba}{\ensuremath{\bar{v}(p_2)}}

\newcommand{\unb}{\ensuremath{u(p_1)}}

\newcommand{\gam}{\ensuremath{\gamma^{\mu}}}

\newcommand{\spence}{{\rm Li}_2}

\begin{document}
\begin{flushright}
  HEPHY-PUB 807/05 \\
  hep-ph/0506021
\end{flushright}

\title{Full {\boldmath{$\cal{O}(\a)$}} corrections to {\boldmath{ $e^+ e^-
   \rightarrow \tilde{f}_i\ {\bar{\!\!\tilde{f}}}_{\!j}$}}}

\author{K.~Kova\v{r}\'{\i}k}
\email{kovarik@hephy.oeaw.ac.at} \affiliation{Institut f\"ur
Hochenergiephysik der \"Osterreichischen Akademie der
Wissenschaften,\\ A-1050 Vienna, Austria\\ and Department of
Theoretical Physics FMFI UK Comenius University, SK-84248
Bratislava, Slovakia}
\author{C.~Weber}
\email{weber@hephy.oeaw.ac.at}
\author{H.~Eberl}
\email{helmut@hephy.oeaw.ac.at}

\author{W.~Majerotto}%
\email{majer@hephy.oeaw.ac.at}

\affiliation{
        Institut f\"ur Hochenergiephysik der \"Osterreichischen Akademie
        der Wissenschaften,\\ A-1050 Vienna, Austria}

\date{June 3, 2005}

\begin{abstract}
We present a complete precision analysis of the sfermion pair
production process $e^+ e^- \rightarrow \tilde{f}_i\
{\bar{\!\!\tilde{f}}}_{\!j}\;(f = t, b, \tau, \nu_\tau)$ in the
Minimal Supersymmetric Standard Model. Our results extend the
previously calculated weak corrections by including all one-loop
corrections together with higher order QED corrections. We present
the details of the analytical calculation and discuss the
renormalization scheme. The numerical analysis shows the results
for total cross-sections, forward-backward and left-right
asymmetries. It is based on the SPS1a' point from the SPA project.
The complete corrections are about 10\% and have to be taken into
account in a high precision analysis.
\end{abstract}


\maketitle

\section{Introduction}
\vspace{2mm} The Minimal Supersymmetric Standard Model (MSSM)
provides the most attractive extension of the Standard Model (SM).
Among other particles it includes supersymmetric partners of the
fermions. These scalar states $\sf_L$, $\sf_R$ (sfermions)
correspond to the two chirality states of each fermion $f$. The
mass eigenstates $\sf_1$ and $\sf_2$ though are not identical with
$\sf_L$, $\sf_R$ and are rather a linear combination of them. The
mixing terms are proportional to the mass of the corresponding
fermion. Hence the sfermions of the third generation play a
special r{$\hat{\rm o}$}le. As a consequence, one eigenstate
($\sf_1$) can be much lighter than the other one.
\newline %
The sfermions, especially the strongly interacting ones ($\st_i,
\sb_i$), are likely to be detected at the LHC or the Tevatron.
Nevertheless, to extract the fundamental parameters one must have
a significant accuracy only obtainable at a linear collider. From
sfermion pair production in $e^+e^-$ collisions the sfermion
mixing angle can be extracted. This is one of the reasons why it
has been extensively studied phenomenologically \cite{exp}. To
match the expected precision of the linear collider, theory
predictions must reach a similar accuracy. The effort to calculate
higher order corrections to the sfermion production has begun by
calculating the leading QCD, SUSY-QCD and Yukawa corrections
\cite{QCD1, SUSY-QCD-A, SUSY-QCD-H, Yukawa}. It was further shown
that taking only the leading terms of the one-loop corrections is
not sufficient and so also the full weak corrections were
presented in \cite{letter, hollik}.
\newline %
It is the aim of this paper to extend the existing weak
corrections by including the full $\cal{O}(\alpha)$ contributions
in a similar manner as in the case of the selectrons and the
smuons in \cite{freitas}. In addition, we present the full
analytical results and all the details of the calculation for both
the weak corrections \cite{letter} and the QED contributions.
Moreover, we generalize the results to include also the effects of
polarization of the electron and positron beams. Apart from
cross-sections, we calculate other observables such as the
forward-backward and the left-right asymmetries as well.
\newline %
Although we present the results in the form of cross-sections and
asymmetries, we are well aware of the fact that the precise
predictions have to be used for parameter extraction. As the
definition of the parameters is no longer unique beyond the
tree-level, there has been a recent proposal by the so-called SPA
project (SUSY parameter analysis) which defines these parameters
\cite{SPA}. The SPA project also gives a firm base for calculating
all sorts of observables (masses, decay widths, cross-sections
etc.) and enables the development of tools for extracting the
parameters.
\newline %
The fundamental SUSY parameters in the SPA project are defined
using the \drbar (dimensional reduction) renormalization scheme at
the scale $Q=1\tev$. Specifying the renormalization scheme serves
only to define the parameters uniquely and does not restrict the
use of other schemes in different calculations. In this paper, we
use an on-shell renormalization scheme. To use the parameters from
the SPA project we have to translate them into the on-shell
renormalization scheme. The results for any observable using
different schemes (with correctly translated input parameters)
must agree up to contributions of higher order.
\newline %
The paper is organized as follows. In section~\ref{treelevel} we
give the formulae for the tree-level cross-section for polarized
electron and positron beams. The calculation of the virtual
corrections with a detailed discussion of the applied on-shell
renormalization scheme are outlined in section~\ref{radcor}. All
explicit analytic formulas needed for the calculation are given in
the Appendices \ref{appVertex}, \ref{appBox}, \ref{appSE} and
\ref{Bremint}. In section~\ref{realcorr} we work out the real
radiative corrections where we include the Bremsstrahlung process
$\s(e^+e^- \rightarrow \tilde{f}_i \
{\bar{\!\!\tilde{f}}}_{\!j}\g)$. In section~\ref{numerics} we
present the numerical analysis with some results of the
corrections. Section~\ref{conclusion} summarizes our conclusions.
\section{Tree level}\label{treelevel}
The sfermion mixing is described by the diagonalization of the
sfermion mass matrix given in the left-right basis $(\sfL, \sfR)$
into the mass basis $(\sf_1, \sf_2)$, $f = t,b$ or $\tau$
\cite{GunionHaber},
\begin{eqnarray}
  {\cal M}_{\sf}^{\,2} &=&
   \left(
     \begin{array}{cc}
       m_{\sf_L}^{\,2} & a_f\, m_f
       \\
       a_f\,m_f & m_{\sf_R}^{\,2}
     \end{array}
   \right)
  = \left( R^\sf \right)^\dag
   \left(
     \begin{array}{cc}
       m_{\sf_1}^{\,2} & 0
       \\
       0 & m_{\sf_2}^{\,2}
     \end{array}
   \right) R^\sf \,,
\end{eqnarray}
where $R^\sf_{i\a}$ is a 2 x 2 rotation matrix with rotation angle
$\theta_{\sf}$, which relates the mass eigenstates $\sf_i$, $i =
1, 2$, $(m_{\sf_1} < m_{\sf_2})$ to the weak eigenstates $\sf_\a$,
$\a = L, R$, by $\sf_i = R^\sf_{i\a} \sf_\a$, with $R^{\ti f}_{11}
= R^{\ti f}_{22} = \cos\theta_\sf$ and $R^\sf_{12} = -R^\sf_{21} =
\sin\theta_\sf$, and
\begin{eqnarray}
  m_{\sf_L}^{\,2} &=& M_{\{\ti Q,\,\ti L \}}^2
       + (I^{3L}_f \!-\! e_f\,s_W^2)\cos2\b\,
       m_{Z}^{\,2}
       + m_{f}^2\,, \\ \label{MsD}
  m_{\sf_R}^{\,2} &=& M_{\{\ti U,\,\ti D,\,\ti E \}}^2
       + e_{f}\,s_W^2 \cos2\b\,m_{Z}^{\,2}
       + m_f^2\,, \\
  a_f &=& A_f - \mu \,(\tan\b)^{-2 I^{3L}_f} \,.
\end{eqnarray}
$M_{\ti Q}$, $M_{\ti L}$, $M_{\ti U}$, $M_{\ti D}$ and $M_{\ti E}$
are soft SUSY breaking masses, $A_f$ is the trilinear scalar
coupling parameter, $\mu$ the higgsino mass parameter, $\tan\b =
\frac{v_2}{v_1}$ is the ratio of the vacuum expectation values of
the two neutral Higgs doublet states , $I^{3L}_f$ denotes the
third component of the weak isospin of the fermion $f$, $e_f$ the
electric charge in terms of the elementary charge $e$, and $s_W$
is the sine of the Weinberg angle $\tw$.
\\
The mass eigenvalues and the mixing angle are
\begin{eqnarray}
  \msf{1,2}^2
    &=& \frac{1}{2} \left(
    \msf{L}^2 + \msf{R}^2 \mp
    \sqrt{(\msf{L}^2 \!-\! \msf{R}^2)^2 + 4 a_f^2
    m_f^2}\,\right)\,,
\\
  \cos\t_{\sf}
    &=& \frac{-a_f\,m_f}
    {\sqrt{(\msf{L}^2 \!-\! \msf{1}^2)^2 + a_f^2 m_f^2}}
  \hspace{2cm} (0\leq \t_{\sf} < \pi) \,,
\end{eqnarray}
and the mass of the sneutrino $\ti\nu_\tau$ is given by
$m_{\ti\nu_\tau}^2 = M_{\ti L}^2 + \frac{1}{2}\,m_Z^2 \cos2\beta$.
\newline\newline
The tree-level cross-section of $e^+e^- \rightarrow \tilde{f}_i \
{\bar{\!\!\tilde{f}}}_{\!j}$ for polarized electron and positron
beams is given by
\begin{eqnarray}\label{tree}
\s^{\rm tree} &=& \frac{1}{4}(1-P_-)(1+P_+)\,\s_L^{\rm tree}
+\frac{1}{4}(1+P_-)(1-P_+)\,\s_R^{\rm tree}\,,
\end{eqnarray}
where $P_-, P_+\in(-1,1)$ are the degrees of polarization of the
electron and positron beams (e.~g. $P_- (P_+) = -0.8$ means 80\%
of electrons (positrons) left polarized and 20 \% unpolarized).

As we neglect the electron mass, we have only two terms
contributing (out of 4 possible) where $\s_L^{\rm tree}$ is the
tree-level cross-sections for $e_R^+\,e_L^- \rightarrow
\tilde{f}_i \ {\bar{\!\!\tilde{f}}}_{\!j}$ (below referred to as
the left part of the polarized cross-section) and $\s_R^{\rm
tree}$ stands for $e_L^+e_R^- \rightarrow \tilde{f}_i \
{\bar{\!\!\tilde{f}}}_{\!j}$ (analogously referred to as the right
part of the cross-section). They have the form
\begin{eqnarray}
\s_{L,R}^{\rm tree}(e^+e^- \rightarrow \tilde{f}_i \
{\bar{\!\!\tilde{f}}}_{\!j}) &=& \frac{N_C}{3}\frac{\kappa^3 (s,
m^2_{\sf_i}, m^2_{\sf_j})}{4 \,\pi\,
s^2}\left(T_{L,R}^{\g\g}+T_{L,R}^{\g Z}+T_{L,R}^{ZZ}\right)\,,
\end{eqnarray}
where
\begin{eqnarray}\label{T1tree}
T_{L,R}^{\g\g}&=&\frac{e^4 e_f^2 (\d_{ij})^2}{s^2}\onehfbi
K_{L,R}^2\,,
\\ \label{T2tree}
T_{L,R}^{\g Z}&=& -\frac{g_Z^2 e^2 e_f a_{ij}^\sf\d_{ij}}{4s
(s-m_Z^2)}\, C_{L,R} K_{L,R}\,,
\\ \label{T3tree}
T_{L,R}^{ZZ}&=&\frac{g_Z^4(a_{ij}^\sf)^2}{32\,(s-m_Z^2)^2}
\,C_{L,R}^2\,,
\end{eqnarray}
and $\kappa (x, y, z) = \sqrt{(x-y-z)^2 - 4 y z}$.\newline Here we
use $K_{L,R}$ and $C_{L,R}$ as the left- and right-handed
couplings of the electron to the  photon and $Z$-boson,
respectively,
\begin{equation}
K_L = K_R = 1,\qquad C_L = -\frac{1}{2}+s_W^2, \qquad C_R = s_W^2.
\end{equation}
The matrix elements $a_{ij}^\sf$ come from the coupling of
$Z\sf_{i}\sf_{j}$,
\begin{eqnarray}
a_{ij}^\sf &=& \left(\begin{array}{cc} 4(I^{3L}_f \cos^2
\t_\sf-s_W^2 e_f) & -2I^{3L}_f \sin 2\t_\sf \\ -2I^{3L}_f \sin
2\t_\sf & 4(I^{3L}_f \cos^2\t_\sf-s_W^2 e_f)
\end{array}\right).
\end{eqnarray}
\newline %
Apart from the tree-level cross-section we can calculate other
observables such as the left-right asymmetry and the
forward-backward asymmetry. They are defined by
\begin{eqnarray}\label{asym}
A_{LR} \,=\, \frac{\s_L-\s_R}{\s_L+\s_R}\,,\qquad\quad A_{FB}
\,=\, \frac{\s_F-\s_B}{\s_F+\s_B}\,,
\end{eqnarray}
with
\begin{equation}
\s_F \,=\,
\int_{0}^{2\pi}d\varphi\int_{0}^{\pi/2}\left(\frac{d\s}{d
\Omega}\right)\,d\cos\vartheta \,,\qquad\quad \s_B \,=\,
\int_{0}^{2\pi}d\varphi\int_{\pi/2}^{\pi}\left(\frac{d\s}{d
\Omega}\right)\,d\cos\vartheta\,.
\end{equation}
There is no lowest order (tree-level) contribution to the
$A_{FB}$-asymmetry as the angle distribution is symmetric.
\section{Virtual corrections}\label{radcor}
\begin{figure}[h!]
\begin{picture}(160,25)
    \put(0,3){\mbox{\resizebox{15.5cm}{!}
     {\includegraphics{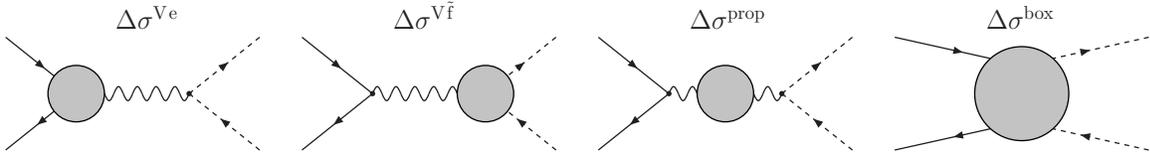}}}}
\end{picture}
\caption{\label{UVfiniteparts}UV-finite parts of the electro-weak
corrections}
\end{figure}

\begin{figure}[h!]
\begin{picture}(160,105)
    \put(0,3){\mbox{\resizebox{15.5cm}{!}
     {\includegraphics{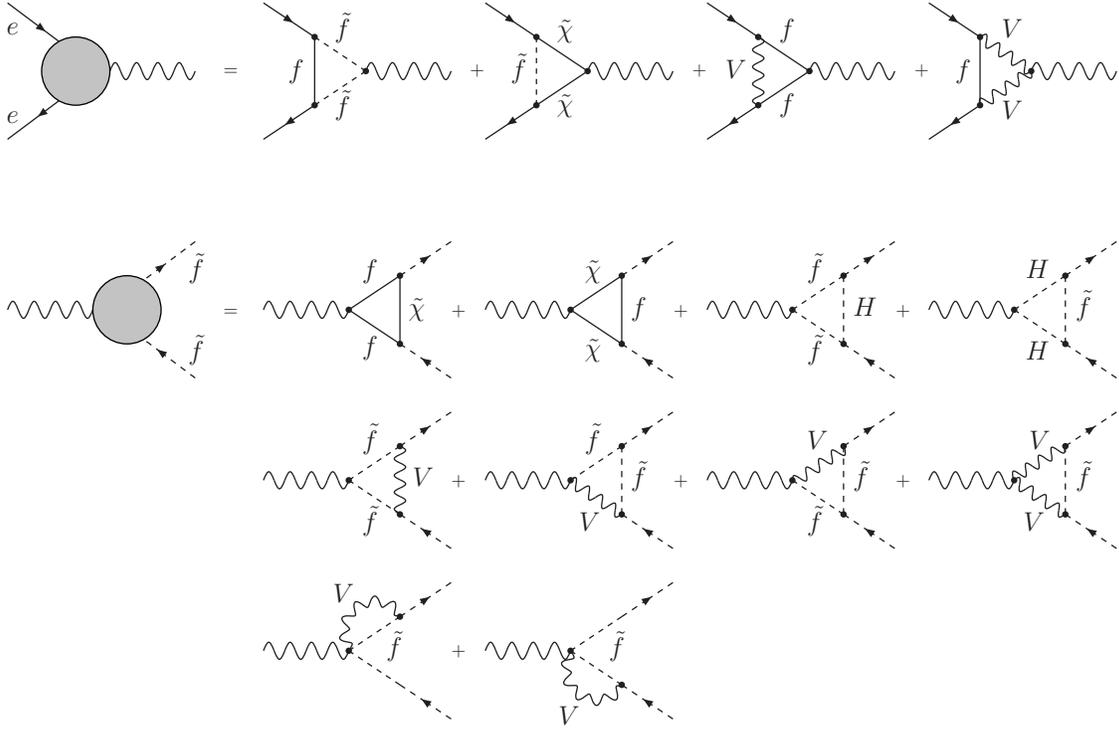}}}}
\end{picture}
\caption{Vertex diagrams contributing to $\d e_{L,R}^{(v)}$ , $\d
a_{L,R}^{(v)}$ , $(\d e_f)_{ij}^{(v)}$ and $(\d a_f)_{ij}^{(v)}$.
\label{vertex-graphs}}
\end{figure}

\begin{figure}[h!]
\begin{picture}(160,75)
    \put(0,35){\mbox{\resizebox{15.5cm}{!}
     {\includegraphics{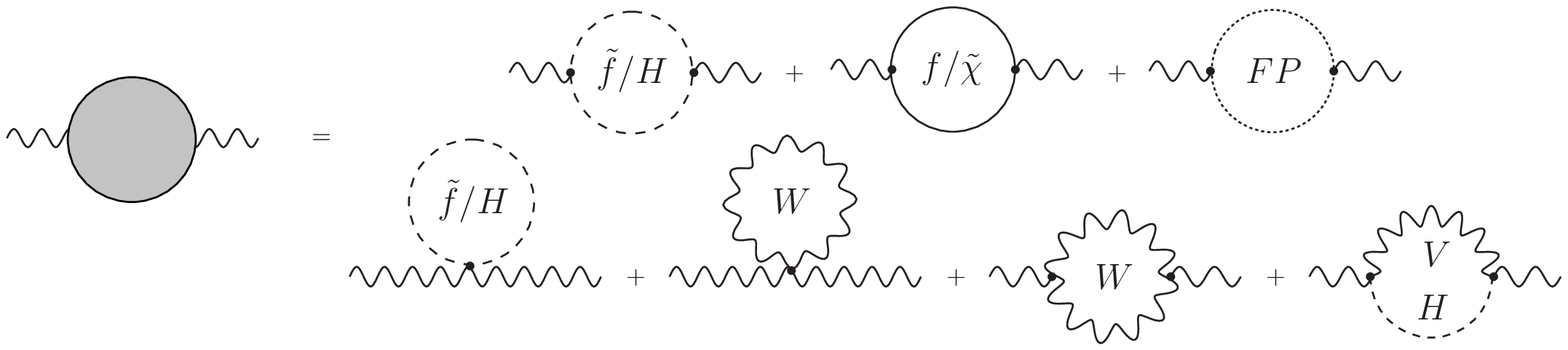}}}}
    \put(0,3){\mbox{\resizebox{15.5cm}{!}
     {\includegraphics{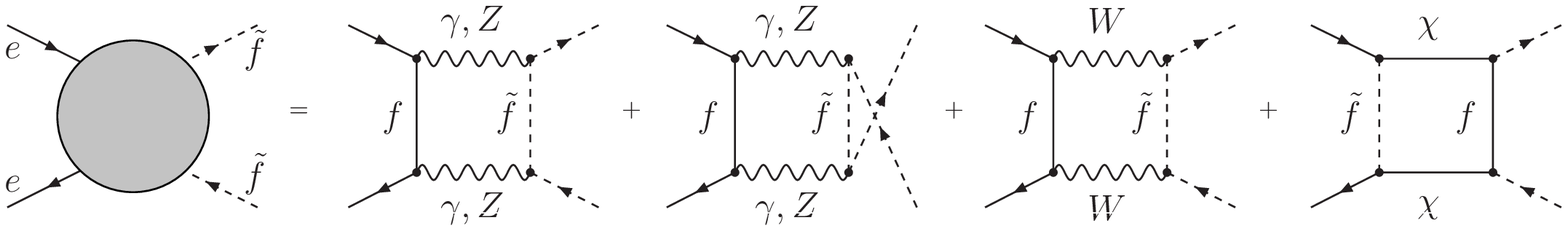}}}}
\end{picture}
\caption{Diagrams contributing to the propagator and the box
corrections.\label{propbox}}
\end{figure}

For a precision analysis of the sfermion production one has to
include also higher order corrections. The calculation of the
higher order corrections is performed analytically in the \drbar
scheme, adopting the $\xi =1$ 'tHooft-Feynman gauge. All necessary
ingredients of the analytical calculation are given in the
Appendices. Furthermore, we neglect the electron mass wherever
possible $(m_e =0)$. For the numerical evaluation of the loop
integrals we use the packages LoopTools and FF \cite{loopFF}. At
the end the whole analytic result was checked with the result
obtained using the computer algebra tools FeynArts and FormCalc
\cite{feyn}.
\newline The virtual corrections receive contributions from
vertex, self-energy and box diagrams depicted generically in
Fig.~\ref{UVfiniteparts} and explicitly in
Figs.~\ref{vertex-graphs} and \ref{propbox}. All these
contributions are summarized in the renormalized cross-section
$\sigma^{\rm ren}$.
\newline The one-loop (renormalized) cross-section $\sigma^{\rm
ren}$ for polarized beams is expressed analogously to
Eq.~(\ref{tree}),
\begin{eqnarray}
\s^{\rm ren}(e^+e^- \rightarrow \tilde{f}_i \
{\bar{\!\!\tilde{f}}}_{\!j}) &=&
\frac{1}{4}(1-P_-)(1+P_+)\,\s_L^{\rm ren}
+\frac{1}{4}(1+P_-)(1-P_+)\,\s_R^{\rm ren}\,,
\end{eqnarray}
where the left/right renormalized cross-sections are defined as
\begin{eqnarray}
\s^{\rm ren}_{L,R} &=& \s^{\rm tree}_{L,R}+\D\s^{\rm
QCD}_{L,R}+\D\s^{\rm EW}_{L,R}
\end{eqnarray}
with the symbol $\D$ denoting UV-finite quantities.
\newline The SUSY-QCD corrections ($\D\s^{\rm QCD}$) have already been
calculated for the unpolarized case in
\cite{SUSY-QCD-A,SUSY-QCD-H}. As the gluon part of $\D\s^{\rm
QCD}$ is proportional to the tree-level cross-section, the
polarized cross-sections are easily obtained using
$\s^{tree}_{L,R}$ instead of $\s^{tree}$. The gluino part of
$\D\s^{\rm QCD}$ is treated analogously to $\D\s^{V\sf}$ (see
Fig.~\ref{UVfiniteparts}).
\newline We have already
presented the unpolarized results for the electro-weak corrections
($\D\s^{\rm EW}$) in \cite{letter}. In this paper, we give the
result for polarized beams and also all formulas needed for the
calculation.\newline The electroweak corrections can be split
further into four UV-finite parts given in
Fig.~\ref{UVfiniteparts},
\begin{eqnarray}
\D\s^{\rm EW}_{L,R} &=& \D\s^{\rm V\hspace{1pt}e}_{L,R}+\D\s^{\rm
V\hspace{1pt}\tilde{f}}_{L,R}+\D\s^{\rm prop}_{L,R}+\D\s^{\rm
box}_{L,R}\,,
\end{eqnarray}
where $\D\s^{\rm V\hspace{1pt}e}_{L,R}$ and $\D\s^{\rm
V\hspace{1pt}\tilde{f}}_{L,R}$ stand for the left/right part of
the renormalized electron and sfermion vertex, $\D\s^{\rm
prop}_{L,R}$ and $\D\s^{\rm box}_{L,R}$ for the left/right part of
renormalized propagators and box contribution.\newline The
renormalized electron vertex has the form
\begin{eqnarray}
\D\s^{\rm V\hspace{1pt}e}_{L,R} &=& \frac{N_C}{3}\frac{\kappa^3
(s, m^2_{\sf_i}, m^2_{\sf_j})}{4 \,\pi\, s^2}\left(\left(\D
T^{V\hspace{1pt}e}_{\g\g}\right)_{L,R}+\left(\D
T^{V\hspace{1pt}e}_{\g Z}\right)_{L,R}+\left(\D
T^{V\hspace{1pt}e}_{ZZ}\right)_{L,R}\right),
\end{eqnarray}
where
\begin{eqnarray}
\left(\D T^{V\hspace{1pt}e}_{\g\g}\right)_{L,R}&=&\frac{e^4
e_f^2(\d_{ij})^2}{s^2}\,(\D e_{L,R} K_{L,R})\,,
\\
\left(\D T^{V\hspace{1pt}e}_{\g Z}\right)_{L,R}&=& -\frac{g_Z^2
e^2 e_f a_{ij}^\sf\d_{ij}}{4s (s-m_Z^2)}\,(\D e_{L,R} C_{L,R} +\D
a_{L,R} K_{L,R})\,,
\\
\left(\D T^{V\hspace{1pt}e}_{ZZ}\right)_{L,R}&=&
\frac{g_Z^4(a_{ij}^\sf)^2}{16\,(s-m_Z^2)^2}\,(\D a_{L,R} C_{L,R}).
\end{eqnarray}
$\D e_{L,R}$ and $\D a_{L,R}$ consist of 3 parts,
\begin{eqnarray}
\D e_{L,R} &=& \d e_{L,R}^{(v)}+\d e_{L,R}^{(w)}+\d
e_{L,R}^{(c)}\,,
\\ \D a_{L,R} &=& \d a_{L,R}^{(v)}+\d
a_{L,R}^{(w)}+\d a_{L,R}^{(c)}\,.
\end{eqnarray}
$\d e_{L,R}^{(v)}$, $\d a_{L,R}^{(v)}$ correspond to the vertex
corrections in Fig.~\ref{vertex-graphs}, $\d e_{L,R}^{(w)}$, $\d
a_{L,R}^{(w)}$ are the wave-function corrections, and $\d
e_{L,R}^{(c)}$, $\d a_{L,R}^{(c)}$ correspond to the
counterterms.\newline The renormalized sfermion vertex has a
similar form,
\begin{eqnarray}
\D\s^{\rm V\hspace{1pt}\tilde{f}}_{L,R} &=&
\frac{N_C}{3}\frac{\kappa^3 (s, m^2_{\sf_i}, m^2_{\sf_j})}{4
\,\pi\, s^2}\left(\left(\D
T^{V\hspace{1pt}\tilde{f}}_{\g\g}\right)_{L,R}+\left(\D
T^{V\hspace{1pt}\tilde{f}}_{\g Z}\right)_{L,R}+\left(\D
T^{V\hspace{1pt}\tilde{f}}_{ZZ}\right)_{L,R}\right)\,,
\end{eqnarray}
where
\begin{eqnarray}
\left(\D T^{V\hspace{1pt}\tilde{f}}_{\g\g}
\right)_{L,R}&=&\frac{e^4 e_f (\D e_f)_{ij}}{s^2}\,K_{L,R}^2\,,
\\
\left(\D T^{V\hspace{1pt}\tilde{f}}_{\g Z} \right)_{L,R}&=&
-\frac{g_Z^2 \,e^2}{4s (s-m_Z^2)}\,K_{L,R} C_{L,R}\,((\D e_f)_{ij}
a_{ij}^\sf + \d_{ij}(\D a_f)_{ij})\,,
\\
\left(\D
T^{V\hspace{1pt}\tilde{f}}_{ZZ}\right)_{L,R}&=&\frac{g_Z^4\,
a_{ij}^\sf (\D a_f)_{ij}}{16\,(s-m_Z^2)^2}\,C_{L,R}^2\,.
\end{eqnarray}
$(\D e_f)_{ij}$ and $(\D a_f)_{ij}$ can also be split into vertex
corrections (see Fig.~\ref{vertex-graphs}), wave-function
corrections and counterterms,
\begin{eqnarray}
(\D e_f)_{ij} &=& (\d e_f)_{ij}^{(v)}+(\d e_f)_{ij}^{(w)}+(\d
e_f)_{ij}^{(c)}\,,\\ (\D a_f)_{ij} &=& (\d a_f)_{ij}^{(v)}+(\d
a_f)_{ij}^{(w)}+(\d a_f)_{ij}^{(c)}\,.
\end{eqnarray}
The diagrams contributing to the vertex corrections are shown in
Fig.~\ref{vertex-graphs} and the explicit form of the corrections
are given in Appendix \ref{appVertex}. The wave-function
corrections and the counterterms to both vertices are listed in
detail in sections \ref{wavefunction} and \ref{counterterm}. The
$(\d e_f)_{ij}$ and $(\d a_f)_{ij}$ corresponding to gluino
corrections can be found in \cite{SUSY-QCD-H}.
\newline
The correction $\D\s^{\rm prop}_{L,R}$ which comes from inserting
the self-energies of the $\g$ and $Z$-boson, in the propagator,
see Fig.~\ref{propbox}, can be expressed as
\begin{eqnarray}\label{propform}
\D\s^{\rm prop}_{L,R}&=& \frac{N_C}{3}\frac{\kappa^3 (s,
m^2_{\sf_i}, m^2_{\sf_j})}{4 \,\pi\, s^2}\times\non
\\[7pt]
&& \hspace{-1cm}\times\, 2\,\Re\left[\left(
-\frac{\rpi{\g\g}}{s}\right)\,T^{\g\g}_{L,R}+\left(
-\frac{\rpi{\g\g}}{s}-\frac{\rpi{ZZ}}{s-m_Z^2}\right)\,T^{\g
Z}_{L,R}+\left(-\frac{\rpi{ZZ}}{s-m_Z^2}\right)\,T^{ZZ}_{L,R}\right.
\\
&& \hspace{-1cm} \left.+\left(s_W
c_W\frac{\rpi{Z\g}}{s}\right)\,\left(
(T_{Z\g}^\g)_{L,R}+(T_{ZZ}^\g)_{L,R}\right)+\left(\frac{1}{s_W
c_W}\frac{\rpi{\g Z}}{s-m_Z^2}\right)\,\left((T_{\g\g}^Z)_{L,R}
+(T_{\g Z}^Z)_{L,R}\right) \right]\,,\non
\end{eqnarray}
where $T^{\g\g}_{L,R}$, $T^{\g Z}_{L,R}$, $T^{ZZ}_{L,R}$ are
defined in Eqs.~(\ref{T1tree}-\ref{T3tree}) and
\begin{eqnarray}
(T_{Z\g}^\g)_{L,R} &=& -\frac{g_Z^2 e^2 e_f a_{ij}^\sf\d_{ij}}{4s
\,(s-m_Z^2)}\,\onehfbi K_{L,R}^2\,,
\\
(T_{ZZ}^\g)_{L,R} &=&
\frac{g_Z^4(a_{ij}^\sf)^2}{16\,(s-m_Z^2)^2}\,\onehfbi C_{L,R}
K_{L,R}\,,
\\
(T_{\g\g}^Z)_{L,R} &=& \frac{e^4 e_f^2 (\d_{ij})^2}{s^2}\onehfbi
C_{L,R} K_{L,R}\,,
\\
(T_{\g Z}^Z)_{L,R} &=& -\frac{g_Z^2 e^2 e_f a_{ij}^\sf\d_{ij}}{4s
\,(s-m_Z^2)}\,\onehfbi C_{L,R}^2\,.
\end{eqnarray}
The $\rpi{VV}$ in Eq.~(\ref{propform}) are the transverse parts of
the renormalized self-energies of the vector bosons $\g$ and $Z$.
The unrenormalized self-energies are given in Appendix
\ref{appvector-SE} and the renormalization is done following
\cite{Denner}.\newline The box corrections are obtained by adding
up the diagrams shown in Fig.~\ref{propbox} and are given by
\begin{eqnarray}
\D\s^{\rm box}_{L,R} &=& \frac{N_C}{4}\,\frac{\kappa^3 (s,
m^2_{\sf_i}, m^2_{\sf_j})}{4 \,\pi\, s^2}\!
{\displaystyle{\int_0^\pi}} \!T_{L,R}^{\rm box} \,
\sin^2\vartheta\;{\rm d}\vartheta\,,
\end{eqnarray}
where
\begin{eqnarray}
T_{L,R}^{\rm box}&=& \left(-\frac{1}{(4\pi)^2} \frac{e^2 e_f
\d_{ij}}{s} \frac{1}{2} K_{L,R} + \frac{1}{(4\pi)^2} \frac{g_Z^2\,
a^\sf_{ij}} {4(s-m_Z^2)} \frac{1}{2} C_{L,R}\right) \,B_{L,R}\,.
\end{eqnarray}
The $B_{L,R}$ are the form-factors defined in Appendix
\ref{appBox}, where one can find the analytic expressions as well.
%
%
\subsection{Renormalization scheme}\label{renschem}
In order to make the result finite we have to introduce the
wave-function renormalization constants and counterterms. We fix
them following the on-shell renormalization scheme. The parameters
already occurring in the Standard Model (SM) are renormalized
according to \cite{Denner}. We assume the CKM matrix to be
diagonal and so have no flavour mixing among the SM fermions at
one-loop level.
\subsubsection{Wave-function renormalization}\label{wavefunction}
The wave-function corrections are due to a shift from
unrenormalized (bare) fields to the renormalized (physical) ones.
For the fields relevant here we have
\begin{eqnarray}
\sf_i^0 = (\d_{ij}+\onehf\d Z_{ij})\sf_j\,, \qquad && \left(
\begin{array}{c}
 f_L^0 \\ f_R^0
      \end{array}\right)= \left(\begin{array}{cc}
1+\onehf\d Z_L & 0 \\ 0 & 1+\onehf\d Z_R
     \end{array}\right)\left(
\begin{array}{c}
 f_L \\ f_R
     \end{array}\right),\nonumber
\\[1mm] \left(
\begin{array}{c}
 A_\mu^0 \\ Z_\mu^0
     \end{array}\right)&=& \left(\begin{array}{cc}
1+\onehf\d Z_{\g\g} & \onehf\d Z_{\g Z} \\ \onehf\d Z_{Z\g} &
1+\onehf\d Z_{ZZ}
     \end{array}\right)\left(
\begin{array}{c}
 A_\mu \\ Z_\mu
     \end{array}\right).
\end{eqnarray}
The form of the corrections for the left vertex is
\begin{eqnarray}
\d e_{L,R}^{(w)} &=& (\d Z_{L,R}+\onehf\d
Z_{\g\g})K_{L,R}-\onehfbi\frac{g_Z}{e} \d Z_{Z\g}C_{L,R}\,,
\\
\d a_{L,R}^{(w)} &=& (\d Z_{L,R}+\onehf\d
Z_{ZZ})C_{L,R}-\onehfbi\frac{e}{g_Z}\d Z_{\g Z}K_{L,R}\,.
\end{eqnarray}
The wave-function corrections for the right vertex are
\begin{eqnarray}
(\d e_f)_{ij}^{(w)}&=& \onehfbi e_f (\d Z_{ij}+\d Z_{ji})
+\onehfbi e_f \d Z_{\g\g}\,\d_{ij} + \frac{1}{8s_W
c_W}a^{\sf}_{ij}\,\d Z_{Z\g}\,,
\\
(\d a_f)_{ij}^{(w)}&=& \onehfbi \sum_{k=1}^2\left(\d Z_{ki}
\,a^{\sf}_{kj} + \d Z_{kj}\, a^{\sf}_{ik}\right) + 2 s_Wc_W e_f\,
\d_{ij}\,\d Z_{\g Z}+\onehfbi a^{\sf}_{ij}\,\d Z_{ZZ}\,,
\end{eqnarray}
where $s_W = \sin\tw$ and $c_W = \cos\tw$.
\newline The wave-function renormalization constants are
determined by imposing the on-shell renormalization conditions as
in \cite{onshellren,sche2} such that the on-shell masses are the
real parts of the poles of the propagator and the fields are
properly normalized,
\begin{eqnarray}\label{offdiag}
\d Z_{ii} &=& - \Re\,\dot\Pi_{ii}^{\sf} (m_{\sf_i}^2)\,,\qquad \d
Z_{ij}~=~\frac{2\,\Re\,\Pi_{ij}^{\sf}(m_{\sf_j}^2)}{\msf{i}^2-\msf{j}^2}\,,\;\qquad
\d Z_{\g\g}~=~ - \Re\,\dot\Pi_{\g\g}(0)\,,
\\
\d Z_{ZZ} &=& -\Re\,\dot\Pi_{ZZ}(m_Z^2)\,,\;\quad \d Z_{\g Z}~=~ -
\frac{2\,\Re\,\Pi_{\g Z}(m_Z^2)}{m_Z^2}\,,\quad \d Z_{Z\g}~=~
\frac{2\,\Re\,\Pi_{Z\g}(0)}{m_Z^2}\,,
\\
\d Z_L &=& \Re\,\left[ -\Pi_{L} (m_e^2)- m_e^2(\dot\Pi_{L}
(m_e^2)+ \dot\Pi_{R} (m_e^2))+\frac{1}{2 m_e}(\Pi_{SL}
(m_e^2)-\Pi_{SR} (m_e^2))\right.\nonumber
\\
&& \left.  - m_e(\dot\Pi_{SL} (m_e^2)+\dot\Pi_{SR}
(m_e^2))\right]\,,
\\
\d Z_R &=& \Re\,\left[ -\Pi_{R} (m_e^2)- m_e^2(\dot\Pi_{R}
(m_e^2)+ \dot\Pi_{L} (m_e^2))+\frac{1}{2 m_e}(\Pi_{SR}
(m_e^2)-\Pi_{SL} (m_e^2)) \right.\nonumber
\\
&& \left. - m_e(\dot\Pi_{SR} (m_e^2)+\dot\Pi_{SL}
(m_e^2))\right]\,,
\end{eqnarray}
where $\dot\Pi (m^2)=\left[\frac{\partial}{\partial k^2}\Pi
(k^2)\right]_{k^2=m^2 }$. We use the self-energies given in
Appendix \ref{appSE} and in \cite{chrislet} where we adopted the
conventions from.\newline A remark should be made at this point.
We include the wave-functions renormalization constants of the
vector bosons although they are not external particles. By
introducing them into the wave-function renormalization of the
vertices, we have additional checks that can be made. First of
all, the renormalization constants of the vector bosons must drop
out in the final result. Secondly, the vertex corrections and the
propagators can be both made UV-finite separately.
%
%
\subsubsection{Counterterms}\label{counterterm} The counterterms
come from the shift from the bare to the physical parameters in
the lagrangian. It includes the shifting of $e,m_W,m_Z,\tsf$
defined by
\begin{eqnarray}
e^0=e+\d e\,, \qquad m_W^0=m_W+\d m_W\,,\qquad m_Z^0=m_Z+\d
m_Z\,,\qquad \tsf^0=\tsf+\d \tsf\,.
\end{eqnarray}
The counterterm contributions for both vertices are
\begin{eqnarray}
\d e_{L,R}^{(c)}&=& \frac{\d e}{e}\, K_{L,R}\,,
\\
\d a_{L,R}^{(c)}&=& \left[\frac{\d e}{e}-\left( \frac{\d
m_W}{m_W}-\frac{\d m_Z}{m_Z}\right)+\frac{1-2 s_W}{t_W^2}\left(
\frac{\d m_W}{m_W}-\frac{\d m_Z}{m_Z}\right)\right]\,C_{L,R}\,,
\end{eqnarray}
\begin{eqnarray}
(\d e_f)_{ij}^{(c)}&=& \frac{\d e}{e}\,e_f\,\d_{ij}\,,
\\
\label{symwave} (\d a_f)_{ij}^{(c)}&=& \left[\frac{\d
e}{e}+\frac{c_W^2-s_W^2}{s_W^2}\left( \frac{\d m_W}{m_W}-\frac{\d
m_Z}{m_Z}\right)\right]\,a^{\sf}_{ij}+ 8 e_f c_W^2\left( \frac{\d
m_W}{m_W}-\frac{\d m_Z}{m_Z}\right)\,\d_{ij}\,,
\end{eqnarray}
where the contributions containing $\d\tsf$ were intentionally
left out and will be discussed below.\\
%
%
\subsubsection{Renormalization of electric charge}
The standard on-shell input value for the electric charge is the
one in the Thomson limit $\alpha \equiv e^2/(4 \pi) = 1/137.036$.
This corresponds to a counterterm
\begin{eqnarray}
\frac{\d e}{e} &=& \frac{1}{2}\,\d Z_{\g \g}-\frac{s_W}{2 c_W}\,\d
Z_{Z\g}\,.
\end{eqnarray}
In this way of fixing the electric charge has a significant
theoretical uncertainty coming from the light quarks which we
circumvent by using as input parameter for $\a$ the $\overline{\rm
MS}$ value at the $Z$-pole, $\a \equiv \a(m_{Z})|_{\overline{\rm
MS}} = e^2/(4\pi)$. The counterterm then is given by
\cite{chrislet, 0111303,wcharge}
\begin{eqnarray}\non
  \frac{\d e}{e} &=& \frac{1}{(4\pi)^2}\,\frac{e^2}{6} \Bigg[
  \,4 \sum_f N_C^f\, e_f^2 \bigg(\D + \log\frac{Q^2}{x_f^2} \bigg)
  + \sum_{\sf} \sum_{m=1}^2 N_C^f\, e_f^2
  \bigg( \D + \log\frac{Q^2}{m_{\sf_m}^2} \bigg)
  \\
  && \hspace{18mm}
  +\,4 \sum_{k=1}^2 \bigg( \D + \log\frac{Q^2}{m_{\chp_k}^2} \bigg)
  + \bigg( \D + \log\frac{Q^2}{m_{H^+}^2} \bigg)
  - 21 \bigg( \D + \log\frac{Q^2}{m_{W}^2} \bigg)
  \Bigg]\,,
\end{eqnarray}
with $x_f = m_Z \ \forall\ m_f < m_{Z}$ and $x_t = m_t$.  $N_C^f$
is the color factor, $N_C^f = 1, 3$ for (s)leptons and (s)quarks,
respectively. $\D$ denotes the UV divergence factor, \mbox{$\D =
2/\epsilon - \g + \log 4\pi$}.
%
%
\subsubsection{Renormalization of $m_W$ and $m_Z$}
The masses of the $Z$-boson and the $W$-boson are fixed as the
physical (pole) masses, i.~e.
\begin{eqnarray}
\d m_Z^2 \,=\, \Re\,\Pi^T_{ZZ}(m_Z^2)\,,\qquad \d m_W^2 \,=\,
\Re\,\Pi^T_{WW}(m_W^2)\,,
\end{eqnarray}
where
\begin{eqnarray}
\frac{\d m_Z}{m_Z} \,=\, \onehfb\frac{\d m_Z^2}{m_Z^2}\,,\qquad
\frac{\d m_W}{m_W} \,=\, \onehfb\frac{\d m_W^2}{m_W^2}\,.
\end{eqnarray}
\\
The formulas for the vector boson self-energies
$\Pi^T_{WW}(m_W^2)$ and $\Pi^T_{ZZ}(m_Z^2)$ are given in Appendix
\ref{appSE} and in \cite{chrislet}. The counterterms for the
intermediate boson masses are used to determine the Weinberg angle
fixing according to \cite{Sirlin}.
%
%
\subsubsection{ Renormalization of $\theta_{\!\sf}$}
The counterterm of the sfermion mixing angle, $\d\theta_{\!\sf}$,
is fixed such that it cancels the anti-hermitian part of the
sfermion wave-function corrections \cite{Yukawa, guasch},
\begin{eqnarray}\label{dthetasf}
   \delta \theta_{\sf} & = & \frac{1}{4}\, \left(
   \d Z_{12} - \d Z_{21}\right)
   \,=\, \frac{1}{2\big(m_{\sf_1}^2 \!-\! m_{\sf_{2}}^2\big)}\, \Re\!
   \left( \Pi_{12}^\sf(m_{\sf_{2}}^2) + \Pi_{21}^\sf
   (m_{\sf_{1}}^2)
    \right) \,.
\end{eqnarray}
Including the terms proportional to $\d\theta_{\!\sf}$ in
Eq.~(\ref{symwave}) is equivalent to symmetrizing the off-diagonal
sfermion wave-function corrections in Eq.~(\ref{offdiag}) as
\cite{sche2,sche3}
\begin{eqnarray}
\d Z_{12}~=~\d Z_{21}~=~ \frac{\Re\,\Pi_{12}^{\sf}
(m_{\sf_2}^2)-\Re\,\Pi_{21}^{\sf}
(m_{\sf_1}^2)}{\msf{1}^2-\msf{2}^2}\,.
\end{eqnarray}
This fixing of the counterterm for the mixing angle is analogous
to the renormalization of the CKM matrix in \cite{CKM} and
similarly has to be made gauge-independent. It was shown in
\cite{yamada} that this can be avoided or, equivalently, the
result in the $\xi = 1$ gauge can be regarded as the
gauge-independent one.

%
%
\section{Real photon corrections}\label{realcorr}
\begin{figure}[th]
\begin{picture}(160,25)(0,0)
     \put(0,0){\mbox{\resizebox{15.5cm}{!}
     {\includegraphics{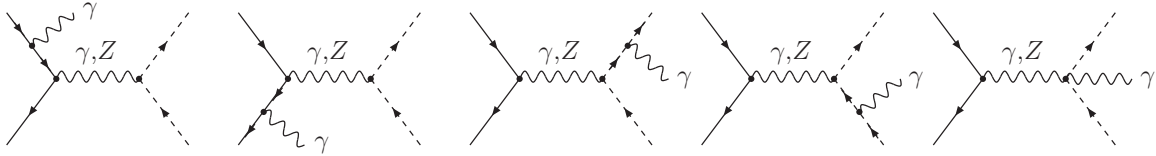}}}}
\end{picture}
\caption{Bremsstrahlung diagrams relevant to the calculation of
the real photon corrections to $e^+ e^- \rightarrow \tilde{f}_i \
{\bar{\!\!\tilde{f}}}_{\!j}$. \label{rads}}
\end{figure}

Similarly to the QCD case where the cross-section was IR-divergent
due to massless gluons \cite{QCD1, QCD2, SUSY-QCD-A, SUSY-QCD-H},
the one-loop cross-section $\s^{\rm ren}(e^+e^- \rightarrow
\tilde{f}_i \ {\bar{\!\!\tilde{f}}}_{\!j})$ is IR-divergent owing
to the diagrams with photon exchange where the photon mass is
zero. This is remedied by introducing a small mass $\lambda$ and
including also the Bremsstrahlung process i.~e. $\s(e^+e^-
\rightarrow \tilde{f}_i \ {\bar{\!\!\tilde{f}}}_{\!j}\g)$, see
Fig.~\ref{rads}. Summing these two contributions yields an
IR-finite result for the physical value $\lambda=0$,
\begin{eqnarray}\label{radiat}
\s^{\rm corr}(e^+e^- \rightarrow \tilde{f}_i \
{\bar{\!\!\tilde{f}}}_{\!j}) &=& \s^{\rm ren}(e^+e^- \rightarrow
\tilde{f}_i \ {\bar{\!\!\tilde{f}}}_{\!j})+\s(e^+e^- \rightarrow
\tilde{f}_i \ {\bar{\!\!\tilde{f}}}_{\!j}\g)\,.
\end{eqnarray}
To calculate the radiative cross-section $\s(e^+e^- \rightarrow
\tilde{f}_i \ {\bar{\!\!\tilde{f}}}_{\!j}\g)$ we use the
phase-space splicing method \cite{slicing1} which splits the
bremsstrahlung phase-space into 3 regions. The corresponding 3
parts are
\begin{eqnarray}
\s(e^+e^- \rightarrow \tilde{f}_i \ {\bar{\!\!\tilde{f}}}_{\!j}\g)
&=& \s^{\rm soft}(\lambda, \D E) + \s^{\rm hard}(\D E, \D \theta)
+ \s^{\rm coll}(\D E, \D \theta)\,.
\end{eqnarray}
In our calculation, we used a soft-photon approximation ($\s^{\rm
soft}$) to reproduce the divergence pattern correctly. However,
this approximation introduces a cut $\D E$ on the energy of the
radiated photon. The dependence on the cut $\D E$ drops out if we
include the full $2\rightarrow 3$ process ($\s^{\rm hard}$). In
order to get simpler expressions for $\s^{\rm hard}$ we neglect
the electron mass but then a collinear divergence occurs when the
photon is radiated in the direction of the electron and positron
beams. This collinear divergence can be regulated by introducing
yet another approximation ($\s^{\rm coll}$) for the above
mentioned phase-space region. Another cut $\D \theta$ is hereby
introduced. After summing the 3 contributions the result must be
independent of both the cuts and has to cancel the IR-divergence
of the one-loop cross-section. This is the ultimate test we have
made at the end of the calculation.
%
%
%
\subsection{Soft-photon approximation}
The soft-photon approximation supposes that the 4-momentum of the
photon is small compared to other momenta (for details see
e.~g.\cite{Denner}). Using this assumption the differential
cross-section $\left(\frac{{\rm d} \sigma}{{\rm d}
\Omega}\right)_{\rm soft}$ is proportional to the tree-level
differential cross-section. The full cross-section for polarized
beams is
\begin{eqnarray}
\s^{\rm soft}(\lambda, \D E) &=&
\frac{1}{4}(1-P_-)(1+P_+)\,\s_L^{\rm soft}
+\frac{1}{4}(1+P_-)(1-P_+)\,\s_R^{\rm soft}\,,
\end{eqnarray}
where
\begin{eqnarray}
\s_{L,R}^{\rm soft} &=& \int\left(\frac{{\rm d} \sigma_{L,R}}{{\rm
d} \Omega}\right)_{\rm soft}\,{\rm d} \Omega \,=\,
\int\left(\frac{{\rm d} \sigma_{L,R}}{{\rm d} \Omega}\right)_{\rm
tree}\delta_s\,{\rm d} \Omega\, \,.
\end{eqnarray}
The factor $\d_s$ is defined as
\begin{eqnarray}
\delta_s &=& - \frac{\alpha}{4 \pi^2}\left(I_{p_1^2} + I_{p_2^2} -
2 I_{p_1 p_2} + e_f^2 (I_{k_1^2} + I_{k_2^2} - 2 I_{k_1 k_2}) + 2
e_f (I_{p_1 k_1} + I_{p_2 k_2} - I_{p_1 k_2} - I_{p_2
k_1})\right)\,,
\end{eqnarray}
where the integrals $I_{a b}$ are defined in \cite{Denner} and
were worked out e.~g. in \cite{tHooft}. The explicit formula for
$\delta_s$ can be found in Appendix \ref{Bremint}.
%
%
\subsection{Hard and collinear photon radiation}
The cross-section for the full bremsstrahlung process $e^+(p_2)\,
e^-(p_1) \rightarrow \tilde{f}_i(k_1)\, \
{\bar{\!\!\tilde{f}}}_{\!j}(k_2)\,\gamma (k_3)$ is given by
\begin{eqnarray}
\sigma^{\rm hard}(\D E, \D \theta) &=&
\frac{1}{2s}\frac{1}{8\,(2\pi)^4}\int |\mathcal{M}|^2\,{\rm
d}k_1^0\,{\rm d}k_3^0\,{\rm d}\eta\ {\rm d}\!\cos\theta\,,
\end{eqnarray}
where the cuts $\D E$ and $\D \theta$ appear in the integration
bounds of ${\rm d}k_3^0$ and ${\rm d}\!\cos\theta$. The angle
$\eta$ is defined as in \cite{feyn}. The explicit form of the
squared matrix element is given in Appendix \ref{Bremint} and the
integral is evaluated numerically using the routines from the CUBA
library \cite{cuba}.
\newline As we have neglected the electron mass in the calculation
of $\sigma^{\rm hard}$, we have to take an another approach in the
collinear region of the phase-space. We follow the approach of
\cite{slicing1, slicing2} and get for the collinear cross-section
the following expression\,,
\begin{eqnarray}\non
\s^{\rm coll}(\D E, \D \theta) &=& \frac{1}{4}\left[
(1-P_-)(1-P_+)\,\s_{LL}^{\rm coll} + (1-P_-)(1+P_+)\,\s_{LR}^{\rm
coll} \right. \\ &&\left. \hspace{2cm} +
(1+P_-)(1-P_+)\,\s_{RL}^{\rm coll} + (1+P_-)(1+P_+)\,\s_{RR}^{\rm
coll} \right]\,,
\end{eqnarray}
where all polarization states $(\s_{LL}^{\rm coll}, \s_{LR}^{\rm
coll}, \s_{RL}^{\rm coll}, \s_{RR}^{\rm coll})$ appear. This is
due to the radiation of an additional photon and the fact that the
electron is massive ($\s_{LR}$ stands for the cross-section with
left-handed electrons and right-handed positrons etc.). The single
polarization states are given by
\begin{eqnarray}
\sigma_{LL,RR}^{\rm coll} &=&
\frac{e^2}{8\pi^2}\int_{x_{min}}^{x_{max}} {\rm d}x\;
\,x\,\left[\s^{\rm tree}_{L}\big( (1-x)s \big) + \s^{\rm
tree}_{R}\big( (1-x)s \big)\right]\,,
\\ \
\sigma_{LR,RL}^{\rm coll} &=&
\frac{e^2}{8\pi^2}\int_{x_{min}}^{x_{max}} {\rm d}x\; 2\,
\frac{x^2-2x+2}{x}\,\left[ \log\left(
\frac{s\D\theta^2}{4m_e^2}\right)-1\right]\s^{\rm tree}_{L,R}\big(
(1-x)s \big)\,,
\end{eqnarray}
with
\begin{eqnarray}
x_{min} \,=\, 2\D E/\sqrt{s}\,,\qquad\qquad x_{max} \,=\, 1-
\frac{(m_{\sf_i}+m_{\sf_j})^2}{s}\,.
\end{eqnarray}
After including all the above-mentioned contributions we arrive at
a cut-independent result.
%
\subsection{Higher order corrections}
Substantial correction from the collinear photon radiation is due
to the smallness of the electron mass compared to a typical energy
scale in the process. This effect is such that to reach the
collider precision one has to include also the leading higher
order corrections (i.~e. beyond ${\cal O}(\alpha)$). Owing to the
mass-factorization theorem, one can factorize the corrections in
the leading-log (LL) approximation as
\begin{eqnarray}\label{univ}
\int\!d\s^{\rm tree}+\int\!d\s^{\rm LL} &=& \int_0^1\!
dx_1\!\int_0^1\!dx_2\,\G_{ee}^{\rm LL}(x_1,Q^2) \G_{ee}^{\rm
LL}(x_2,Q^2)\,\int\!\!d\s^{\rm tree}(x_1p_1,x_2p_2)\,.
\end{eqnarray}
where $x_1$, $x_2$ are the momentum fractions of the electron and
the positron carried after the radiation of the photon(s).
\newline
The $\G^{\rm LL}_{ee}(x,Q^2)$ is the leading-log structure
function up to ${\cal O}(\a^3)$, given in ref.\cite{Sk90},
\begin{eqnarray}\label{GLL}
\G^{\rm LL}_{ee}(x,Q^2) &=&
\frac{\exp(-\frac{1}{2}\b\,\g_E+\frac{3}{8}\b)}{\G(1+\frac{\b}{2})}
\frac{\b}{2}(1-x)^{\frac{\b}{2}-1}\nonumber
\\
&&-\frac{\b}{4}(1+x)+\frac{\b^2}{16}\Big(-2(1+x)
\log(1-x)-\frac{2\log x}{1-x}+\frac{3}{2}(1+x)\log
x-\frac{x}{2}-\frac{5}{2}\Big)\nonumber
\\
&&+ \frac{\b^3}{8} \left[-\frac{1}{2}(1+x) \left(\frac{9}{32}-
\frac{\pi^2}{12}+
\frac{3}{4}\log(1-x)+\frac{1}{2}\log^2(1-x)-\frac{1}{4} \log
x\log(1-x) \right. \right.\nonumber
\\
&&\left. \left.+\frac{1}{16}\log^2x-\frac{1}{4}{\rm
Li}_2(1-x)\right)+\frac{1}{2}\frac{1+x^2}{1-x}
\Big(-\frac{3}{8}\log x+\frac{1}{12}\log^2x - \frac{1}{2}\log
x\log(1-x)\Big)\right.\nonumber
\\
&&\left.-\frac{1}{4}(1-x)\Big(\log(1-x) +
\frac{1}{4}\Big)+\frac{1}{32}(5-3x)\log x\right]\,,
\end{eqnarray}
with the gamma function $\G$, the Euler constant $\g_E\sim
0.577216$, and
\mbox{$\b=\frac{2\a}{\pi}(\log\frac{Q^2}{m_e^2}-1)$}. For the free
scale $Q^2$ we take the typical energy of the process $s$. The
soft-photon contributions were summed up to all orders in the
perturbation series.\newline The structure function (\ref{GLL})
contains not only the higher orders beginning with ${\cal
O}(\a^2)$ but also parts of terms ${\cal O}(\a)$ already included
elsewhere. To avoid double counting we subtract these terms as in
\cite{slicing2}.
\noindent
\section{Numerical analysis}\label{numerics}
\begin{figure}[h!]
\begin{center}
\mbox{\mbox{\resizebox{80mm}{!}{\includegraphics{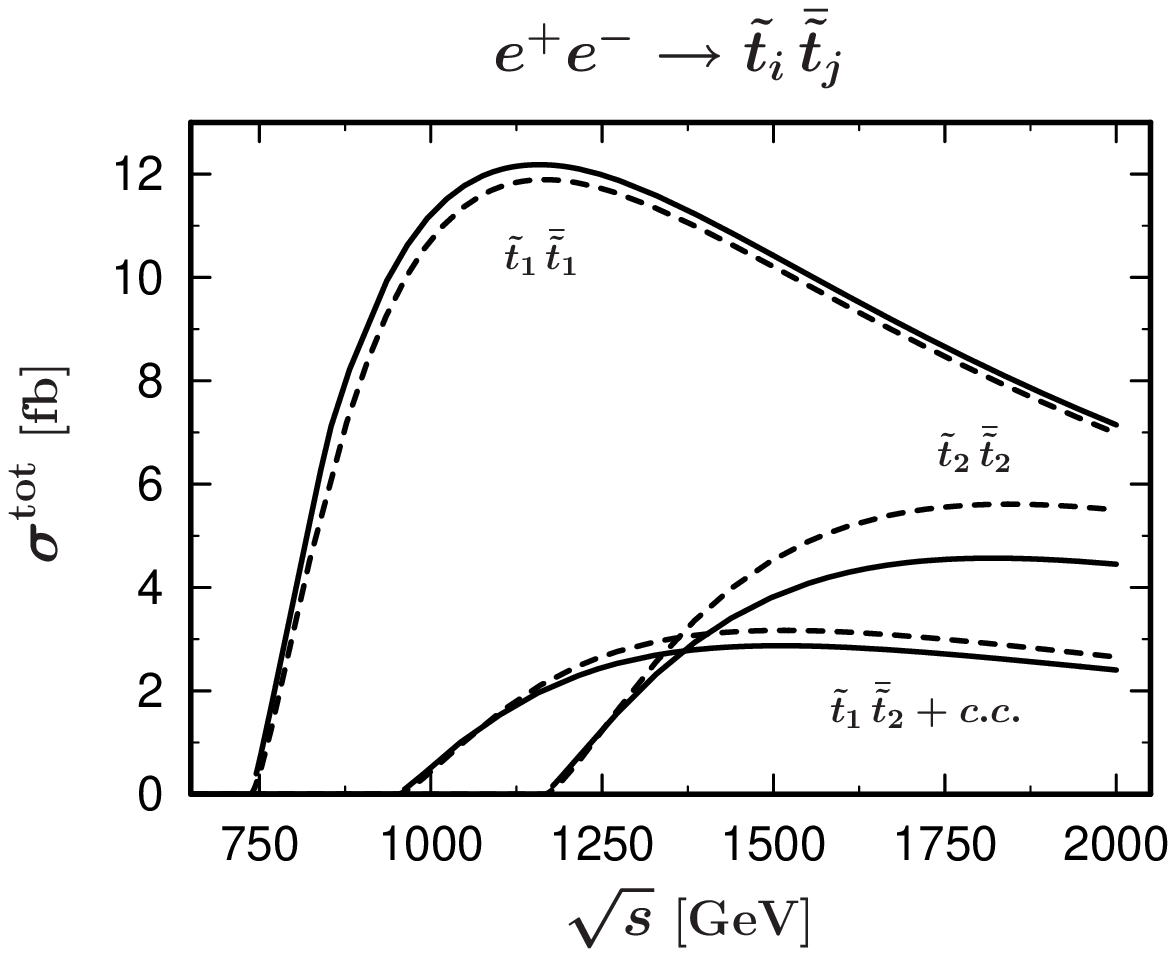}}}\hspace{2mm}
\mbox{\resizebox{80mm}{!}{\includegraphics{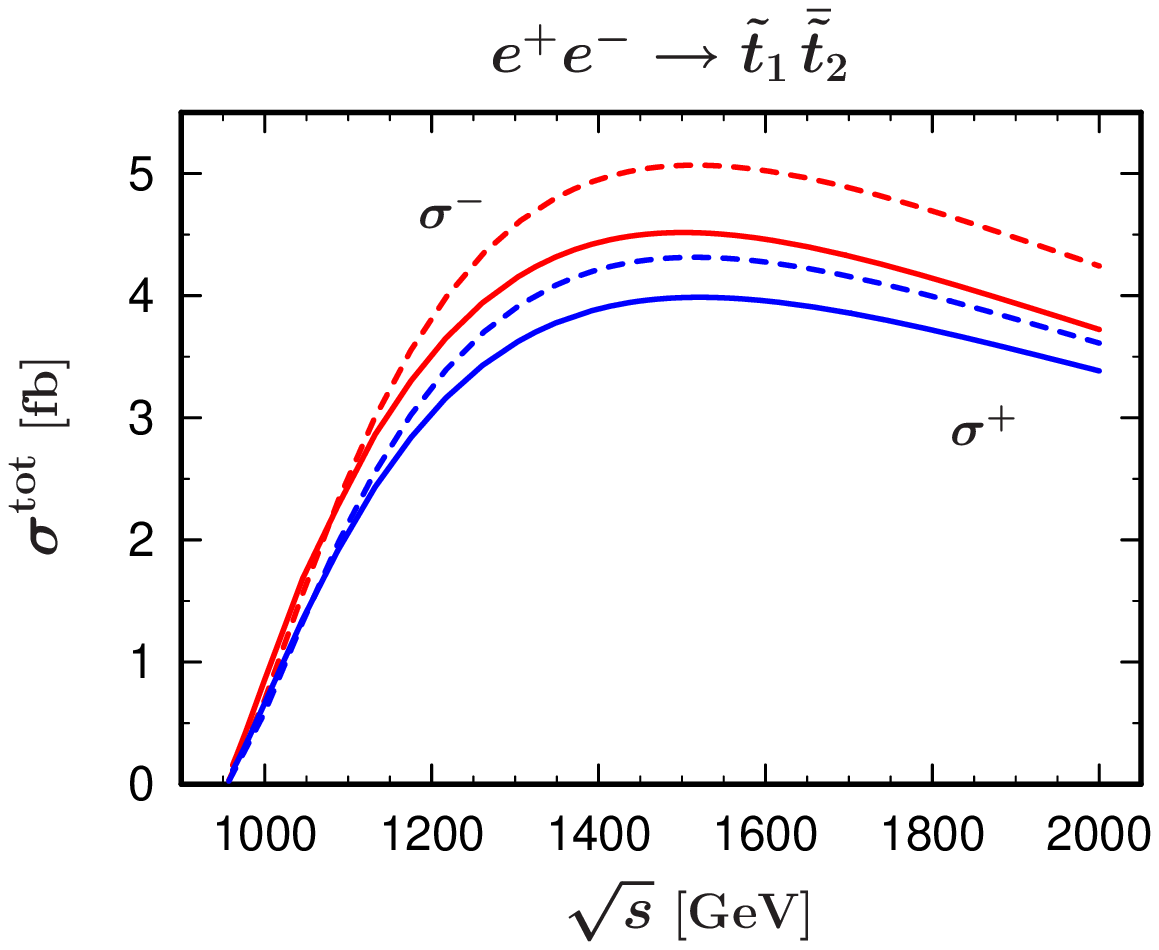}}}}
\end{center}
\caption{Left: The total cross-sections for stop pair production
(all channels) at tree-level \{dashed\} and with complete
corrections \{solid\}. Right: The total cross-sections for
polarized beams: $\sigma^-$ stands for $P_- = -0.8,\, P_+ = 0.6$
and $\sigma^+$ for $P_- = 0.8,\, P_+ = -0.6$.}\label{figstop}
\end{figure}

\begin{figure}[h!]
\begin{center}
\mbox{\mbox{\resizebox{80mm}{!}{\includegraphics{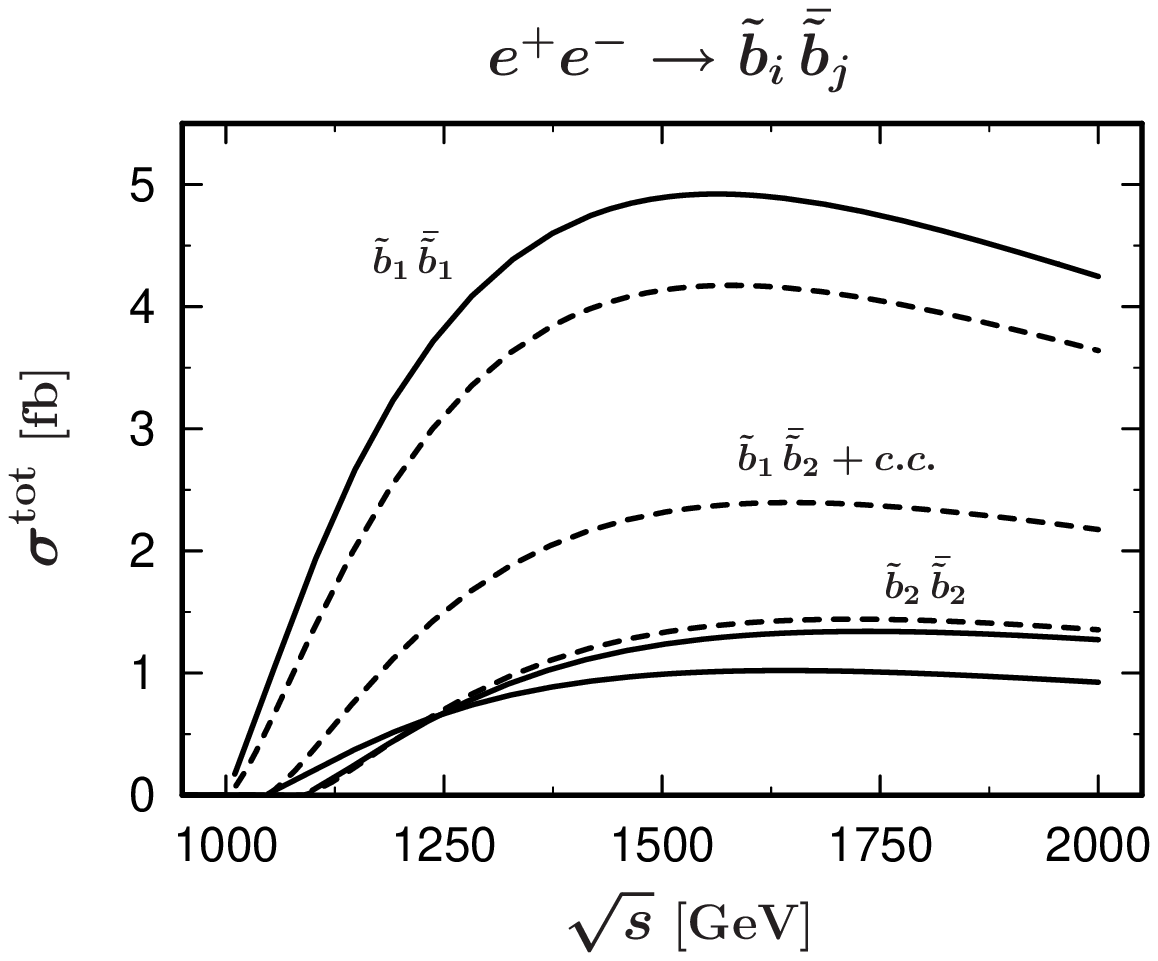}}}\hspace{2mm}
\mbox{\resizebox{80mm}{!}{\includegraphics{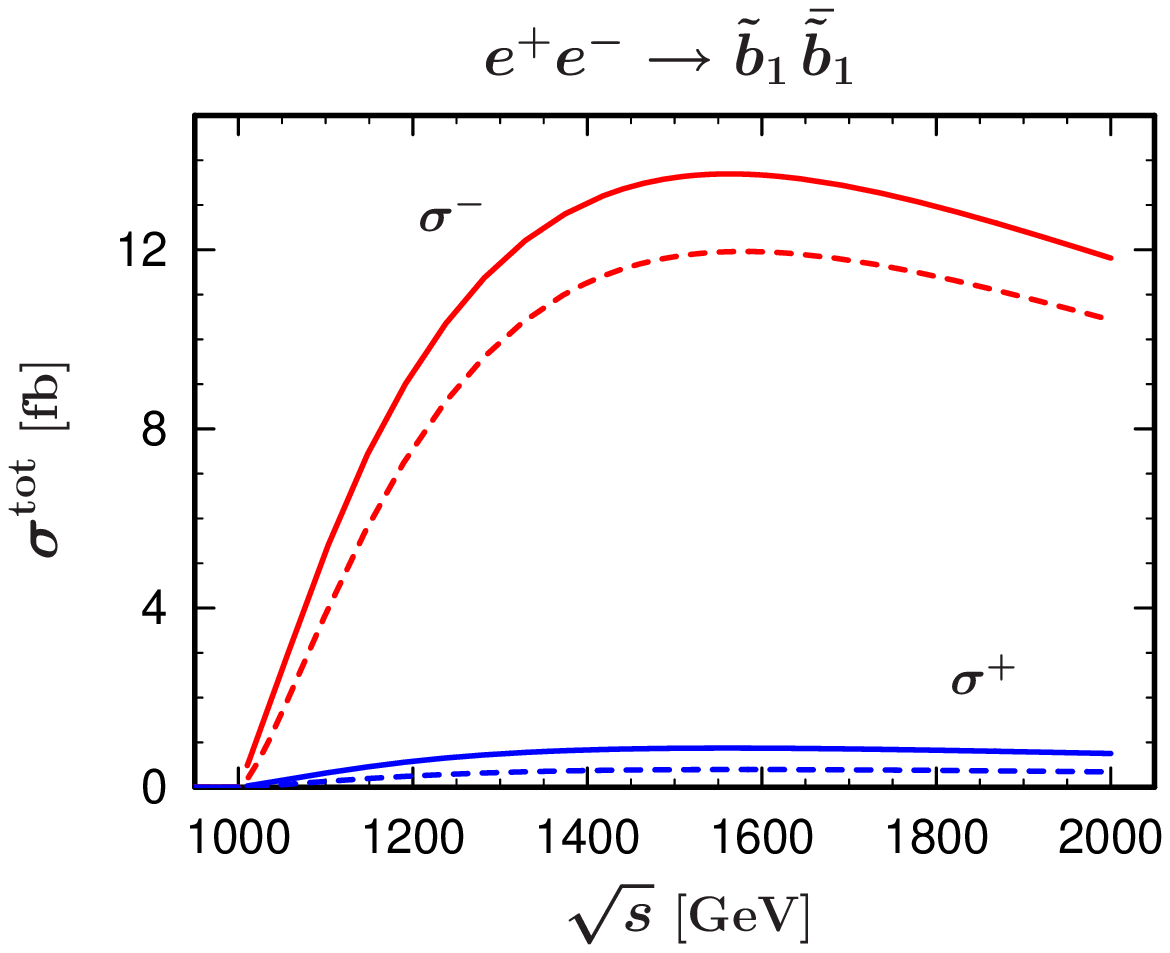}}}}
\end{center}
\caption{Left: The total cross-sections for sbottom pair
production (all channels) at tree-level \{dashed\} and with
complete corrections \{solid\}. Right: The total cross-sections
for polarized beams $\sigma^-$ stands for $P_- = -0.8,\, P_+ =
0.6$ and $\sigma^+$ for $P_- = 0.8,\, P_+ = -0.6$.
}\label{figsbot}
\end{figure}

\begin{figure}[h!]
\begin{center}
\mbox{\mbox{\resizebox{80mm}{!}{\includegraphics{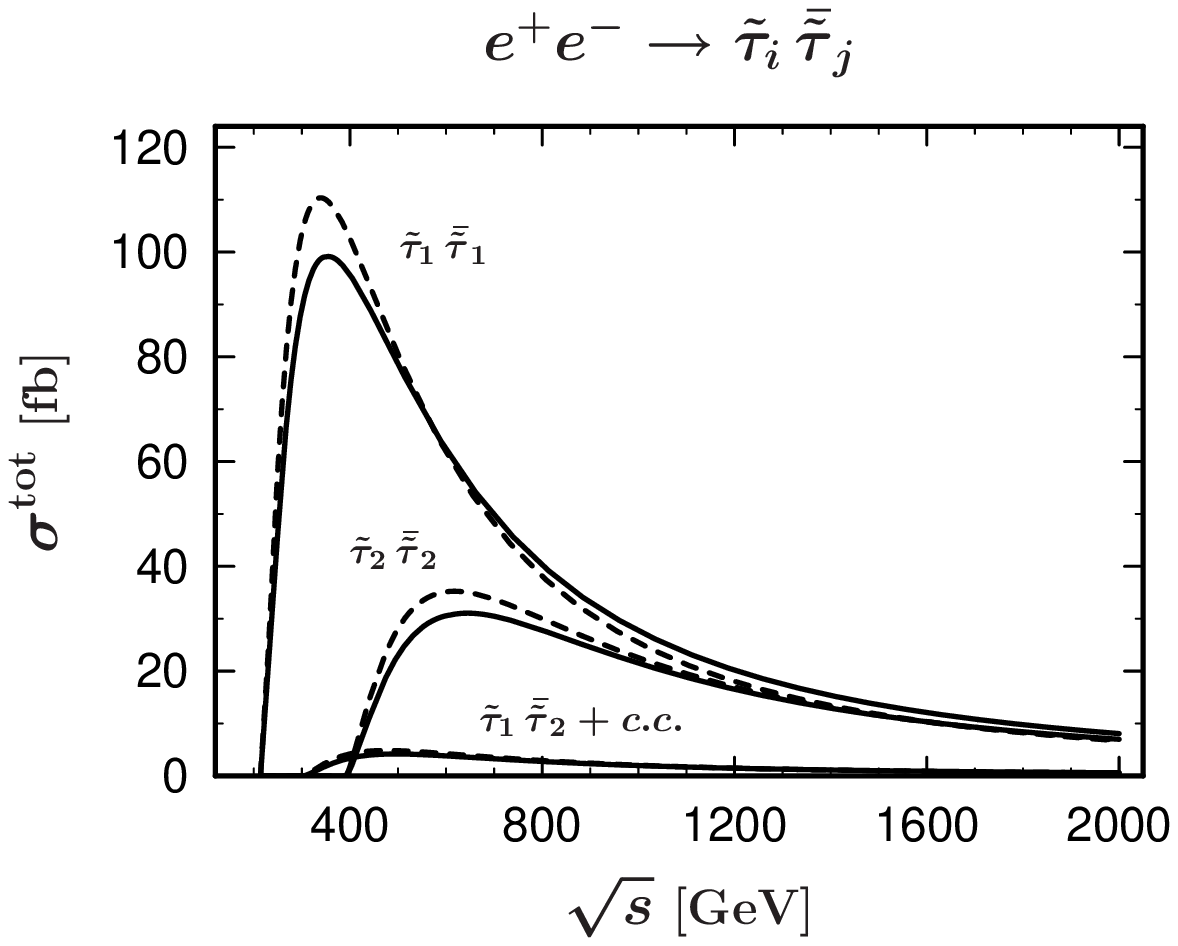}}}\hspace{2mm}
\mbox{\resizebox{80mm}{!}{\includegraphics{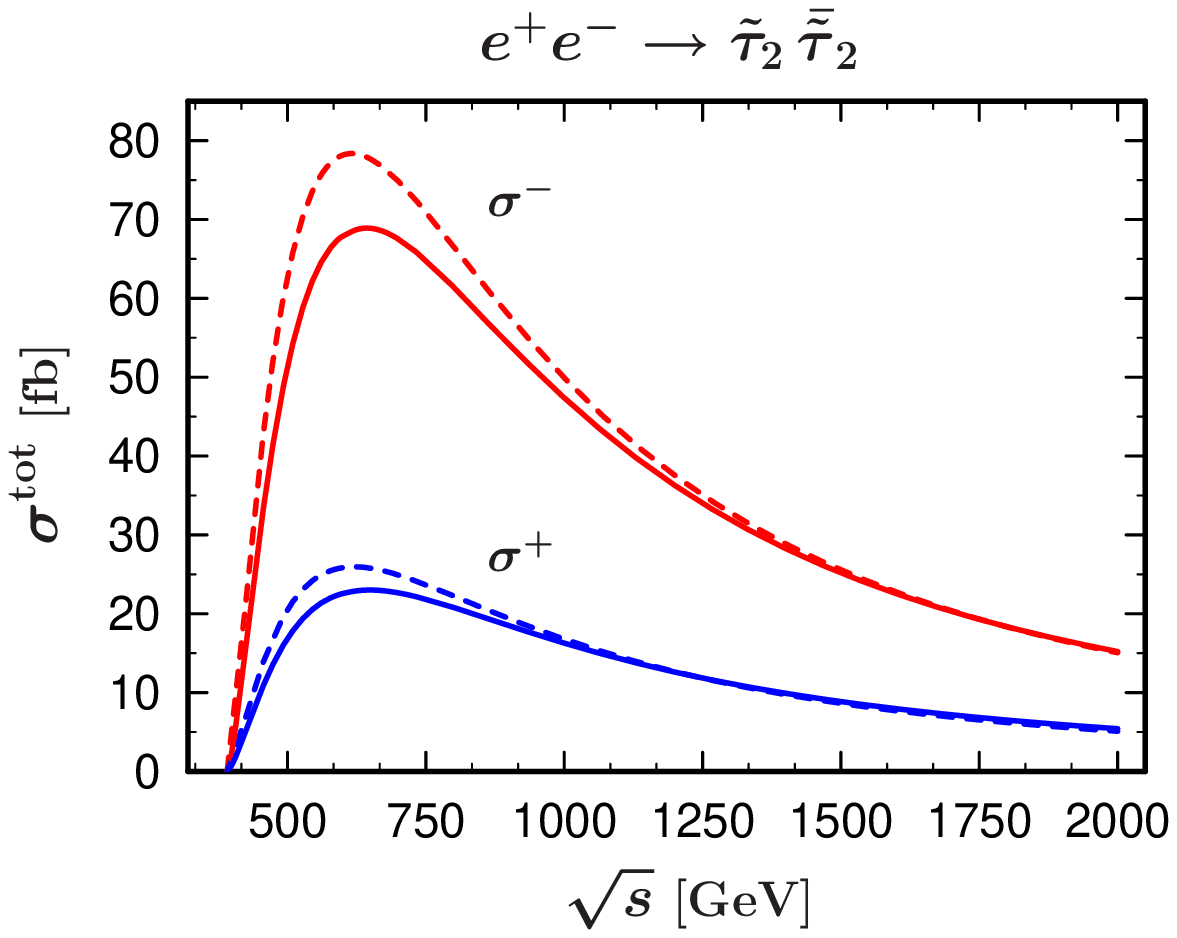}}}}
\end{center}
\caption{Left: The total cross-sections for stau pair production
(all channels) at tree-level \{dashed\} and with complete
corrections \{solid\}. Right: The total cross-sections for
polarized beams $\sigma^-$ stands for $P_- = -0.8,\, P_+ = 0.6$
and $\sigma^+$ for $P_- = 0.8,\, P_+ = -0.6$.}\label{figstau}
\end{figure}

\begin{figure}[h!]
\begin{center}
\mbox{\mbox{\resizebox{80mm}{!}{\includegraphics{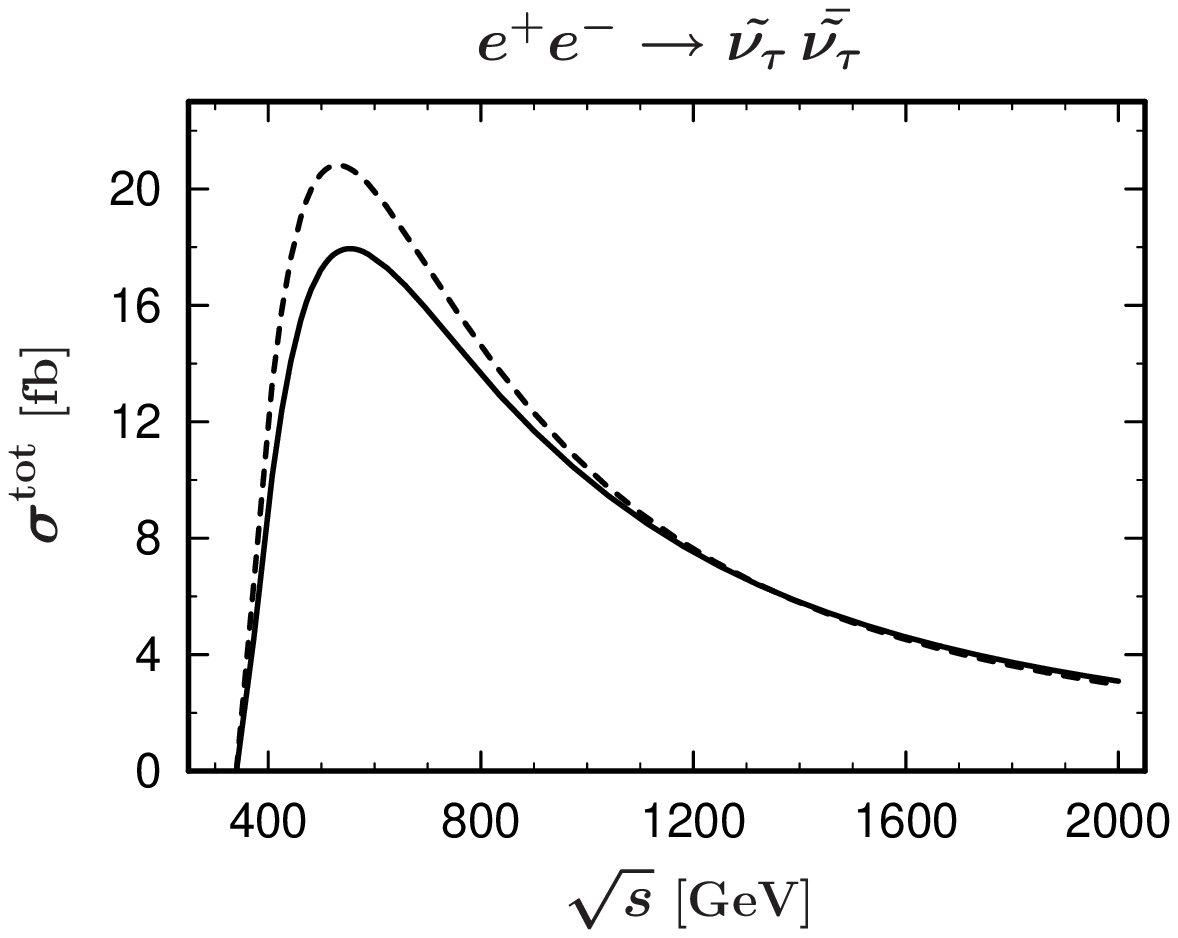}}}\hspace{2mm}
\mbox{\resizebox{80mm}{!}{\includegraphics{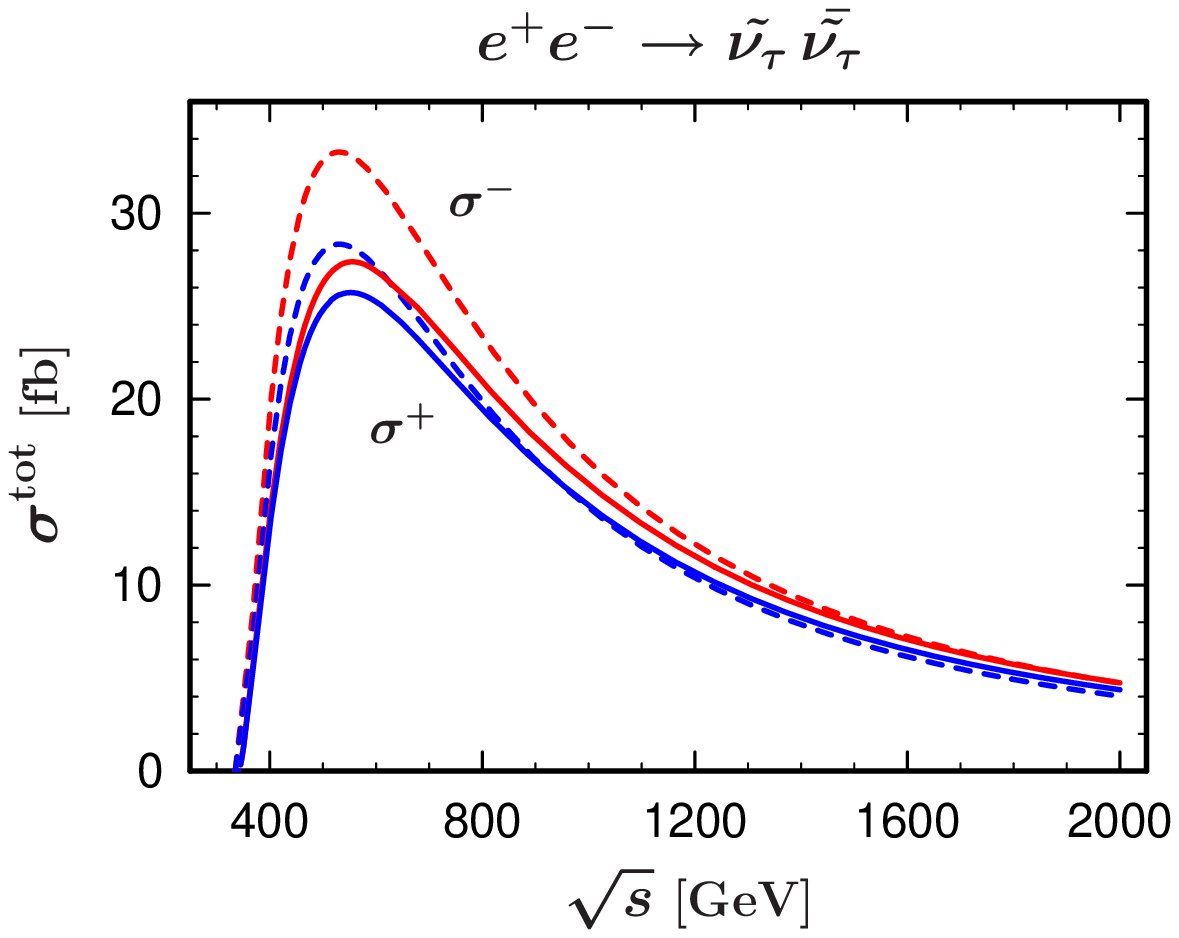}}}}
\end{center}
\caption{Left: The total cross-sections for tau-sneutrino pair
production at tree-level \{dashed\} and with complete corrections
\{solid\}. Right: The total cross-sections for polarized beams
$\sigma^-$ stands for $P_- = -0.8,\, P_+ = 0.6$ and $\sigma^+$ for
$P_- = 0.8,\, P_+ = -0.6$.}\label{figstaune}
\end{figure}

\begin{figure}[h!]
\begin{center}
\mbox{\mbox{\resizebox{80mm}{!}{\includegraphics{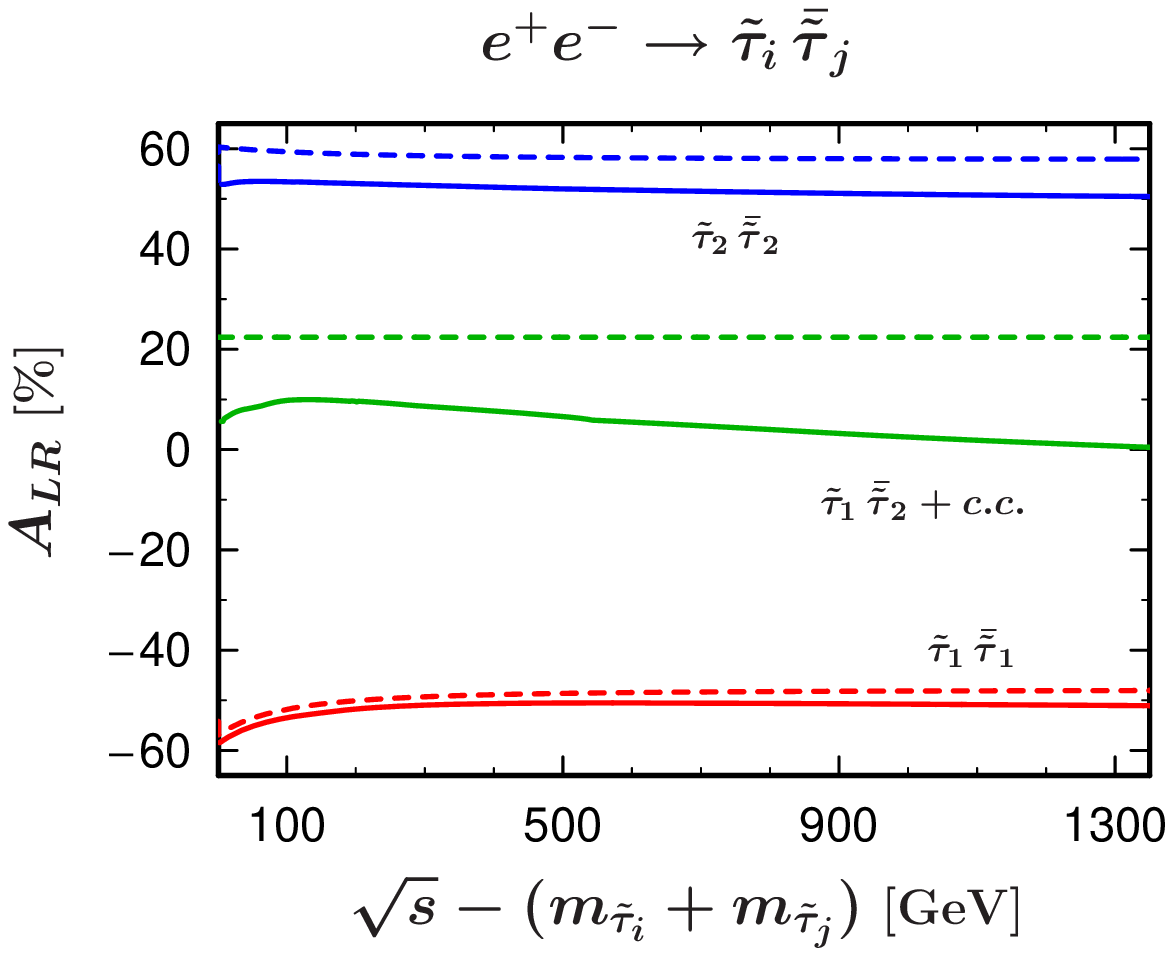}}}\hspace{2mm}
\mbox{\resizebox{80mm}{!}{\includegraphics{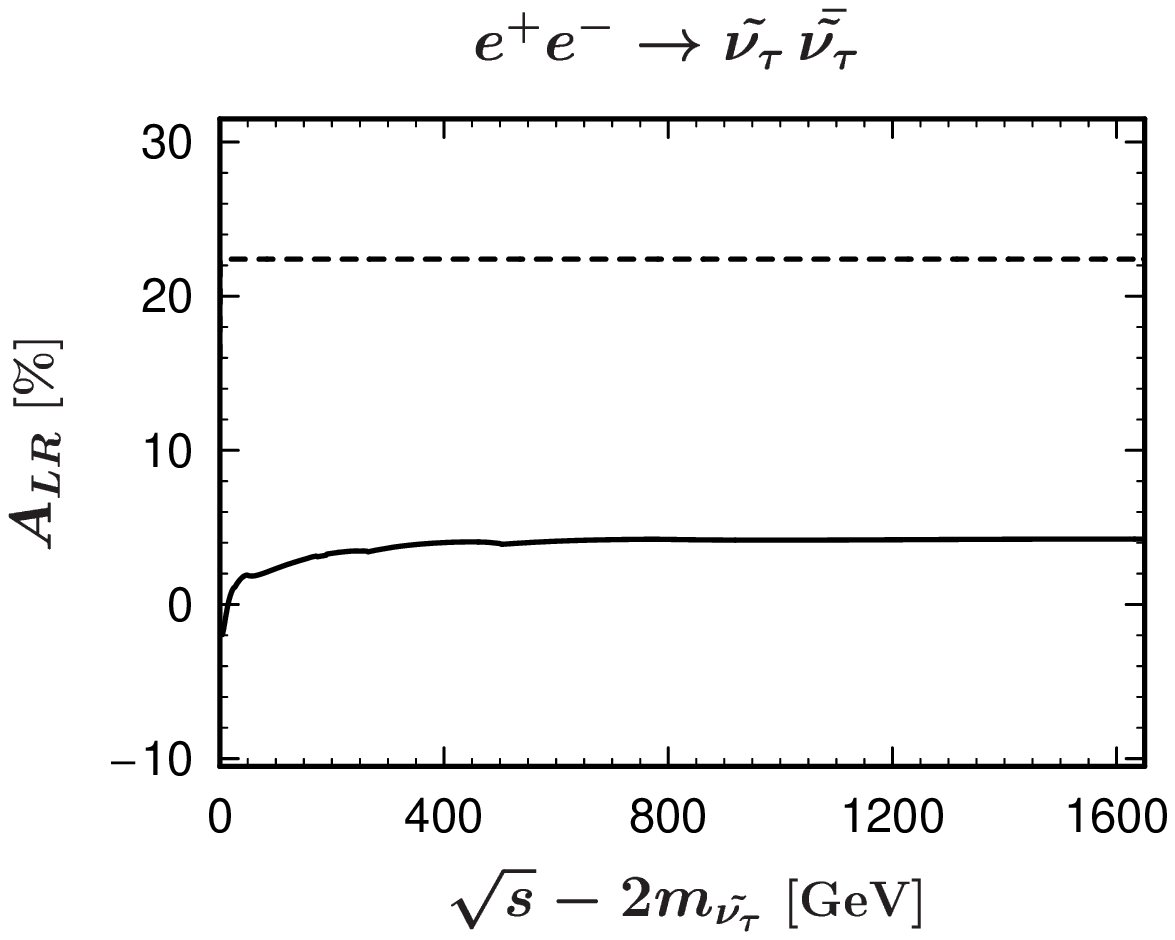}}}}
\end{center}
\caption{Left: The left-right asymmetry for stau pair production
at tree-level \{dashed\} and with complete corrections \{solid\}.
Right: The left-right asymmetry for tau-sneutrino pair production
at tree-level \{dashed\} and with complete corrections
\{solid\}.}\label{alrstau_ne}
\end{figure}

\begin{figure}[h!]
\begin{center}
\mbox{\mbox{\resizebox{80mm}{!}{\includegraphics{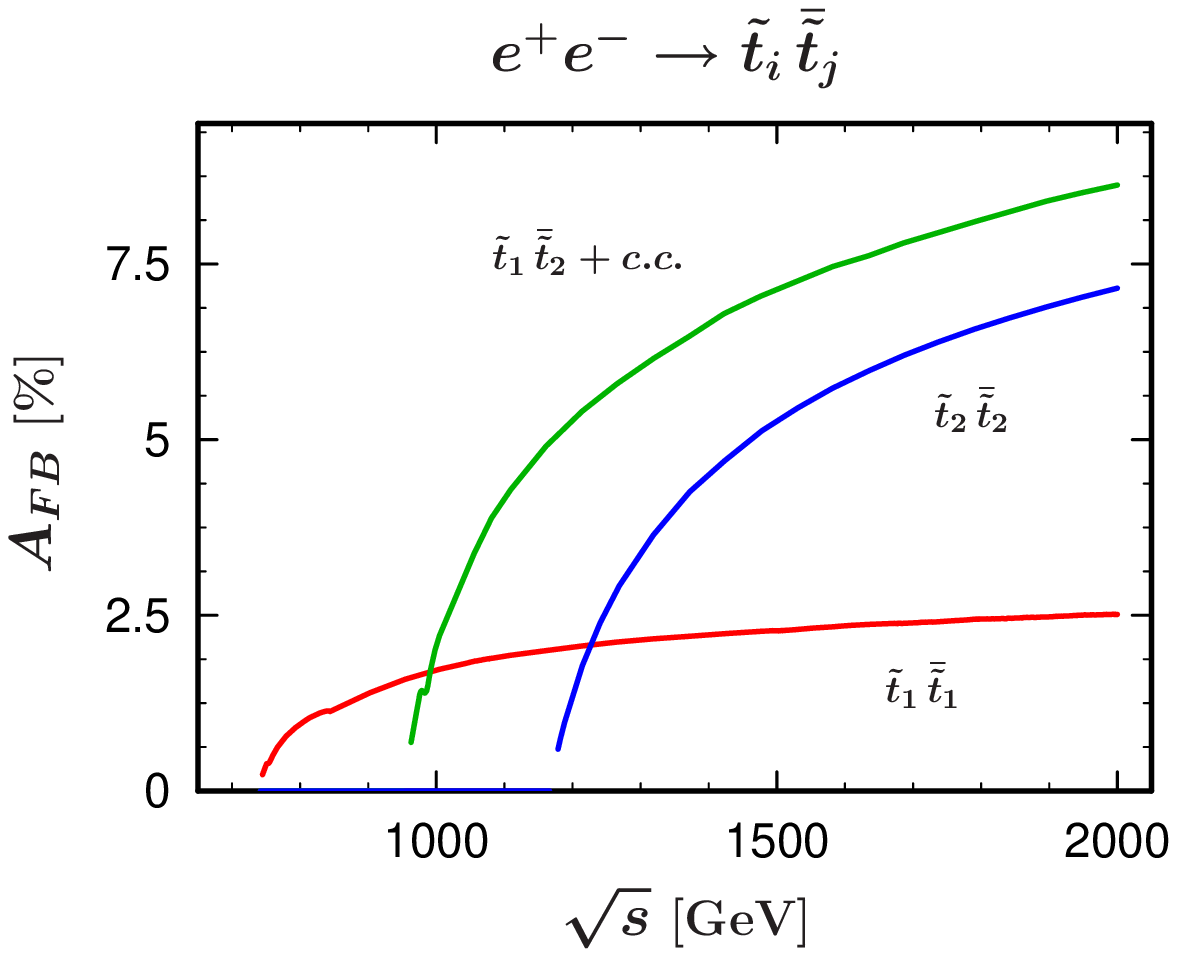}}}\hspace{2mm}
\mbox{\resizebox{80mm}{!}{\includegraphics{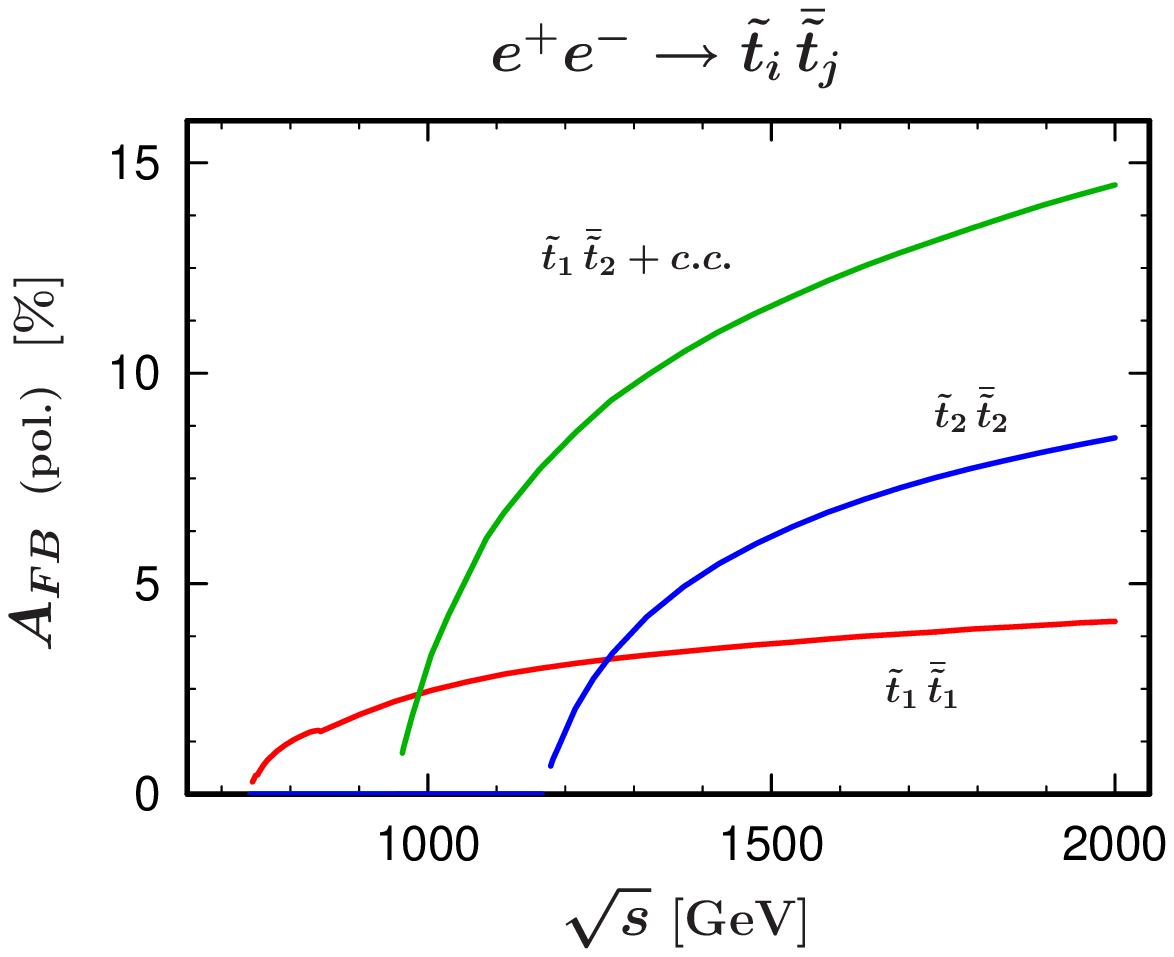}}}}
\end{center}
\caption{Left: The forward-backward asymmetry for stop pair
production (no tree-level contribution) for unpolarized beams.
Right: The forward-backward asymmetry for stop pair production
with polarized beams $P_- = -0.8,\, P_+ = 0.6$.}\label{afbstop}
\end{figure}

\begin{figure}[h!]
\begin{center}
\mbox{\mbox{\resizebox{80mm}{!}{\includegraphics{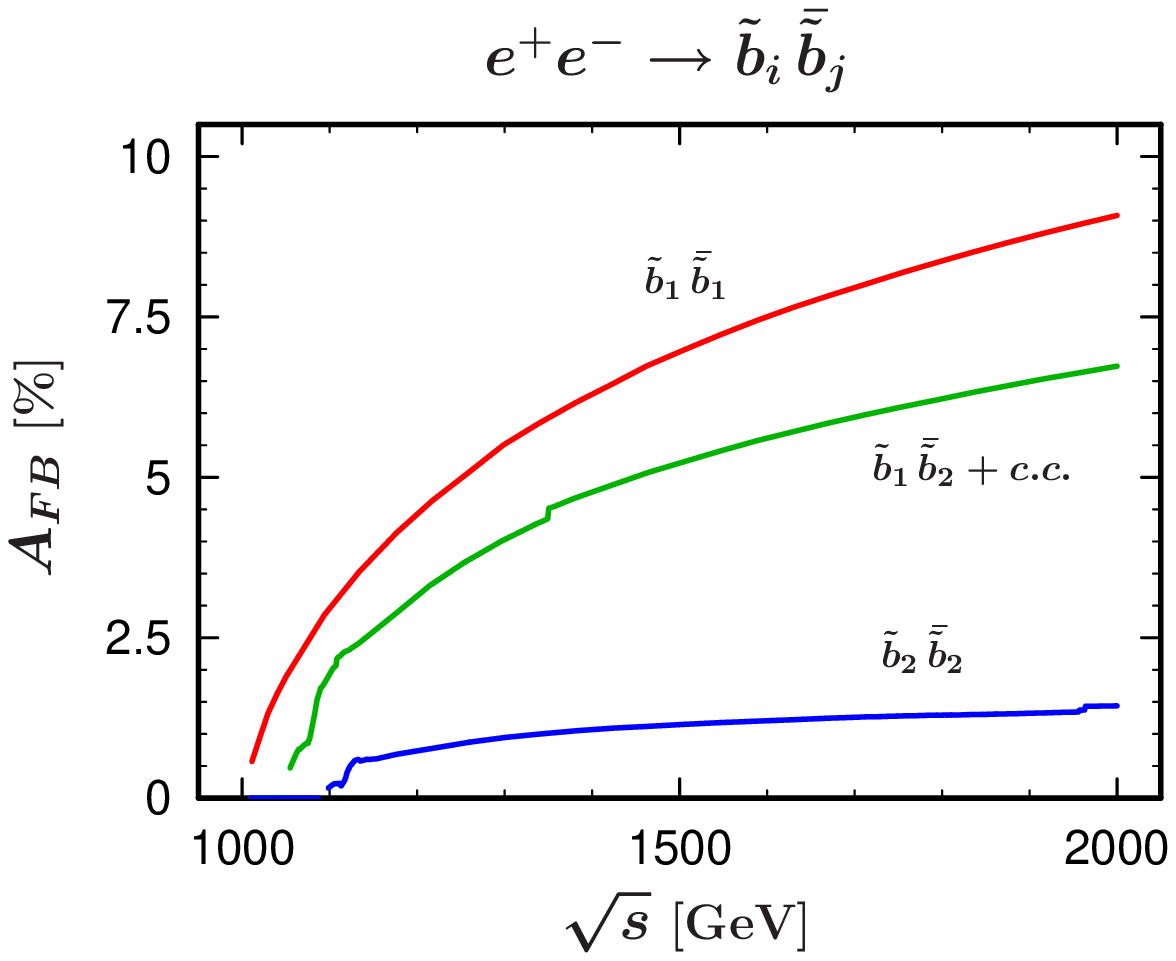}}}\hspace{2mm}
\mbox{\resizebox{80mm}{!}{\includegraphics{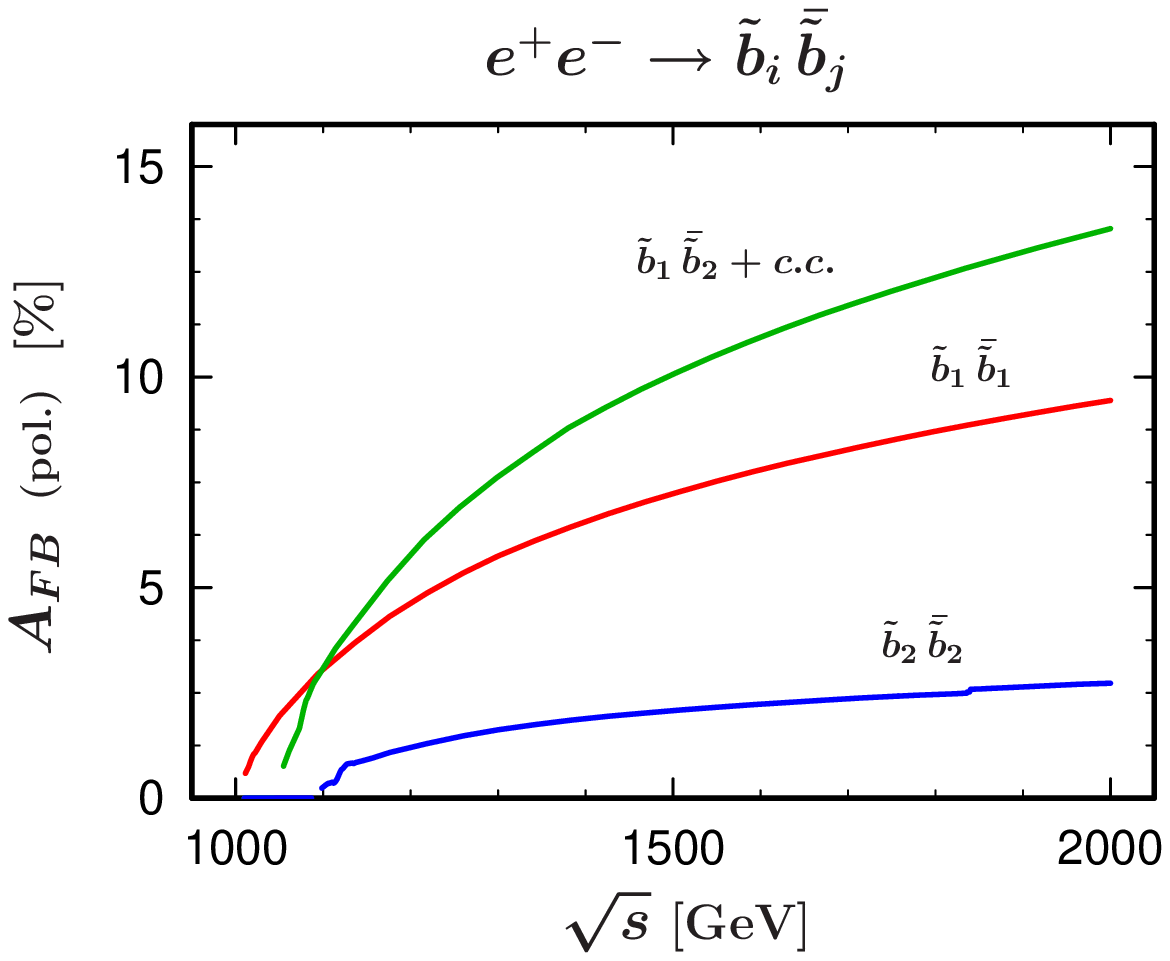}}}}
\end{center}
\caption{Left: The forward-backward asymmetry for sbottom pair
production (no tree-level contribution) for unpolarized beams.
Right: The forward-backward asymmetry for sbottom pair production
with polarized beams $P_- = -0.8,\, P_+ = 0.6$.}\label{afbsbot}
\end{figure}

In contrast to \cite{letter}, we do not attempt to make a scan
over a large area of MSSM parameter space but rather consider only
one benchmark point in the numerical analysis. It is the SPS1a'
point we use as input which is defined in the SPA project
\cite{SPA}. The point is chosen such that it satisfies all the
precision data and both the bounds for the masses of the SUSY
particles and the bounds from cosmology.
\newline The input parameters for the SPS1a' point are defined in
the \drbar scheme at the scale $Q=1\tev$. As we use the on-shell
renormalization scheme, we have to transform the SPS1a' input
parameters ${\cal P}$ into on-shell parameters ${\cal P}^{\rm
OS}$. This transformation is simply performed by subtracting the
corresponding counterterms i.~e. $ {\cal P}^{\rm OS} = {\cal
P}(Q)-\d {\cal P}(Q)$ and the results for the relevant parameters
are listed in Table~\ref{tab:SPS1ap}. All other parameters do not
enter in the calculation at tree-level and so the differences when
using the on-shell or the \drbar value are of a higher order. A
further remark is necessary here. One of the parameters not
entering the tree-level directly is the infamous $A_b$ parameter.
Fortunately, the parameter set taken here causes all the on-shell
input parameters to be insensitive to the problems of the $A_b$
on-shell definition.
\renewcommand{\arraystretch}{1.3}
\begin{table}[h!]
\begin{center}
\begin{tabular}{|c||c|c|}
   \hline
  ${\cal P}$ & \drbar & {\rm OS} \\ \hline \hline
  $\mu$ & 402.87 & 399.94\\
  $\tan\b$ & 10 & 10.31\\
  $M_{\tilde{Q}_3}$ & 471.26 & 507.23\\
  $M_{\tilde{U}_3}$ & 384.59 & 410.11\\
  $M_{\tilde{D}_3}$ & 501.35 & 538.92\\
  $M_{\tilde{E}_3}$ & 109.87 & 111.58\\
  $M_{\tilde{L}_3}$ & 179.49 & 181.78\\
  \hline
\end{tabular}
\hspace{1cm}
\begin{tabular}{|c||c|}
   \hline
  ${\cal M}$ & {\rm OS} \\ \hline \hline
  $\mst{1}$ & 368.6 \\
  $\mst{2}$ & 583.1 \\
  $\msb{1}$ & 499.9 \\
  $\msb{2}$ & 543.7 \\
  $\mstau{1}$ & 107.5 \\
  $\mstau{2}$ & 195.4 \\
  $m_{\tilde{\nu}_{\tau}}$ & 170.7 \\
  \hline
\end{tabular}
\end{center}
\caption[SPS1ap]{Parameters of the SPS1a' scenario in the \drbar
and on-shell scheme and particle masses.}\label{tab:SPS1ap}
\end{table}
\newline The Figs.~\ref{figstop}-\ref{figstaune} show the total
cross-sections for the pair production of the sfermions of the
third generation. In general, we show the complete corrections and
the tree-level where the tree-level is defined according to the
SPA project. According to the SPA project, all masses are taken
on-shell and all parameters in the couplings are given in the
\drbar scheme. The virtue of using this tree-level definition is
that not only the total corrected cross-sections are directly
comparable to other calculations using the SPA conventions but one
can also compare the relative corrections.
\newline
For each sfermion type we show an unpolarized case (left) and a
case where the beams are polarized (right). We take two sets of
polarizations, either $P_- = -0.8$ and $P_+ = 0.6$ or $P_- = 0.8$
and $P_+ = -0.6$. The difference to the earlier calculations
\cite{letter,hollik} are the QED corrections which give a negative
contribution near the threshold due to the known soft-photon
behaviour. The QED corrections are substantial (as can be checked
when comparing the results of this paper with those of
\cite{letter}) and cannot to be neglected. The plots on the
right-hand side of Fig.~\ref{figstop}-\ref{figstaune} show the
effect of beams polarization on the radiative corrections.
Polarization and its effects are best seen in other observables
which we discuss in the following.
\begin{figure}[h!]
\begin{center}
\mbox{\mbox{\resizebox{80mm}{!}{\includegraphics{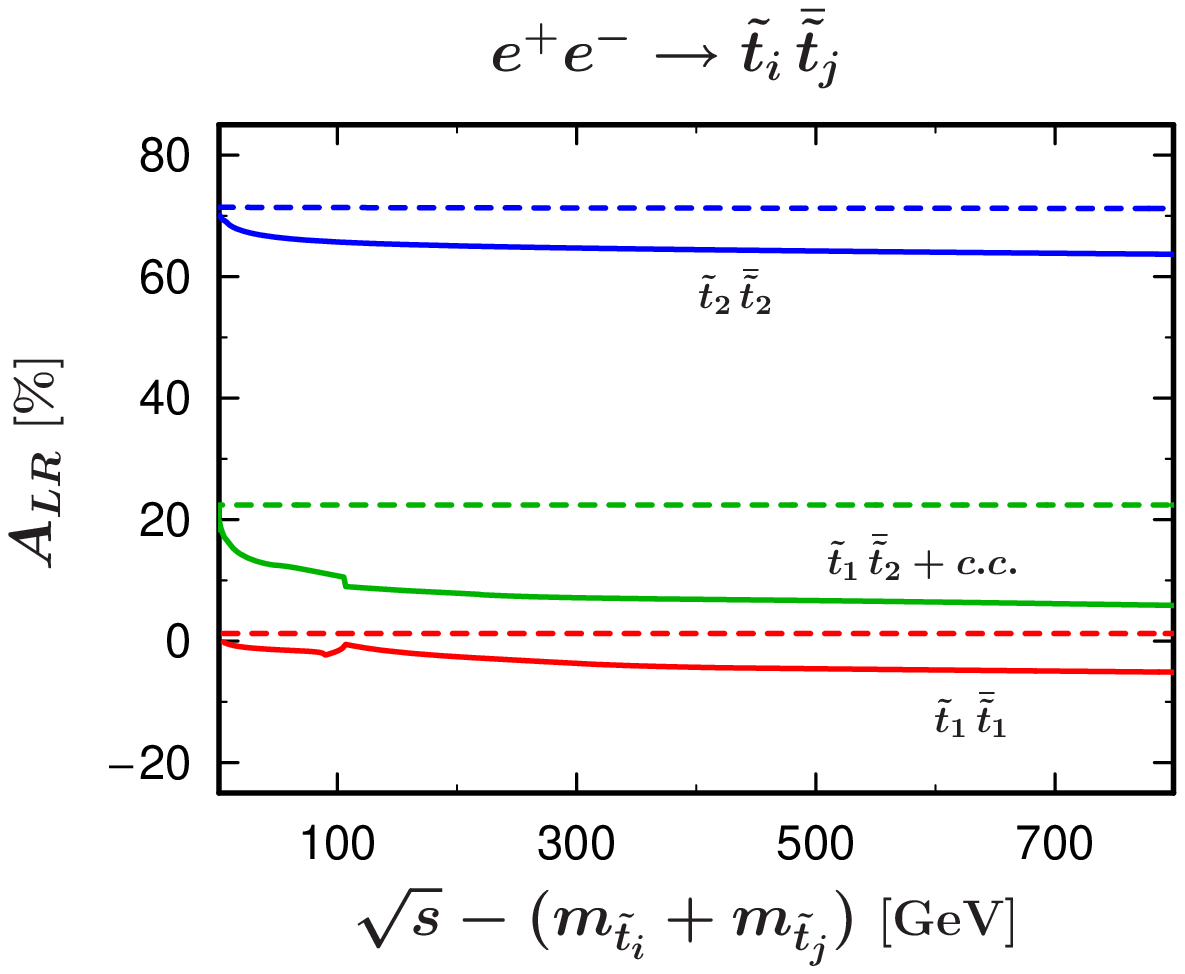}}}\hspace{2mm}
\mbox{\resizebox{80mm}{!}{\includegraphics{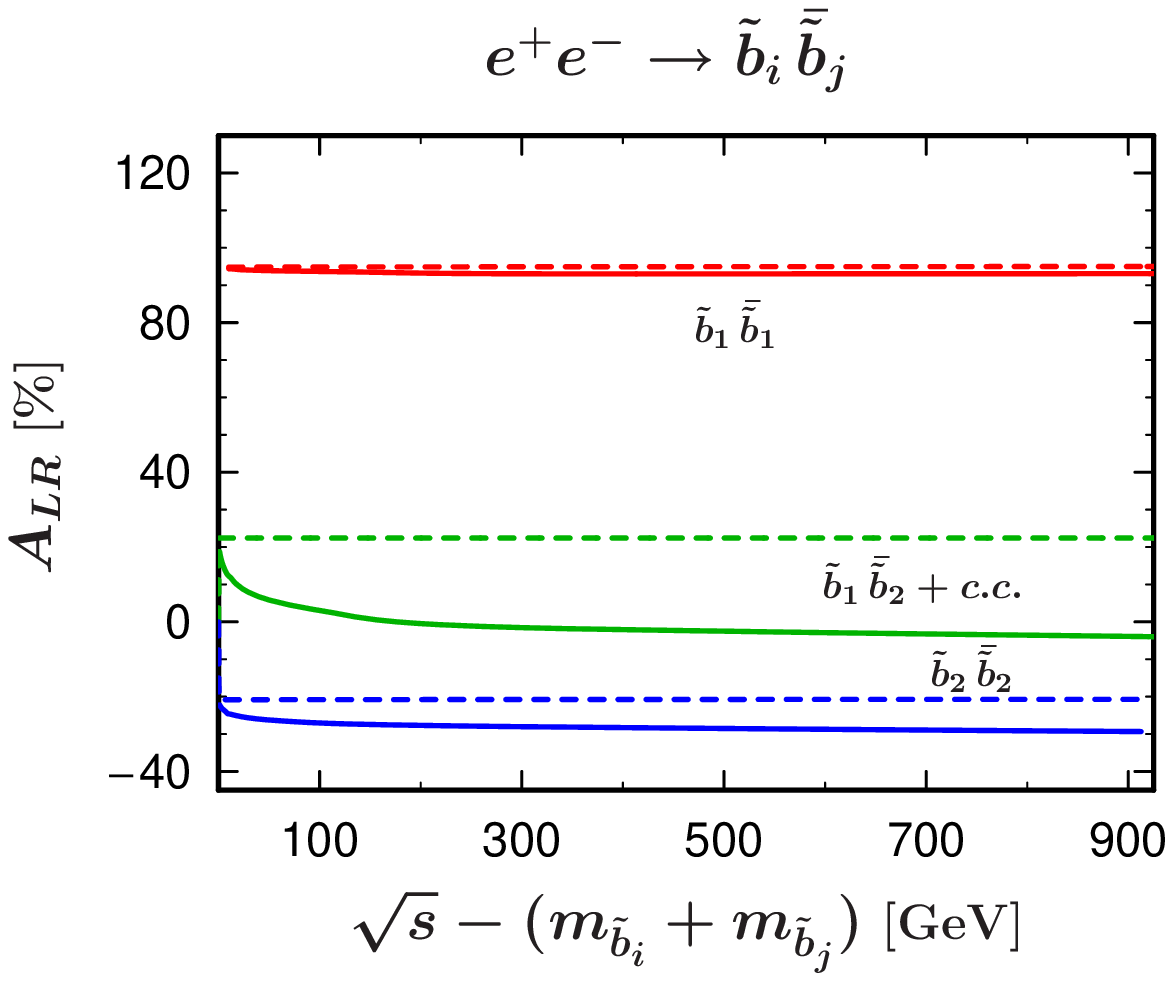}}}}
\end{center}
\caption{Left: The left-right asymmetry for stop pair production
at tree-level \{dashed\} and with complete corrections \{solid\}.
Right: The left-right asymmetry for sbottom pair production at
tree-level \{dashed\} and with complete corrections
\{solid\}.}\label{alrstopbot}
\end{figure}
\newline%
The Figs.~\ref{alrstopbot}-\ref{afbsbot} show the left-right and
forward-backward asymmetries for different final states as defined
in Eq.~(\ref{asym}). Owing to the fact that at tree level there is
only a s-channel contribution the $\sqrt{s}$-dependence drops out
in the left-right asymmetry, making it to a good approximation
constant. The $\sqrt{s}$-dependence is then a result of the
one-loop corrections. Notice that the corrections are substantial
especially in the  $\st_1 \st_2$, $\sb_1 \sb_2$, $\stau_1
\stau_2$, as well as in the $\ti\nu_\tau \ti\nu_\tau$ channel,
where there is only a $Z$ exchange at tree-level.
\newline%
As we have already mentioned, there is no tree-level contribution
to the forward-backward asymmetry and thus the asymmetry is
loop-induced. In the calculation of the forward-backward asymmetry
at one-loop one has to define the forward direction, in particular
for the contributions coming from the photon radiation . We define
it by $\s_{F}\equiv\s (\cos\theta_{\vec{p_1}\vec{k_{1,2}}}\geq 0)$
where $\theta_{\vec{p_1}\vec{k_{1,2}}}$ is the angle between the
incoming electron and the outgoing sfermion with negative isospin.
As an additional feature, we also show the forward-backward
asymmetry for polarized beams where the polarizations are $P_- =
-0.8$ and $P_+ = 0.6$. In general, one sees that the asymmetries
receive sizeable corrections and thus justify the higher-order
calculation.
\newline
\section{Conclusion}\label{conclusion}
We have calculated the full $\cal{O}(\alpha)$ corrections to stop,
sbottom, stau and tau-sneutrino production in the MSSM. We have
presented the details of our analytical calculation which was also
checked by the computer algebra tools FeynArts and FormCalc
\cite{feyn}. The results extend our previous calculations
\cite{SUSY-QCD-H, Yukawa, letter} by including also QED
contributions and the real photon radiation. We have also used the
structure function approach \cite{slicing2} to include some higher
order effects. Moreover, the whole calculation was extended to the
case of polarized $e^\pm$-beams.
\newline %
In the numerical analysis, we have studied only one specific
scenario based on the SPS1a' benchmark point defined in the SPA
project. We have transformed the input parameters into the
on-shell renormalization scheme which we have used throughout the
paper. The numerical results show the total cross-sections and
asymmetries with the effect of the $\cal{O}(\alpha)$ corrections.
These are found to be sizeable (in some cases up to 15\% and
larger), and in particular the forward-backward asymmetry is only
due to higher order corrections.
\newline\newline
\noindent {\bf Acknowledgements}\\ \noindent We thank W.~\"{O}ller
for useful discussions. The authors acknowledge support from EU
under the HPRN-CT-2000-00149 network programme and the ``Fonds zur
F\"orderung der wissenschaftlichen Forschung'' of Austria, project
No. P16592-N02.

\newpage
\appendix

\section{Vertex corrections}\label{appVertex}
Here we give the explicit form of the electroweak contributions to
the vertex corrections which are depicted in
Fig.~\ref{vertex-graphs}. For SUSY-QCD contributions we refer to
\cite{SUSY-QCD-H}. All couplings used in this paper can be found
in \cite{chrislet}.
\newline\newline
The vertex corrections $(\d e_f)_{ij}^{(v)}$ and $(\d
a_f)_{ij}^{(v)}$ (or $\d e_{L,R}^{(v)}$ and $\d a_{L,R}^{(v)}$)
originate from the diagrams in Fig.~\ref{vertex-graphs} with $\g$,
$Z$-boson exchange, respectively.
\newline The vertex corrections to the right vertex $(\d
e_f)_{ij}^{(v)}$ and $(\d a_f)_{ij}^{(v)}$ are defined on the
amplitude level from the corresponding diagrams. The general form
of the amplitude is given by
\begin{equation}\label{vert1}
\MM = \frac{i}{(4\pi)^2}\frac{X_V}{s-m_V^2}\vba\,\gam(Y^V_L\,P_L +
Y^V_R\,P_R)\,\unb \left[A_{ij}\,(k_1-k_2)_\mu +
B_{ij}\,(k_1+k_2)_\mu \right]\,.
\end{equation}
For $X_V$ and $Y^V_{L,R}$ with $V=\gamma ,Z$ we have
\begin{eqnarray}
X_{\gamma} = e\,,\qquad Y^{\gamma}_{L,R}= K_{L,R}\,,\qquad X_{Z} =
-g_Z\,,\qquad Y^Z_{L,R}= C_{L,R}\,.
\end{eqnarray}
We use the form-factor $A_{ij}$ (the other form-factor vanishes in
$|\MM|^2$) to define the vertex correction $(\d e_f)_{ij}^{(v)}$
and $(\d a_f)_{ij}^{(v)}$ as
\begin{equation}
(\d e_f)_{ij}^{(v)} =
-\frac{1}{(4\pi)^2}\frac{1}{e}\,A_{ij}\,,\qquad\quad (\d
a_f)_{ij}^{(v)} = -\frac{1}{(4\pi)^2}\frac{4\,c_W}{g}\,A_{ij}\,.
\end{equation}
The explicit formulas for $(\d e_f)_{ij}^{(v)}$ and $(\d
a_f)_{ij}^{(v)}$ are given below in the Appendices
\ref{gsfisfjvertex} and \ref{Zsfisfjvertex}.
\newline The vertex corrections to the left vertex, $\d e_{L,R}^{(v)}$ and
$\d a_{L,R}^{(v)}$, include corrections to the two chiral parts of
the electron photon/$Z$-boson vertex. The generic form is
\begin{eqnarray}
\MM =
\frac{i}{(4\pi)^2}\frac{(X_V)_{ij}}{s-m_V^2}\,\vba\left[\,A_L\gam
\,P_L + A_R\gam\,P_R+ B_L P_L (p_1-p_2)^\mu+ B_R P_R (p_1-p_2)^\mu
\right.\nonumber\\ \left. + C_L P_L (p_1+p_2)^\mu+ C_R P_R
(p_1+p_2)^\mu\,\right]\unb (k_1-k_2)_\mu\,,
\end{eqnarray}
where $(X_V)_{ij}$ ($V=\gamma ,Z$) stands for
\begin{equation}
(X_{\gamma})_{ij} = -e\, e_f\d_{ij}\,,\qquad\qquad (X_{Z})_{ij} =
-\frac{g_Z}{4} a^f_{ij}\,.
\end{equation}
From this generic structure only the form-factors $A_{L,R}$
survive. We define the vertex corrections as
\begin{equation}
\d e_{L,R}^{(v)} =
\frac{1}{(4\pi)^2}\frac{1}{e}\,A_{L,R}\,,\qquad\quad \d
a_{L,R}^{(v)} = -\frac{1}{(4\pi)^2}\frac{c_W}{g}\,A_{L,R}\,.
\end{equation}
The contributions to the vertex corrections $\d e_{L,R}^{(v)}$ and
$\d a_{L,R}^{(v)}$ are given in the Appendices \ref{geevertex} and
\ref{Zeevertex}.
\subsection{Corrections to {\boldmath{$\g \sf_i \sf_j$}} vertex\label{gsfisfjvertex}}
The vertex correction $(\d e_f)_{ij}^{(v)}$ is composed of
contributions from different classes of diagrams as follows,
\begin{equation}\label{vercomp1}
(\d e_f)_{ij}^{(v)} = (\d e_f)_{ij}^{(v, \ti\chi)} + (\d
e_f)_{ij}^{(v, SSS)} + (\d e_f)_{ij}^{(v, V\!\sf\!\sf)} + (\d
e_f)_{ij}^{(v, S\!S\!V+S\!V\!S)} + (\d e_f)_{ij}^{(v, S\!V\!V)} +
(\d e_f)_{ij}^{(v, S\!V)}\,.
\end{equation}
In the following we use the standard two- and three-point
functions $B_i$ and $C_i$  from \cite{PaVe} in the conventions of
\cite{Denner}. We introduce the following standard set of
arguments $C_i \equiv C_i\big( m_{\sf_i}^2, s, m_{\sf_j}^2, M_0^2,
M_1^2, M_2^2 \big)$ to be used in the generic functions ${\cal
S}$.\newline

The first contribution coming from the exchange of one or two
gauginos is
\begin{eqnarray}\non
(\d e_f)_{ij}^{(v, \ti\chi)} &=& \frac{1}{(4\pi)^2} \sum_{k,l =
1}^2 {\cal S}^{F\!F\!F}_{ij} \Big( m_{f'}, m_{\chp_k}, m_{\chp_l};
2 I_f^{3L} \d_{kl}, 2 I_f^{3L} \d_{kl}, k^\sf_{ik}, l^\sf_{ik},
l^\sf_{jl}, k^\sf_{jl} \Big)
\\ \non
&& + \frac{1}{(4\pi)^2} \sum_{k = 1}^4 {\cal S}^{F\!F\!F}_{ij}
\Big( m_{\nt_k}, m_f, m_f; e_f, e_f, b^\sf_{ik}, a^\sf_{ik},
a^\sf_{jk}, b^\sf_{jk} \Big)
\\
&& + \frac{1}{(4\pi)^2} \sum_{k = 1}^2 {\cal S}^{F\!F\!F}_{ij}
\Big( m_{\chp_k}, m_{f'}, m_{f'}; e_{f'}, e_{f'}, k^\sf_{ik},
l^\sf_{ik}, l^\sf_{jk}, k^\sf_{jk} \Big) \,,
\end{eqnarray}
with the generic vertex function
\begin{eqnarray}\label{SFFF} \non
&&\hspace{-1cm}{\cal S}^{F\!F\!F}_{ij} \big( M_0, M_1, M_2; g_0^R,
g_0^L, g_1^R, g_1^L, g_2^R, g_2^L \big) ~=~ M_0 M_2 \big(g_0^L
g_1^L g_2^L + g_0^R g_1^R g_2^R\big) \big(C_0  + C_1 + C_2\big)
\\ \non
&&+ M_0 M_1 \big(g_0^L g_1^R g_2^R + g_0^R g_1^L g_2^L\big)
\big(C_0  + C_1 + C_2\big) + M_1 M_2 \big(g_0^L g_1^R g_2^L +
g_0^R g_1^L g_2^R\big) \big(C_1 + C_2\big)
\\
&&+ \big(g_0^L g_1^L g_2^R + g_0^R g_1^R g_2^L\big) \Big(
B_0\big(s, M_1^2, M_2^2\big) + M_0^2 \big(2 C_0 + C_1 + C_2\big) +
m_{\sf_i}^2 C_1 + m_{\sf_j}^2 C_2\Big) \,.
\end{eqnarray}
\\
The corrections due to graphs with 3 scalar particles in the loop
are given by
\begin{eqnarray}\non
(\d e_f)_{ij}^{(v, SSS)} &=& -\frac{1}{(4\pi)^2}\, 2 I_f^{3L}
\sum_{k,m = 1}^2 \! G^{\sf\sf'}_{imk} G^{\sf\sf'}_{jmk} \big( C_0
+ C_1 + C_2 \big) \big( m_{\sf_i}^2, s, m_{\sf_j}^2, m_{\sf'_m}^2,
m_{H_k^+}^2, m_{H_k^+}^2 \big)
\\ \non
&& - \frac{1}{(4\pi)^2}\, e_f \sum_{k,m = 1}^2 \! G^{\sf}_{imk}
G^{\sf}_{mjk} \big( C_0 + C_1 + C_2 \big) \big( m_{\sf_i}^2, s,
m_{\sf_j}^2, m_{H_k^0}^2, m_{\sf_m}^2, m_{\sf_m}^2 \big)
\\
&& - \frac{1}{(4\pi)^2}\, e_{f'} \sum_{k,m = 1}^2 \!
G^{\sf\sf'}_{imk} G^{\sf\sf'}_{jmk} \big( C_0 + C_1 + C_2 \big)
\big( m_{\sf_i}^2, s, m_{\sf_j}^2, m_{H_k^+}^2, m_{\sf'_m}^2,
m_{\sf'_m}^2 \big) \,.
\end{eqnarray}
The graphs with one vector particle ($\g, Z^0, W^+$) and two
sfermions in the loop yield
\begin{eqnarray}\non
(\d e_f)_{ij}^{(v, V\!\sf\!\sf)} &=& \frac{1}{(4\pi)^2}\, e^2
e_f^3 \, \d_{ij}\, {\cal S}^{V\!\sf\!\sf}_{ij} \big( \l,
m_{\sf_i}, m_{\sf_i} \big) + \frac{1}{(4\pi)^2}\, g_Z^2\, e_f
\sum_{m = 1}^2 z^\sf_{im}\, z^\sf_{mj}\, {\cal
S}^{V\!\sf\!\sf}_{ij} \big( m_Z, m_{\sf_m}, m_{\sf_m} \big)
\\
&& + \frac{1}{(4\pi)^2}\, \frac{g^2}{2}\, e_{f'} R^\sf_{i1}
R^\sf_{j1} \sum_{m = 1}^2 \big( R^{\sf'}_{m1} \big)^2 \, {\cal
S}^{V\!\sf\!\sf}_{ij} \big( m_W, m_{\sf'_m}, m_{\sf'_m} \big) \,,
\end{eqnarray}
where ${\cal S}^{V\!\sf\!\sf}_{ij}(\ldots)$ is a short form for
\begin{eqnarray}\non
&&\hspace{-1cm}{\cal S}^{V\!\sf\!\sf}_{ij} \big( M_0, M_1, M_2
\big) ~=~ 4 C_{00} + \big(M_0^2 - 2 s + 2 m_{\sf_i}^2 + 2
m_{\sf_j}^2 \big) \big(C_0 + C_1 + C_2\big)
\\ \non
&& + \big( 2 m_{\sf_i}^2 + 2 m_{\sf_j}^2 - s \big) \big(C_1 + C_2
+ C_{11} + 2 C_{12} + C_{22}\big) + \big(m_{\sf_i}^2 - m_{\sf_j}^2
\big) \big(C_1 - C_2 + C_{11} - C_{22}\big)\,.
\\
\end{eqnarray}
From the diagrams with one $W^+$-boson, one Goldstone boson $G^+$
and a sfermion we obtain
\begin{eqnarray}\non
(\d e_f)_{ij}^{(v, S\!S\!V+S\!V\!S)} &=& \frac{1}{(4\pi)^2}\,
\frac{g\, m_W}{2 \sqrt 2} \sum_{m = 1}^2 \left( R^\sf_{i1}
G^{\sf\sf'}_{jm2} + R^\sf_{j1} G^{\sf\sf'}_{im2} \right)
R^{\sf'}_{m1}
\\
&& \hspace{75pt} \times~\big( C_0 - C_1 - C_2 \big) \big(
m_{\sf_i}^2, s, m_{\sf_j}^2, m_{\sf'_m}^2, m_W^2, m_W^2 \big) \,,
\end{eqnarray}
and with the generic scalar--vector--vector vertex function
\begin{eqnarray}\label{SSVV} \non
&&\hspace{-1.5cm}{\cal S}^{S\!V\!V}_{ij} \big( M_0, M_1, M_2 \big)
~=~ 2 B_0 \big( s, M_1^2, M_2^2 \big) - 2 C_{00} +
\Big(\frac{s}{2} + 2 M_0^2 \Big) C_0 - s \big(C_1 + C_2 \big)
\\
&& - \frac{1}{2} \big( m_{\sf_i}^2 - m_{\sf_j}^2 \big) \big( C_1 -
C_2 + C_{11} - C_{22} \big)  -\frac{1}{2} \big(2 m_{\sf_i}^2 + 2
m_{\sf_j}^2 - s \big) \big(C_{11} + 2 C_{12} + C_{22} \big)
\end{eqnarray}
the correction due to the exchange of two $W^+$-bosons and one
sfermion reads
\begin{eqnarray}
(\d e_f)_{ij}^{(v, S\!V\!V)} &=& \frac{1}{(4\pi)^2}\, g^2\,
I_f^{3L} R^\sf_{i1} R^\sf_{j1} \sum_{m = 1}^2 \big( R^{\sf'}_{m1}
\big)^2 \, {\cal S}^{S\!V\!V}_{ij} \big( m_{\sf'_m}, m_W, m_W
\big) \,.
\end{eqnarray}
The contributions from one sfermion and one vector particle ($\g,
Z^0, W^+$) can be expressed as
\begin{eqnarray}\non
(\d e_f)_{ij}^{(v, S\!V)} &=& -\frac{1}{(4\pi)^2}\, 2 e^2 e_f^3 \,
\d_{ij} \big( 2 B_0 + B_1 \big) \big( m_{\sf_i}^2, \l^2,
m_{\sf_i}^2 \big)
\\ \non
&& -\frac{1}{(4\pi)^2}\, g_Z^2\, e_f \sum_{m=1}^2 z^\sf_{im}
z^\sf_{jm} \big( 2 B_0 + B_1 \big) \big( m_{\sf_i}^2, m_Z^2,
m_{\sf_m}^2 \big) \qquad + (i \leftrightarrow j)
\\
&& -\frac{1}{(4\pi)^2}\, \frac{g^2}{4}\, Y^f_L R^\sf_{i1}
R^\sf_{j1} \sum_{m=1}^2 \big( R^{\sf'}_{m1} \big)^2  \big( 2 B_0 +
B_1 \big) \big( m_{\sf_i}^2, m_W^2, m_{\sf'_m}^2 \big) \qquad + (i
\leftrightarrow j)
\end{eqnarray}
with $Y_{L}^f = 2 (I_f^{3L} - e_f)$. The symbol $(i
\leftrightarrow j)$ denotes the previous term with the indices $i$
and $j$ interchanged.
\subsection{Corrections to {\boldmath{$Z^0 \sf_i \sf_j$}} vertex \label{Zsfisfjvertex}}
The corrections to the $Z^0 \sf_i \sf_j$ vertex have the same
components as in Eq.~(\ref{vercomp1}). Using the same
abbreviations for the generic vertex functions as in the previous
section we get for the single contributions:
\begin{eqnarray}\non
(\d a_f)_{ij}^{(v, \ti\chi)} &=& -\frac{4}{(4\pi)^2} \sum_{k,l =
1}^4 {\cal S}^{F\!F\!F}_{ij} \Big( m_{f}, m_{\nt_k}, m_{\nt_l};
O_{kl}^{''R}, O_{kl}^{''L}, b^\sf_{ik}, a^\sf_{ik}, a^\sf_{jl},
b^\sf_{jl} \Big)
\\ \non
&& - \frac{4}{(4\pi)^2}\, 2I^{3L}_f \sum_{k,l = 1}^2 {\cal
S}^{F\!F\!F}_{ij} \Big( m_{f'}, m_{\chp_k}, m_{\chp_l}; \widetilde
O_{kl}^{'R}, \widetilde O_{kl}^{'L}, k^\sf_{ik}, l^\sf_{ik},
l^\sf_{jl}, k^\sf_{jl} \Big)
\\ \non
&& + \frac{4}{(4\pi)^2} \sum_{k = 1}^4 {\cal S}^{F\!F\!F}_{ij}
\Big( m_{\nt_k}, m_f, m_f; C^f_R, C^f_L, b^\sf_{ik}, a^\sf_{ik},
a^\sf_{jk}, b^\sf_{jk} \Big)
\\
&& + \frac{4}{(4\pi)^2} \sum_{k = 1}^2 {\cal S}^{F\!F\!F}_{ij}
\Big( m_{\chp_k}, m_{f'}, m_{f'}; C^{f'}_R, C^{f'}_L, k^\sf_{ik},
l^\sf_{ik}, l^\sf_{jk}, k^\sf_{jk} \Big)\,,
\end{eqnarray}
with $\widetilde O_{kl}^{'L/R} = O_{kl}^{'L/R}$ for up-type
sfermions (up-squarks and sneutrinos) and $\widetilde
O_{kl}^{'L/R} = O_{kl}^{'R/L}$ for down-type sfermions
(down-squarks and sleptons),

\begin{eqnarray}\non
(\d a_f)_{ij}^{(v, SSS)} &=& \frac{2i}{(4\pi)^2} \sum_{k,m = 1}^2
\sum_{l = 3}^4 \! R_{k,l-2}(\b-\a)\, G^{\sf}_{imk} G^{\sf}_{mjl}
\big( C_0 + C_1 + C_2 \big) \big( m_{\sf_i}^2, s, m_{\sf_j}^2,
m_{\sf_m}^2, m_{H_k^0}^2, m_{H_l^0}^2 \big)
\\ \non
&& -\frac{2i}{(4\pi)^2} \sum_{k = 3}^4 \sum_{l,m = 1}^2 \!
R_{l,k-2}(\b-\a)\, G^{\sf}_{imk} G^{\sf}_{mjl} \big( C_0 + C_1 +
C_2 \big) \big( m_{\sf_i}^2, s, m_{\sf_j}^2, m_{\sf_m}^2,
m_{H_k^0}^2, m_{H_l^0}^2 \big)
\\ \non
&& -\frac{4}{(4\pi)^2}\, I_f^{3L} \cos 2\theta_W \sum_{k,m = 1}^2
\! G^{\sf\sf'}_{imk} G^{\sf\sf'}_{jmk} \big( C_0 + C_1 + C_2 \big)
\big( m_{\sf_i}^2, s, m_{\sf_j}^2, m_{\sf'_m}^2, m_{H_k^+}^2,
m_{H_k^+}^2 \big)
\\ \non
&& -\frac{4}{(4\pi)^2} \sum_{k,m,n = 1}^2 \! z^\sf_{mn}
G^{\sf}_{imk} G^{\sf}_{njk} \big( C_0 + C_1 + C_2 \big) \big(
m_{\sf_i}^2, s, m_{\sf_j}^2, m_{H_k^0}^2, m_{\sf_m}^2, m_{\sf_n}^2
\big)
\\
&& -\frac{4}{(4\pi)^2} \sum_{k,m,n = 1}^2 \! z^{\sf'}_{mn}
G^{\sf\sf'}_{imk} G^{\sf\sf'}_{jnk} \big( C_0 + C_1 + C_2 \big)
\big( m_{\sf_i}^2, s, m_{\sf_j}^2, m_{H_k^+}^2, m_{\sf'_m}^2,
m_{\sf'_n}^2 \big) \,,
\end{eqnarray}

\begin{eqnarray}\non
(\d a_f)_{ij}^{(v, V\!\sf\!\sf)} &=& \frac{4}{(4\pi)^2}\, e^2
e_f^2 \, z^\sf_{ij}\, {\cal S}^{V\!\sf\!\sf}_{ij} \big( \l,
m_{\sf_i}, m_{\sf_j} \big) + \frac{4}{(4\pi)^2}\, g_Z^2 \sum_{m,n
= 1}^2\! z^\sf_{im}\, z^\sf_{mn}\, z^\sf_{nj}\, {\cal
S}^{V\!\sf\!\sf}_{ij} \big( m_Z, m_{\sf_m}, m_{\sf_n} \big)
\\
&& + \frac{4}{(4\pi)^2}\, \frac{g^2}{2}\, R^\sf_{i1} R^\sf_{j1}
\sum_{m,n = 1}^2\! R^{\sf'}_{m1} R^{\sf'}_{n1} \, {\cal
S}^{V\!\sf\!\sf}_{ij} \big( m_W, m_{\sf'_m}, m_{\sf'_n} \big) \,,
\end{eqnarray}

\begin{eqnarray}\non
(\d a_f)_{ij}^{(v, S\!S\!V+S\!V\!S)} &=&
\\ \non
&&\hspace{-2cm} -\frac{1}{(4\pi)^2} \sqrt 2 \, g\, s_W^2 m_W
\!\sum_{m = 1}^2\! \left( R^\sf_{i1} G^{\sf\sf'}_{jm2} +
R^\sf_{j1} G^{\sf\sf'}_{im2} \right) R^{\sf'}_{m1} \big( C_0 - C_1
- C_2 \big) \big( m_{\sf_i}^2, s, m_{\sf_j}^2, m_{\sf'_m}^2,
m_W^2, m_W^2 \big)
\\ \non
&&\hspace{-2cm} +\frac{1}{(4\pi)^2}\, 2\, g_Z\, m_Z \!\sum_{k,m =
1}^2 \! R_{k_2}(\b-\a) \,z^\sf_{im} G^{\sf}_{mjk} \big( C_0 - C_1
- C_2 \big) \big( m_{\sf_i}^2, s, m_{\sf_j}^2, m_{\sf_m}^2, m_Z^2,
m_{H_k^0}^2 \big)
\\
&&\hspace{-2cm} +\frac{1}{(4\pi)^2}\, 2\, g_Z\, m_Z \!\sum_{k,m =
1}^2 \! R_{k_2}(\b-\a) \,z^\sf_{mj} G^{\sf}_{imk} \big( C_0 - C_1
- C_2 \big) \big( m_{\sf_i}^2, s, m_{\sf_j}^2, m_{\sf_m}^2,
m_{H_k^0}^2, m_Z^2 \big) \,,
\\
(\d a_f)_{ij}^{(v, S\!V\!V)} &=& \frac{4}{(4\pi)^2}\, g^2\,
c_W^2\, I_f^{3L} R^\sf_{i1} R^\sf_{j1} \sum_{m = 1}^2 \big(
R^{\sf'}_{m1} \big)^2 \, {\cal S}^{S\!V\!V}_{ij} \big( m_{\sf'_m},
m_W, m_W \big) \,,
\\ \non
(\d a_f)_{ij}^{(v, S\!V)} &=& -\frac{4}{(4\pi)^2}\, e^2 e_f^2 \,
z^\sf_{ij} \big( 2 B_0 + B_1 \big) \big( m_{\sf_i}^2, \l^2,
m_{\sf_i}^2 \big) \qquad + (i \leftrightarrow j)
\\ \non
&& -\frac{4}{(4\pi)^2}\, g_Z^2\, \sum_{m=1}^2 z^\sf_{im} \ti
z^\sf_{jm} \big( 2 B_0 + B_1 \big) \big( m_{\sf_i}^2, m_Z^2,
m_{\sf_m}^2 \big) \qquad + (i \leftrightarrow j)
\\
&& +\frac{1}{(4\pi)^2}\, e^2\, Y^f_L \,R^\sf_{i1} R^\sf_{j1}
\sum_{m=1}^2 \big( R^{\sf'}_{m1} \big)^2  \big( 2 B_0 + B_1 \big)
\big( m_{\sf_i}^2, m_W^2, m_{\sf'_m}^2 \big) \qquad + (i
\leftrightarrow j)\,.
\end{eqnarray}
\subsection{Corrections to {\boldmath{$\g e^+ e^-$}} vertex\label{geevertex}}
In the following we list the analytic formulas of the vertex
corrections to the electron--positron--photon vertex. We only give
the right-handed coefficients of the generic vertex functions,
${\cal F}_{\!R}^{(\cdots)}$ as the coefficients ${\cal
F}_{\!L}^{(\cdots)}$ can be obtained by exchanging the indices $R$
and $L$, i.~e. ${\cal F}_{\!L}^{(\cdots)} = {\cal
F}_{\!R}^{(\cdots)} (R \leftrightarrow L)$. \newline In the
remaining vertex corrections we use the standard set of arguments
for the whole class of $C$-functions $C \equiv C \big( m_e^2, s,
m_e^2, M_0^2, M_1^2, M_2^2 \big)$.
\newline%
The vertex correction $\d e_{L,R}^{(v)}$ is split into the
following classes:
\begin{equation}
\d e_{L,R}^{(v)} = \d e_{L,R}^{(v, \sf \ti\chi \ti\chi)} + \d
e_{R}^{(v, \ti\chi \sf \sf)} + \d e_{L,R}^{(v, V\!F\!F)} + \d
e_{L,R}^{(v, F\!V\!V)}
\end{equation} The contribution from one sfermion
and two gauginos in the loop is given by
\begin{eqnarray}
\d e_{L,R}^{(v, \sf \ti\chi \ti\chi)} &=& \frac{1}{(4\pi)^2}
\sum_{k = 1}^2 {\cal F}_{\!L,R}^{S\!F\!F} \Big( m_{\ti\nu_e},
m_{\chp_k}, m_{\chp_k}; 1, 1, l^{\ti\nu_e}_{1k},
k^{\ti\nu_e}_{1k}, k^{\ti\nu_e}_{1k}, l^{\ti\nu_e}_{1k} \Big) \,,
\end{eqnarray}
where we have used the generic vertex function
\begin{eqnarray}\label{FSFF} \non
&&\hspace{-1cm}{\cal F}_{\!R}^{S\!F\!F} \big( M_0, M_1, M_2;
g_0^R, g_0^L, g_1^R, g_1^L, g_2^R, g_2^L \big) ~=~ -g_0^L g_1^L
g_2^R \Big( 2 C_{00} - B_0\big(s, M_1^2, M_2^2 \big) \Big)
\\ \non
&&  - \big(g_0^R h_1^{RL} h_2^{LR} - g_0^L g_1^L g_2^R M_0^2\big)
C_0 - \big(g_0^R g_1^R h_2^{LR} - g_0^L g_2^R h_1^{LR}\big) m_e
C_1 + \big(g_0^L g_1^L h_2^{RL} - g_0^R g_2^L h_1^{RL}\big) m_e
C_2
\\
\end{eqnarray}
and the abbreviations (no sum over $i$) $h_i^{jk} = g_i^j m_e +
g_i^k M_i$ for $i = 1,2$ and $(j,k) = L,R$.
\\
The corrections due to the exchange of one gaugino and two
sfermions are
\begin{eqnarray}
\d e_{R}^{(v, \ti\chi \sf \sf)} &=& \frac{2}{(4\pi)^2} \sum_{k =
1}^4 \sum_{m = 1}^2 \big( b^{\ti e}_{mk} \big)^2 C_{00} \big(
m_e^2, s, m_e^2, m_{\nt_k}^2, m_{\ti e_m}^2, m_{\ti e_m}^2
\big)\,,
\end{eqnarray}
and $\d e_{L}^{(v, \ti\chi \sf \sf)} = \d e_{R}^{(v, \ti\chi \sf
\sf)} (b^{\ti e}_{mk} \rightarrow a^{\ti e}_{mk})$.
\\
Using the generic vertex function for one vector particle and two
fermions in the loop,
\begin{eqnarray}\label{FVFF} \non
&&\hspace{-5mm}{\cal F}_{\!R}^{V\!F\!F} \big( M_0, M_1, M_2;
g_0^R, g_0^L, g_1^R, g_1^L, g_2^R, g_2^L \big) ~=~ -2 \Big[ g_0^R
g_1^R g_2^R \Big( 2 C_{00} - B_0\big(s, M_1^2, M_2^2\big) - m_e^2
\big (C_1 + C_2 \big)
\\ \non
&&\hspace{5mm} - M_0^2 C_0 + \big(s - 2 m_e^2\big) \big(C_{0} +
C_{1} + C_{2}\big) + \frac{r}{2} \Big) + g_0^L g_1^R g_2^R M_1
M_2\, C_0 - g_0^L g_1^L g_2^L m_e^2 \big(C_0 + C_1 + C_2\big)\Big]
\,,
\\
\end{eqnarray}
with $r=0$ in the \drbar renormalization scheme we get the
corrections stemming from one vector boson ($\g, Z^0$) and two
electrons given by
\begin{eqnarray}\non
\d e_{L,R}^{(v, V\!F\!F)} &=& \frac{1}{(4\pi)^2} e^2\, {\cal
F}_{\!L,R}^{V\!F\!F} \Big( \l, m_e, m_e; 1, 1, 1, 1, 1, 1 \Big)
\\
&& + \frac{1}{(4\pi)^2} g_Z^2\, {\cal F}_{\!L,R}^{V\!F\!F} \Big(
m_Z, m_e, m_e; 1, 1, C^e_R, C^e_L, C^e_R, C^e_L\Big) \,.
\end{eqnarray}
For the graphs with one electron-neutrino and two $W$-bosons we
obtain
\begin{eqnarray}
\d e_{L,R}^{(v, F\!V\!V)} &=& \frac{1}{(4\pi)^2} \frac{g^2}{2}\,
{\cal F}_{\!L,R}^{F\!V\!V} \Big( 0, m_W, m_W; 1, 0, 1, 0, 1 \Big)
\,,
\end{eqnarray}
where we have used the function
\begin{eqnarray}\label{FFVV} \non
&&\hspace{-5mm}{\cal F}_{\!R}^{F\!V\!V} \big( M_0, M_1, M_2; g_0,
g_1^R, g_1^L, g_2^R, g_2^L \big) ~=~ g_0 \Big[ g_1^R g_2^R \Big( 2
B_0 \big(s, M_1^2, M_2^2\big) - r + 4 C_{00} + 2 M_0^2 C_0
\\
&&\hspace{5mm} + \big( 5 m_e^2 - 2 s\big) \big( C_1 + C_2 \big)
\Big) + 3 \big(g_1^L g_2^R + g_1^R g_2^L\big) m_e M_0 C_0 + 3
g_1^L g_2^L m_e^2 \big(C_1 + C_2\big) \Big] \,.
\end{eqnarray}
\subsection{Corrections to {\boldmath{$Z^0 e^+ e^-$}} vertex\label{Zeevertex}}
In the following we list the single contributions to the
electron--positron--$Z^0$ vertex. The generic vertex functions
used in this section can be looked up in Appendix~\ref{geevertex}.

\begin{eqnarray}\non
\d a_{L,R}^{(v, \sf \ti\chi \ti\chi)} &=& -\frac{1}{(4\pi)^2}
\sum_{k,l = 1}^4 \sum_{m = 1}^2 {\cal F}_{\!L,R}^{S\!F\!F} \Big(
m_{\ti e_m}, m_{\nt_k}, m_{\nt_l}; O_{kl}^{''R}, O_{kl}^{''L},
a^{\ti e}_{mk}, b^{\ti e}_{mk}, b^{\ti e}_{ml}, a^{\ti e}_{ml}
\Big)
\\
&& +\frac{1}{(4\pi)^2} \sum_{k,l = 1}^2 {\cal F}_{\!L,R}^{S\!F\!F}
\Big( m_{\ti\nu_e}, m_{\chp_k}, m_{\chp_l}; O_{kl}^{'L},
O_{kl}^{'R}, l^{\ti\nu_e}_{1k}, k^{\ti\nu_e}_{1k},
k^{\ti\nu_e}_{1l}, l^{\ti\nu_e}_{1l} \Big) \,,
\\ \non
\d a_{R}^{(v, \ti\chi \sf \sf)} &=& \frac{2}{(4\pi)^2} \sum_{k =
1}^4 \sum_{m,n = 1}^2 z^\sf_{mn} b^{\ti e}_{mk} b^{\ti e}_{nk}
C_{00} \big( m_e^2, s, m_e^2, m_{\nt_k}^2, m_{\ti e_m}^2, m_{\ti
e_n}^2 \big)
\\
&& + \frac{1}{(4\pi)^2} \sum_{k = 1}^2 \big( k^{\ti\nu_e}_{1k}
\big)^2 C_{00} \big( m_e^2, s, m_e^2, m_{\nt_k}^2, m_{\ti\nu_e}^2,
m_{\ti\nu_e}^2 \big)\,,
\end{eqnarray}
and $\d a_{L}^{(v, \ti\chi \sf \sf)} = \d a_{R}^{(v, \ti\chi \sf
\sf)} (b^{\ti e}_{\ldots} \rightarrow a^{\ti e}_{\ldots},
k^{\ti\nu_e}_{\ldots} \rightarrow l^{\ti\nu_e}_{\ldots})$.
\begin{eqnarray}\non
\d a_{L,R}^{(v, V\!F\!F)} &=& \frac{1}{(4\pi)^2} e^2\, {\cal
F}_{\!L,R}^{V\!F\!F} \Big( \l, m_e, m_e; C^e_R, C^e_L, 1, 1, 1, 1
\Big)
\\ \non
&& + \frac{1}{(4\pi)^2}\, g_Z^2\, {\cal F}_{\!L,R}^{V\!F\!F} \Big(
m_Z, m_e, m_e; C^e_R, C^e_L, C^e_R, C^e_L, C^e_R, C^e_L\Big)
\\
&& + \frac{1}{(4\pi)^2}\, \frac{g^2}{4}\, {\cal
F}_{\!L,R}^{V\!F\!F} \Big( m_W, 0, 0; 0, 1, 0, 1, 0, 1 \Big)\,,
\\
\d a_{L,R}^{(v, F\!V\!V)} &=& -\frac{1}{(4\pi)^2} \frac{g^2
c_W^2}{2}\, {\cal F}_{\!L,R}^{F\!V\!V} \Big( 0, m_W, m_W; 1, 0, 1,
0, 1 \Big)\,.
\end{eqnarray}
%
%
\section{Box contributions}\label{appBox}
In this section we give the explicit form of the radiative
corrections which stem from box diagrams with two different
topologies. The matrix element is parameterized as
\begin{eqnarray}
\MM_{\rm box} &=& \frac{i}{(4 \pi)^2} \vba\left[A_L\,P_L +
A_R\,P_R + B_L\,\onehfbi\,(\kslash_{1}-\kslash_{2})\,P_L +
B_R\,\onehfbi\,(\kslash_{1}-\kslash_{2})\,P_R\right]\,\unb\,,
\end{eqnarray}
where the form-factors $A_{L,R}$ do not contribute to the squared
matrix element. The form-factors $B_{L,R}$ depend on the
Mandelstam variables $t$ and $u$ which are defined as
\begin{eqnarray}
t &=& \frac{1}{2} (m^2_{\sf_i} +m^2_{\sf_j} -s) +\frac{1}{2}
\kappa(s, m^2_{\sf_i}, m^2_{\sf_j}) \cos\vartheta \,,
\\
u &=& 2 m_e^2 + m_{\sf_i}^2 + m_{\sf_j}^2 - s - t \,.
\end{eqnarray}
The single contributions to the form-factors
\begin{eqnarray}
B_{L,R} &=& B_{L,R}^{\g\g} + B_{L,R}^{\g Z} + B_{L,R}^{ZZ} +
B_{L,R}^{WW} + B_{L,R}^{\chp\chp} + B_{L,R}^{\nt\nt}
\end{eqnarray}
correspond to the diagrams with two vector bosons, where the
particles in the loop are indicated by a superscript, and
similarly $B_{L,R}^{\chp\chp}$ and $B_{L,R}^{\nt\nt}$ denote the
contributions from charginos and neutralinos, respectively.
\subsection{Vector bosons in the loop}
In the case of two vector bosons, we use the generic functions
\begin{eqnarray}\non
{\cal B}^{V\!V} \big( M_0, M_1, M_2, M_3 \big) &=& 4 C_0 + C_2 + 4
M_0^2 D_0 + M_0^2 D_3 + 4 \big[(m_{\sf_j}^2 \!-\! u) D_1 - s D_1 +
(t\!-\!m_{\sf_j}^2) D_2 + t D_3 \big]
\\
\end{eqnarray}
for a diagram with vector bosons $V$ in the loop and
\begin{eqnarray}\non
{\cal B}^{V\!V\!x} \big( M_0, M_1, M_2, M_3 \big) &=& - \left( 4
C_0^x + C_1^x + 4 M_0^2 D_0^x + 4 (u \!-\! m_{\sf_j}^2) D_1^x + 4
(u \!-\! m_{\sf_i}^2) D_2^x + (M_0^2 \!+\! 4u) D_3^x\right)
\\
\end{eqnarray}
for the corresponding crossed diagram. In case of a $W$-boson
there is no crossed diagram and we use ${\cal B}^{V\!V}$ or ${\cal
B}^{V\!V\!x}$ depending on the charge of the final state
particle.\newline The scalar three-point and four-point functions
used above have a standard set of arguments defined as
\begin{eqnarray}
\begin{array}{rll@{\,,\qquad}rll}
C_i &=& C_i\big( s, m_{\sf_j}^2, m_{\sf_i}^2, M_1^2, M_2^2, M_3^2
\big) & D_i &=& D_i\big( m_e^2, s, m_{\sf_j}^2, t, m_e^2,
m_{\sf_i}^2, M_0^2, M_1^2, M_2^2, M_3^2 \big)\,,
\\
C_i^x &=& C_i\big( m_{\sf_j}^2, m_{\sf_i}^2, s, M_1^2, M_3^2,
M_2^2 \big) & D_i^x &=& D_i\big( m_e^2, s, m_{\sf_i}^2, u, m_e^2,
m_{\sf_j}^2, M_0^2, M_1^2, M_2^2, M_3^2 \big)\,.
\end{array}
\end{eqnarray}
The contributions from 2 photons, one photon and one $Z$-boson, 2
$Z$-bosons and 2 $W$-bosons are
\begin{eqnarray}
B_{L,R}^{\g\g} &=& e^4 e_f^2 \d_{ij} \left(\, {\cal B}^{V\!V} +
{\cal B}^{V\!V\!x} \,\right) \big( m_e, 0, 0, m_{\sf_i} \big) \,,
\\ \non
B_{L,R}^{\g Z} &=& -e^2 e_f g_Z^2 \,C_{L,R} \sum_{m=1}^2 \d_{im}
z^\sf_{mj} \left[\, {\cal B}^{V\!V} \big( m_e, 0, m_Z, m_{\sf_m}
\big) + {\cal B}^{V\!V\!x} \big( m_e, m_Z, 0, m_{\sf_m} \big)
\,\right]
\\
&& -e^2 e_f g_Z^2 \,C_{L,R} \sum_{m=1}^2 z^\sf_{im} \d_{mj}
\left[\, {\cal B}^{V\!V} \big( m_e, m_Z, 0, m_{\sf_m} \big) +
{\cal B}^{V\!V\!x} \big( m_e, 0, m_Z, m_{\sf_m} \big) \,\right]\,,
\\
B_{L,R}^{ZZ} &=& g_Z^4 \,C_{L,R}^2 \sum_{m=1}^2 z^\sf_{im}
z^\sf_{mj} \left(\, {\cal B}^{V\!V} + {\cal B}^{V\!V\!x} \,\right)
\big( m_e, m_Z, m_Z, m_{\sf_m} \big)\,,
\\
B_{L}^{WW} &=& \frac{g^4}{4} R^\sf_{i1} R^\sf_{j1} \sum_{m=1}^2
\big( R^{\sf'}_{m1} \big)^2 {\cal B}^{V\!V\!(x)} \big( m_e, m_W,
m_W, m_{\sf'_m} \big) \,, \qquad B_{R}^{WW} = 0\,,
\end{eqnarray}
where ${\cal B}^{V\!V\!(x)}$ denotes ${\cal B}^{V\!V\!x}$ for
up-type sfermions and ${\cal B}^{V\!V}$ for down-type sfermions in
the final state.
\subsection{Scalars and fermions in the loop}
In analogy to the case with vector bosons in the loop we define
the following generic function for box diagrams with fermions and
sfermions in the loop
\begin{eqnarray}\non
&&{\cal B}^{F\!F}_{L} \big( M_0, M_1, M_2, M_3; g_0^R, g_0^L,
g_1^R, g_1^L, g_2^R, g_2^L, g_3^R, g_3^L \big) ~=~
\\ \non
&& \hspace{1cm}g_0^L g_1^R g_2^L g_3^R \left( C_0 \big( s,
m_{\sf_j}^2, m_{\sf_i}^2, M_1^2, M_2^2, M_3^2 \big) + M_0^2 D_0-
(M_3^2\!-\!t) D_3 \right  )
\\
&& \hspace{1cm}- M_1 M_2\, g_0^L g_1^L g_2^R g_3^R (D_0 + D_3) -
M_1 M_3\, g_0^L g_1^L g_2^L g_3^R D_3 - M_2 M_3\, g_0^L g_1^R
g_2^R g_3^R D_3 \,,
\end{eqnarray}
and for the crossed counterpart
\begin{eqnarray}\non
&&{\cal B}^{F\!F\!x}_{L} \big( M_0, M_1, M_2, M_3; g_0^R, g_0^L,
g_1^R, g_1^L, g_2^R, g_2^L, g_3^R, g_3^L \big) ~=~
\\ \non
&& \hspace{1cm} g_0^L g_1^R g_2^L g_3^R \left( C_0 \big(
m_{\sf_j}^2, m_{\sf_i}^2, s, M_1^2, M_3^2, M_2^2 \big) + M_0^2
D_0^x - (M_3^2\!-\!u) D_3^x \right)
\\
&& \hspace{1cm} - M_1 M_2\, g_0^L g_1^L g_2^R g_3^R (D_0^x +
D_3^x) - M_1 M_3 \, g_0^L g_1^L g_2^L g_3^R D_3^x - M_2 M_3\,
g_0^L g_1^R g_2^R g_3^R D_3^x \,,
\end{eqnarray}
with ${\cal B}^{F\!F(x)}_{R} = {\cal B}^{F\!F(x)}_{L} (L
\leftrightarrow R )$.
\newline\newline
As in the case of two $W$-bosons in the loop, for the graphs with
charginos we use either
\begin{eqnarray}
{\cal B}^{\chp\!\chp}_{L,R} &=& \sum_{k,l=1}^2 {\cal
B}^{F\!F\!x}_{L,R} \big( m_{\ti\nu_e}, m_{\chp_k},  m_{\chp_l},
m_{f'}; k^{\ti\nu_e}_{1k}, l^{\ti\nu_e}_{1k}, l^{\sf}_{jk},
k^{\sf}_{jk}, k^{\sf}_{il}, l^{\sf}_{il}, l^{\ti\nu_e}_{1l},
k^{\ti\nu_e}_{1l} \big)
\end{eqnarray}
for up-type sfermions or
\begin{eqnarray}
{\cal B}^{\chp\!\chp}_{L,R} &=& \sum_{k,l=1}^2 {\cal
B}^{F\!F}_{L,R} \big( m_{\ti\nu_e}, m_{\chp_k},  m_{\chp_l},
m_{f'}; k^{\ti\nu_e}_{1k}, l^{\ti\nu_e}_{1k}, k^{\sf}_{ik},
l^{\sf}_{ik}, l^{\sf}_{jl}, k^{\sf}_{jl}, l^{\ti\nu_e}_{1l},
k^{\ti\nu_e}_{1l} \big)
\end{eqnarray}
for down-type sfermions in the final state.
\newline %
The contributions from 2 neutralinos in the loop have the
following explicit form:
\begin{eqnarray}\non
{\cal B}^{\nt\!\nt}_{L,R} &=& \sum_{k=1}^4 \sum_{l=1}^4
\sum_{m=1}^2 {\cal B}^{F\!F}_{L,R} \big( m_{\ti e}, m_{\nt_k},
m_{\nt_l}, m_{f}; b^{\ti e}_{mk}, a^{\ti e}_{mk}, b^{\sf}_{ik},
a^{\sf}_{ik}, a^{\sf}_{jl}, b^{\sf}_{jl}, a^{\ti\nu_e}_{ml},
b^{\ti\nu_e}_{ml} \big)
\\
+\hspace{-6mm} && \sum_{k=1}^4 \sum_{l=1}^4 \sum_{m=1}^2 {\cal
B}^{F\!F\!x}_{L,R} \big( m_{\ti e}, m_{\nt_k},  m_{\nt_l}, m_{f};
b^{\ti e}_{mk}, a^{\ti e}_{mk}, a^{\sf}_{jk}, b^{\sf}_{jk},
b^{\sf}_{il}, a^{\sf}_{il}, a^{\ti e}_{ml}, b^{\ti e}_{ml} \big)
\end{eqnarray}
%
%
\section{Self-energies}\label{appSE}
Here we give the explicit form of the self-energies needed for the
computation of some wave-function renormalization constants and
various counterterms. We omit the sfermion self-energies already
given in \cite{chrislet}. All fermion, sfermion and vector
self-energy diagrams are shown in Figs. \ref{propbox} and
\ref{fermion-SE}.
\subsection{Fermion self-energies}\label{appfermion-SE}
\begin{figure}[h!]
\begin{picture}(160,65)(0,0)
     \put(0,0){\mbox{\resizebox{15.5cm}{!}
     {\includegraphics{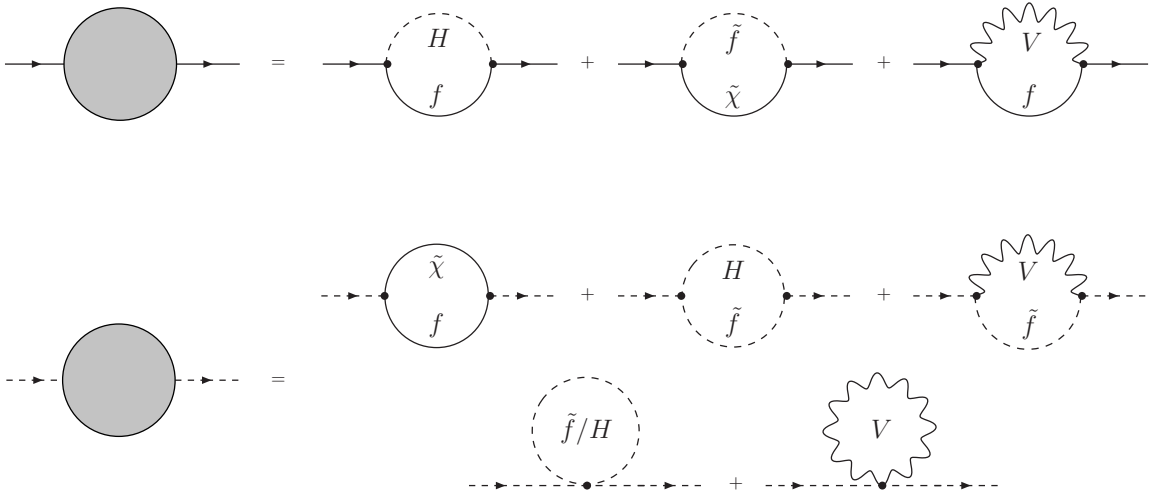}}}}
\end{picture}
\caption{Fermion and sfermion self-energies\label{fermion-SE}}
\end{figure}

In our notation, the fermion self-energy is defined as
\\
\begin{picture}(140,25)(0,0)
     \put(15,0){\mbox{\resizebox{11cm}{!}
     {\includegraphics{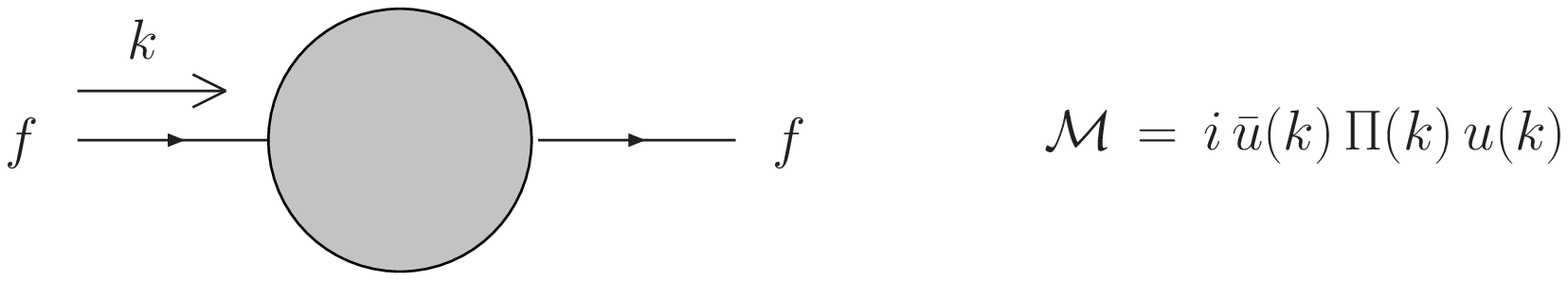}}}}
\end{picture}\\
with
\begin{eqnarray}\label{Piferm}
\Pi (k) & = & \not\!k \, P_L \, \Pi^L (k^2) + \not\!k \, P_R \,
\Pi^R (k^2) + \Pi^{SL} (k^2) P_L + \Pi^{SR} (k^2) P_R \,.
\end{eqnarray}
Below we list the contributions to the left- and right-handed
parts $\Pi^{L,R}$ and $\Pi^{SL,SR}$ from the single diagrams. The
form-factor $\Pi$ is defined as a sum of the contributions coming
from the diagrams in Fig.~\ref{fermion-SE}.
\begin{eqnarray}
\Pi = \Pi^{e\, H_k^0} + \Pi^{\nu_e H_k^+} + \Pi^{{\ti e}\,\nt} +
\Pi^{{\ti\nu_e}\,\chp} + \Pi^{e\,\g} + \Pi^{e\, Z^0} + \Pi^{\nu_e
W^+}\,.
\end{eqnarray}
We give the full formulas for the electron self-energy without
neglecting the electron mass (although it is being neglected in
the actual calculation).
\newline Note that for quarks and leptons (contrary to charginos),
the left- and right-handed scalar parts of $\Pi (k)$ are equal,
i.~e. $\Pi^{SL}(k) = \Pi^{SR}(k)$.
\begin{eqnarray}
\Pi^{e\, H_k^0} (k^2) & = & \frac{1}{(4\pi)^2}\,\Bigg[ \not\!k
\left( -\sum_{l=1}^2 (s^e_l)^2 B_1 + \sum_{l=3}^4 (s^e_l)^2 \, B_1
\right) + \sum_{l=1}^4 (s^e_l)^2\, m_e \,B_0 \Bigg] (k^2, m_e^2,
m_{H_l^0}^2)\,,
\\
\Pi^{\nu_e H_k^+} (k^2) & = & -\frac{1}{(4\pi)^2}\, \not\!k\,  P_R
\sum_{l=1}^2 (y^e_l)^2\, B_1  (k^2, 0, m_{H_l^+}^2)\,,
\\ \non
\Pi^{{\ti e}\,\nt} (k^2) & = & -\frac{1}{(4\pi)^2}\,\sum_{l=1}^4
\sum_{m=1}^2 \Bigg[ \not\!k\, P_L (a^{\ti e}_{ml})^2 \,B_1 +
\not\!k\, P_R (b^{\ti e}_{ml})^2 \,B_1 - m_{\nt_l} \,a^{\ti
e}_{ml} b^{\ti e}_{ml} \,B_0 \Bigg] (k^2, m_{\nt_l}^2, m_{{\ti
e}_m}^2)\,,
\\
\\ \non
\Pi^{{\ti\nu_e}\,\chp} (k^2) & = &
-\frac{1}{(4\pi)^2}\,\sum_{l=1}^2 \sum_{m=1}^2 \Bigg[ \not\!k\,
P_L (l^{\ti\nu_e}_{ml})^2 \,B_1 + \not\!k\, P_R
(k^{\ti\nu_e}_{ml})^2 \,B_1 - m_{\chp_l} \,k^{\ti\nu_e}_{ml} \,
l^{\ti\nu_e}_{ml} \,B_0 \Bigg] (k^2, m_{\chp_l}^2,
m_{\ti\nu_e}^2)\,,
\\
\\
\Pi^{e\, \g} (k^2) & = & -\frac{1}{(4\pi)^2}\,\Bigg[ \not\!k\, 2
e^2 \left( B_1 (k^2, m_e^2, \l^2) + \frac{r}{2} \right) + 4 e^2\,
m_e \left( B_0 (k^2, m_e^2, \l^2) - \frac{r}{2} \right) \Bigg]\,,
\\ \non
\Pi^{e\, Z^0} (k^2) & = & -\frac{1}{(4\pi)^2}\,\Bigg[ \not\!k\, 2
g_Z^2 \, (C^e_{L,R})^2 \left( B_1 (k^2, m_e^2, \l^2) + \frac{r}{2}
\right) + 4 g_Z^2 \, m_e\, C^e_{L} C^e_{R} \left( B_0 (k^2, m_e^2,
\l^2) - \frac{r}{2} \right) \Bigg]\,,
\\
\\
\Pi^{\nu_e W^+} (k^2) & = & -\frac{1}{(4\pi)^2}\,\Bigg[ \not\!k\,
P_L g^2 \left( B_1 (k^2, m_e^2, \l^2) + \frac{r}{2} \right)
\Bigg]\,.
\end{eqnarray}
%
%
\subsection{Vector self-energies}\label{appvector-SE}
Here we give the explicit form of the general gauge boson
self-energies (the transverse parts only) which are then applied
to the cases of the photon and the $Z$-boson (and their mixing).
The corresponding couplings are given in a table after each
generic formula. We do not list the contributions to the
counterterms of the $Z$- and $W$-bosons as they can be found in
\cite{chrislet}.\newline The self-energy of a vector boson is
defined as follows\\
\begin{picture}(160,30)(0,0)
     \put(5,5){\mbox{\resizebox{15cm}{!}
     {\includegraphics{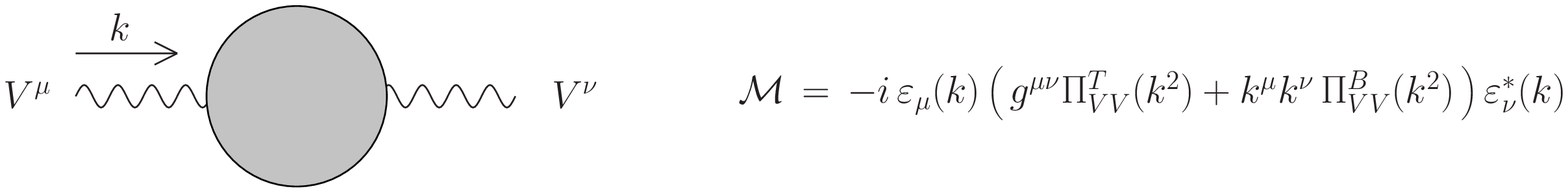}}}}
\end{picture}
The transverse part of the self-energy consists of the following
parts:
\begin{eqnarray}\non
\Pi^{T}_{VV} &=& \big( \Pi^{T}_{VV}\big)^{ff} + \big(
\Pi^{T}_{VV}\big)^{\ch^{0}\ch^{0}}+ \big(
\Pi^{T}_{VV}\big)^{\ch^{\pm}\ch^{\mp}}  + \big(
\Pi^{T}_{VV}\big)^{\sf\sf} + \big(
\Pi^{T}_{VV}\big)^{H^{\pm}H^{\pm}} + \big( \Pi^{T}_{VV}\big)^{H^0
H^0}
\\ \non%
&& +\big( \Pi^{T}_{VV}\big)^{\sf} + \big(
\Pi^{T}_{VV}\big)^{H^{0}} + \big( \Pi^{T}_{VV}\big)^{H^{\pm}} +
\big( \Pi^{T}_{VV}\big)^{Z^0 H^0} + \big(
\Pi^{T}_{VV}\big)^{W^{\pm} H^{\mp}} + \big(
\Pi^{T}_{VV}\big)^{W^{\pm} W^{\mp}}
\\ %
&& +\big( \Pi^{T}_{VV}\big)^{W^{\pm}} + \big(
\Pi^{T}_{VV}\big)^{\rm 2 FP ghosts}
\end{eqnarray}
For the contributions with a fermion loop we define the following
generic function:
\begin{eqnarray}\non
{\cal A}_{VV} \big( k^2, m_1, m_2; g_1^L, g_1^R, g_2^L, g_2^R
\big) &=& 2 \Big[ - \big( g_1^L g_2^L + g_1^R g_2^R \big) \big(
k^2 B_1 + m_1^2 B_0 + A_0(m_2^2) - 2 B_{00} \big)
\\
&&~+\big( g_1^R g_2^L + g_1^L g_2^R \big) m_1 m_2 \,B_0 \Big]
\big(k^2, m_1^2, m_2^2\big)
\end{eqnarray}
Using the generic function we can write the 2 fermion, 2
neutralino and the 2 chargino contributions as
\begin{eqnarray}
\big( \Pi^{T}_{VV}\big)^{ff} (k^2) &=& \frac{1}{(4 \pi)^2}
\sum_{f} N_C^f\, {\cal A}_{VV} \big( k^2, m_f, m_f; g_1^L, g_1^R,
g_2^L, g_2^R \big) \,,
\\
\big( \Pi^{T}_{VV}\big)^{\nt\nt} (k^2) &=& \frac{1}{(4 \pi)^2} \,
\frac{1}{2} \sum_{k,l=1}^4 {\cal A}_{VV} \big( k^2, m_{\nt_l},
m_{\nt_k}; g_1^L, g_1^R, g_2^L, g_2^R \big) \,,
\\
\big( \Pi^{T}_{VV}\big)^{\ch^{\pm}\ch^{\mp}} (k^2) &=& \frac{1}{(4
\pi)^2} \sum_{k,l=1}^2 {\cal A}_{VV} \big( k^2, m_{\chp_l},
m_{\chp_k}; g_1^L, g_1^R, g_2^L, g_2^R \big) \,.
\end{eqnarray}
\renewcommand{\arraystretch}{1.5}
\begin{table}[h!]
\begin{tabular*}{160mm}{|@{\extracolsep{\fill}}c||c|c|c|c||c|c|c|c||c|c|c|c|}
\hline & \multicolumn{4}{c||}{$ff$} &
\multicolumn{4}{c||}{$\nt_k\, \nt_l$} &
\multicolumn{4}{c|}{$\chp_k\, \chm_l$}
\\ %
& $g_1^L$ & $g_1^R$ & $g_2^L$ & $g_2^R$ & $g_1^L$ & $g_1^R$ &
$g_2^L$ & $g_2^R$ & $g_1^L$ & $g_1^R$ & $g_2^L$ & $g_2^R$
\\ %
\hline\hline $\g\g$ & $-e e_f$ & $-e e_f$ & $-e e_f$ & $-e e_f$ &
-- & -- & -- & -- & $-e\,\d_{kl}$ & $-e\,\d_{kl}$ & $-e\,\d_{kl}$
& $-e\,\d_{kl}$
\\ %
\hline $\g Z$ & $-e e_f$ & $-e e_f$ & $-g_Z C_L^f$ & $-g_Z C_R^f$
& -- & -- & -- & -- & $-e\,\d_{kl}$ & $-e\,\d_{kl}$ & $g_Z
O^{''L}_{lk}$ & $g_Z O^{''R}_{lk}$
\\ %
\hline $ZZ$ & $-g_Z C_L^f$ & $-g_Z C_R^f$ & $-g_Z C_L^f$ & $-g_Z
C_R^f$ & $g_Z O^{'L}_{kl}$ & $g_Z O^{'R}_{kl}$ & $g_Z O^{'L}_{lk}$
& $g_Z O^{'R}_{lk}$ & $g_Z O^{''L}_{kl}$ & $g_Z O^{''R}_{kl}$ &
$g_Z O^{''L}_{lk}$ & $g_Z O^{''R}_{lk}$
\\ %
\hline
\end{tabular*}
\caption{Couplings for the 2 fermion, 2 neutralino and the 2
chargino contributions to $\g \g$, $\g Z$ and $Z Z$
self-energies.}
\end{table}
\renewcommand{\arraystretch}{1}
\newline The next set of contributions are the ones with 2 scalar
particles in the loop. For this set we introduce
\begin{eqnarray}
{\cal B}_{VV} \big( k^2, m_1^2, m_2^2; g_1, g_2 \big) &=& -4 g_1
g_2 \,B_{00} \big(k^2, m_1^2, m_2^2\big)
\end{eqnarray}
and get for the sfermion, neutral and charged Higgs in the loop
the following forms:
\begin{eqnarray}
\big( \Pi^{T}_{VV}\big)^{\sf\sf} (k^2) &=& \frac{1}{(4 \pi)^2}
\sum_{f} N_C^f \sum_{m,n=1}^2 {\cal B}_{VV} \big( k^2,
m_{\sf_n}^2, m_{\sf_m}^2; g_1, g_2 \big)
\\
\big( \Pi^{T}_{VV}\big)^{H_k^0 H_l^0} (k^2) &=& \frac{1}{(4
\pi)^2} \sum_{k=1}^2 \sum_{l=3}^4 {\cal B}_{VV} \big( k^2,
m_{H_k^0}^2, m_{H_l^0}^2; g_1, g_2 \big)
\\
\big( \Pi^{T}_{VV}\big)^{H_k^{\pm}H_k^{\pm}} (k^2) &=& \frac{1}{(4
\pi)^2} \sum_{k=1}^2 {\cal B}_{VV} \big( k^2, m_{H_k^+}^2,
m_{H_k^+}^2; g_1, g_2 \big)
\end{eqnarray}
\renewcommand{\arraystretch}{1.5}
\begin{table}[h]
\begin{tabular}{|c||c|c||c|c||c|c|}
\hline %
& \multicolumn{2}{c||}{$\sf_m \sf_n$} &
\multicolumn{2}{c||}{$H_k^0 \,H_l^0$} &
\multicolumn{2}{c|}{$H_k^{\pm}\, H_k^{\pm}$}
\\ %
& $g_1$ & $g_2$ & $g_1$ & $g_2$ & $g_1$ & $g_2$
\\ %
\hline\hline %
$\g\g$ & $-e e_f \,\d_{mn}$ & $-e e_f \,\d_{mn}$ & -- & -- & $-e$
& $-e$
\\ %
\hline %
$\g Z$ & $-e e_f \,\d_{mn}$ & $-g_Z \,z^\sf_{mn}$ & -- & -- & $-e$
& $-\frac{g_Z}{2} \cos 2\theta_W$
\\ %
\hline %
$\quad ZZ \quad$ & $\ -g_Z \,z^\sf_{mn} \ $ & $\ -g_Z \,z^\sf_{mn}
\ $ & $\ \frac{g_Z}{2} R_{l-2,k}(\a\!-\!\b)\ $ & $\ \frac{g_Z}{2}
R_{l-2,k}(\a\!-\!\b)\ $ & $\ -\frac{g_Z}{2} \cos 2\theta_W\ $ & $\
-\frac{g_Z}{2} \cos 2\theta_W\ $
\\ %
\hline %
\end{tabular}
\caption{Couplings for the 2 sfermion, 2 neutral and 2 charged
Higgs contributions to $\g \g$, $\g Z$ and $Z Z$ self-energies.}
\end{table}
\renewcommand{\arraystretch}{1}
\newline
The next class are the self-energies with a single scalar particle
in the loop for which we use the generic form
\begin{eqnarray}
{\cal C}_{VV} \big(m^2; g_1 \big) &=& g_1 \,A_{0} \big(m^2\big)\,.
\end{eqnarray}
The diagrams with 1 sfermion, 1 neutral or charged boson can be
written as
\begin{eqnarray}
\big( \Pi^{T}_{VV}\big)^{\sf} &=& \frac{1}{(4 \pi)^2} \sum_{f}
N_C^f \sum_{m=1}^2 {\cal C}_{VV} \big( m_{\sf_m}^2; g_1 \big)\,,
\\
\big( \Pi^{T}_{VV}\big)^{H_k^0} &=& \frac{1}{(4 \pi)^2}
\sum_{k=1}^4 {\cal C}_{VV} \big( m_{H_k^0}^2; g_1 \big)\,,
\\
\big( \Pi^{T}_{VV}\big)^{H_k^+} &=& \frac{1}{(4 \pi)^2}
\sum_{k=1}^2 {\cal C}_{VV} \big( m_{H_k^+}^2; g_1 \big)\,.
\end{eqnarray}
\renewcommand{\arraystretch}{1.8}
\begin{table}[h]
\begin{tabular}{|c|c|c|c|}
\hline %
& $\sf_m$ & $H_k^0$ & $H_k^+$
\\ %
& $g_1$ & $g_1$ & $g_1$
\\ %
\hline\hline %
$\g\g$ & $2 (e e_f)^2$ & -- & $2 e^2$
\\ %
\hline %
$\g Z$ & $2 e e_f \,g_Z \left[ C_L^f (R^\sf_{m1})^2 + C_R^f
(R^\sf_{m2})^2 \right]$ & -- & $e \,g_Z \cos 2\theta_W$
\\ %
\hline %
$\quad ZZ \quad$ & $\ 2 g_Z^2 \left[ (C_L^f)^2 (R^\sf_{m1})^2 +
(C_R^f)^2 (R^\sf_{m2})^2 \right]\ $ & $\quad g_Z^2/4 \quad$ & $\
\big(g_Z^2/2\big) \, \big(1 - 2 s_W^2 \big)^2\ $
\\ %
\hline %
\end{tabular}
\caption{Couplings for the 1 sfermion, 1 neutral and charged Higgs
contributions to $\g \g$, $\g Z$ and $Z Z$ self-energies.}
\end{table}
\renewcommand{\arraystretch}{1}
\newline
The diagrams with a vector and a scalar particle in the loop use
the simple generic form
\begin{eqnarray}
{\cal D}_{VV} \big(k^2, m_1^2, m_2^2; g_1, g_2 \big) &=& g_1 g_2
\,B_{0} \big(k^2, m_1^2, m_2^2\big)
\end{eqnarray}
and give
\begin{eqnarray}
\big( \Pi^{T}_{VV}\big)^{H_k^0 Z} &=& \frac{1}{(4 \pi)^2}
\sum_{k=1}^2 {\cal D}_{VV} \big( k^2, m_{H_k^0}^2, m_Z^2; g_1, g_2
\big)\,,
\\
\big( \Pi^{T}_{VV}\big)^{G^\pm W^\mp} &=& \frac{1}{(4 \pi)^2}
{\cal D}_{VV} \big( k^2, m_{G^+}^2, m_W^2; g_1, g_2 \big)\,.
\end{eqnarray}
\renewcommand{\arraystretch}{1.5}
\begin{table}[h]
\begin{tabular}{|c||c|c||c|c|}
\hline %
& \multicolumn{2}{c||}{$H_k^0 Z$} & \multicolumn{2}{c|}{$G^\pm
W^\mp$}
\\ %
& $g_1$ & $g_2$ & $g_1$ & $g_2$
\\ %
\hline\hline %
$\g\g$ & -- & -- & $2g\, s_W\, m_W$ & $g\, s_W\, m_W$
\\ %
\hline %
$\g Z$ & -- & -- & $2g\, s_W\, m_W$ & $-g_Z\, m_W\, s_W^2$
\\ %
\hline %
$\quad ZZ \quad$ & $\ g_Z \,m_Z\,R_{2k}(\a\!-\!\b) \ $ & $\ g_Z
\,m_Z\,R_{2k}(\a\!-\!\b) \ $ & $\ -2 g_Z \,m_W s_W^2 $ & $\ - g_Z
\, m_W s_W^2 \ $
\\ %
\hline %
\end{tabular}
\caption{Couplings for the $Z$-boson--neutral Higgs and the
$W$-boson--charged Higgs contributions to $\g \g$, $\g Z$ and $Z
Z$ self-energies.}
\end{table}
\renewcommand{\arraystretch}{1}
\newline
The remaining 3 contributions comprising of 2 $W$-bosons, 2 FP
ghosts and a single $W$-boson in the loop have the following
explicit forms:
\begin{eqnarray}\non
\big( \Pi^{T}_{VV}\big)^{W^+W^-} &=& -\frac{1}{(4 \pi)^2}\;
g_1\,g_2\left[10\,B_{00} + 5\,k^2\,B_{0} + 2\,k^2\,B_{1} + 2
A_0\big(m_W^2 \big)\right.
\\
&& \left. + 2\,m_W^2 B_0 + r \big(\frac{2}{3}\,k^2 -4\,m_W^2
\big)\right]\big( k^2, m_W^2, m_W^2 \big) \,,
\\
\big( \Pi^{T}_{VV}\big)^{W^\pm} &=& \frac{1}{(4
\pi)^2}\;g_1\left[3\,A_0\big(m_W^2\big) - 2\,r\,m_W^2\right] \,,
\\
\big( \Pi^{T}_{VV}\big)^{\rm 2FPghosts} &=& \frac{1}{(4 \pi)^2}\;
g_1\,g_2\,B_{00}\big( k^2, m_W^2, m_W^2 \big) \,.
\end{eqnarray}
\renewcommand{\arraystretch}{1.8}
\begin{table}[h]
\begin{tabular}{|c|c|c|c|}
\hline %
& $W^+ W^-$ & $W^{\pm}$ & ${\rm 2\, FP\, ghosts}$
\\ %
& $g_1\,g_2$ & $g_1$ & $g_1\,g_2$
\\ %
\hline\hline %
$\g\g$ & $e^2$ & $2 e^2$ & $2 e^2$
\\ %
\hline %
$\g Z$ & $e\,g_Z\,(1-s_W^2)$ & $2e\,g_Z\,(1-s_W^2)$ & $2
e\,g_Z\,(1-s_W^2)$
\\ %
\hline %
$\quad ZZ \quad$ & $\quad g_Z^2\,(1-s_W^2)^2\quad$ & $\quad
2\,g_Z^2\,(1-s_W^2)^2\quad$ & $\quad 2\,g_Z^2\,(1-s_W^2)^2\quad$
\\ %
\hline %
\end{tabular}
\caption{Couplings for the 2 $W$-bosons, a single $W$-boson and 2
FP ghosts contributions to $\g \g$, $\g Z$ and $Z Z$
self-energies.}
\end{table}
\renewcommand{\arraystretch}{1}
\section{Bremsstrahlung integrals}\label{Bremint}
\subsection{Soft photon integral}
Using the kinematics of the process we get for $\delta_s$ the
explicit form
\begin{eqnarray}\non
\delta_s  & = &   - \frac{\alpha}{\pi}\Bigg\{\left(1 +
\log\frac{m_e^2}{s}\right)
 \log \frac{4 (\Delta E)^2}{\lambda^2} + \log\frac{m_e^2}{s} +
\frac{1}{2} \log^2\frac{m_e^2}{s} + \frac{\pi^2}{3}
\\ \non
&& +\, e_f^2 \bigg[\frac{s - m_{\sf_i}^2 - m_{\sf_j}^2}{\kappa}
\left( \frac{1}{2}\log\frac{d_i^-}{d_i^+} \frac{d_j^-}{d_j^+} \log
\frac{4 (\Delta E)^2}{\lambda^2} +
\frac{1}{4}\log^2\frac{d_i^-}{d_i^+} +
\frac{1}{4}\log^2\frac{d_j^-}{d_j^+}+ \spence{\frac{2
\kappa}{d_i^+}} + \spence{\frac{2 \kappa}{d_j^+}} \right)
\\ \non
&& + \log \frac{4 (\Delta E)^2}{\lambda^2} + \frac{c_i}{2 \kappa}
\log\frac{d_i^-}{d_i^+} + \frac{c_j}{2 \kappa}
\log\frac{d_j^-}{d_j^+}\bigg] + \, e_f \bigg[
\log\frac{(m_{\sf_i}^2 - t)(m_{\sf_j}^2 - t)}{(m_{\sf_i}^2 -
u)(m_{\sf_j}^2 - u)} \log \frac{4 (\Delta E)^2}{\lambda^2}
\\ \non
&& + \spence\left(1\! -\! \frac{d_i^+}{2(m_{\sf_i}^2\! -
\!u)}\right) + \spence\left(1\! -\! \frac{d_i^-}{2(m_{\sf_i}^2\!
-\! u)}\right) + \spence\left(1\! -\! \frac{d_j^+}{2(m_{\sf_j}^2\!
- \!u)}\right) + \spence\left(1 \!-\! \frac{d_j^-}{2(m_{\sf_j}^2\!
-\! u)}\right)
\\ \non
&& - \spence\left(1\! -\! \frac{d_i^+}{2(m_{\sf_i}^2\! -\!
t)}\right) - \spence\left(1\! -\! \frac{d_i^-}{2(m_{\sf_i}^2\! -\!
t)}\right) - \spence\left(1\! - \!\frac{d_j^+}{2(m_{\sf_j}^2\! -\!
t)}\right) - \spence\left(1\! -\! \frac{d_j^-}{2(m_{\sf_j}^2\! -\!
t)}\right) \bigg] \Bigg\}\,,
\\
\end{eqnarray}
where $s, t$ and $u$ are the Mandelstam variables, $d_i^\pm = c_i
\pm \kappa$ with $c_i$ and $\kappa$ defined as
\begin{eqnarray}
\begin{array}{rcl}
& \kappa ~=~ \kappa(s,m_{\sf_i}^2,m_{\sf_j}^2) ~=~
\sqrt{\lambda(s,m_{\sf_i}^2,m_{\sf_j}^2)} ~=~
\sqrt{(s-m_{\sf_i}^2-m_{\sf_j}^2)^2-4m_{\sf_i}^2m_{\sf_j}^2}\,, &
\\
& c_i ~=~ s+m_{\sf_i}^2-m_{\sf_j}^2\,, \qquad\quad
c_j~=~s+m_{\sf_j}^2-m_{\sf_i}^2\,. &
\end{array}
\end{eqnarray}
\subsection{Hard photon integrals}
The squared matrix element for the hard photon radiation can be
split into 3 parts,
\begin{equation}
|{\cal M}^{\rm hard}|^2 = |{\cal M}^{\rm e}|^2 + |{\cal M}^{\rm
\sf}|^2 + 2\,\Re ({\cal M}^{\rm \sf}{\cal M}^{\rm e \dag})\,,
\end{equation}
where $|{\cal M}^{\rm X}|^2$ stands for the part of the amplitude
where the photon is radiated off the particle indicated.\newline
The squared matrix part corresponding to the photon being radiated
from the electron or positron has the form
\begin{equation}
|{\cal M}^{\rm e}|^2 = \frac{1}{4}(1-P_-)(1+P_+)\,|{\cal M}^{\rm
e}|^2_L +\frac{1}{4}(1+P_-)(1-P_+)\,|{\cal M}^{\rm e}|^2_R\,.
\end{equation}
The chiral $L,R$ parts are
\begin{eqnarray}
|{\cal M}^{\rm e}|^2_{L,R} &=& N_C\,e^2 \left[(T_{\rm
e}^{\g\g})_{L,R}+(T_{\rm e}^{\g Z})_{L,R}+ (T_{\rm
e}^{ZZ})_{L,R}\right]\left({\cal R}_1+ {\cal
R}_1(p_1\leftrightarrow p_2)+ {\cal R}_2\right)\,,
\end{eqnarray}
where
\begin{eqnarray}
(T_{\rm e}^{\g\g})_{L,R}&=&\frac{e^4 e_f^2 (\d_{ij})^2}{s_{\rm
red}^2}\, K_{L,R}^2\,,
\\
(T_{\rm e}^{\g Z})_{L,R}&=& -\frac{g_Z^2 e^2 e_f
a_{ij}^\sf\d_{ij}}{2\,s_{\rm red} (s_{\rm red}-m_Z^2)}\, C_{L,R}
K_{L,R}\,,
\\
(T_{\rm e}^{ZZ})_{L,R} &=&\frac{g_Z^4(a_{ij}^\sf)^2}{16\,(s_{\rm
red}-m_Z^2)^2}\,C_{L,R}^2\,,
\end{eqnarray}
with $s_{\rm red} = (p_1+p_2-k_3)^2$.
\newline%
The functions ${\cal R}_1$ and ${\cal R}_2$ contain only scalar
products of the external momenta and are defined as
\begin{eqnarray}\non
{\cal R}_1 &=& -\frac{1}{(p_1.k_3)}\left[-4\,(k_1.k_3)\,(k_1.p_2)
+ 4\,(k_1.p_2)\,(k_2.k_3) + 4\,(k_1.k_3)\,(k_2.p_2) -
4\,(k_2.k_3)\,(k_2.p_2)\right. \\ && \left. + 2
m_{\sf_i}^2\,(k_3.p_2) + 2m_{\sf_j}^2\,(k_3.p_2) -
4\,(k_1.k_2)\,(k_3.p_2) \right]\,,
\\ \non
{\cal R}_2 &=& - \frac{(p_1.p_2)}{(p_1.k_3)(p_2.k_3)}\left[
2\,(k_1.k_3)\,(k_1.p_1) + 2\,(k_1.k_3)\,(k_1.p_2) -
4\,(k_1.p_1)\,(k_1.p_2) - 2\,(k_1.p_1)\,(k_2.k_3) \right.
\\ \non
&&\left. - 2\,(k_1.p_2)\,(k_2.k_3) - 2\,(k_1.k_3)\,(k_2.p_1) +
4\,(k_1.p_2)\,(k_2.p_1) + 2\,(k_2.k_3)\,(k_2.p_1) \right.
\\ \non
&&\left. - 2\,(k_1.k_3)\,(k_2.p_2) + 4\,(k_1.p_1)\,(k_2.p_2) +
2\,(k_2.k_3)\,(k_2.p_2) - 4\,(k_2.p_1)\,(k_2.p_2) + 2
m_{\sf_i}^2\,(p_1.p_2) \right.
\\ \non
&&\left. + 2m_{\sf_j}^2\,(p_1.p_2) -
4\,(k_1.k_2)\,(p_1.p_2)\right] -\frac{1}{(p_1.k_3)}\left[
-2\,(k_1.p_1)^2 + 2\,(k_1.p_1)\,(k_1.p_2) \right.
\\ \non
&&\left. + 4\,(k_1.p_1)\,(k_2.p_1) - 2\,(k_1.p_2)\,(k_2.p_1) -
2\,(k_2.p_1)^2 - 2\,(k_1.p_1)\,(k_2.p_2) + 2\,(k_2.p_1)\,(k_2.p_2)
\right.
\\ \non
&& \left. - 2 m_{\sf_i}^2\,(p_1.p_2) - 2m_{\sf_j}^2\,(p_1.p_2) +
4\,(k_1.k_2)\,(p_1.p_2)\right] -\frac{1}{(p_2.k_3)}\left[
-2\,(k_1.p_2)^2 \right.
\\ \non
&&\left. + 2\,(k_1.p_2)\,(k_1.p_1) + 4\,(k_1.p_2)\,(k_2.p_2) -
2\,(k_1.p_1)\,(k_2.p_2) - 2\,(k_2.p_2)^2 - 2\,(k_1.p_2)\,(k_2.p_1)
\right.
\\
&& \left. + 2\,(k_2.p_2)\,(k_2.p_1) - 2 m_{\sf_i}^2\,(p_1.p_2) -
2m_{\sf_j}^2\,(p_1.p_2) + 4\,(k_1.k_2)\,(p_1.p_2)\right] \,.
\end{eqnarray}
The radiation off the sfermion can be written as
\begin{equation}
|{\cal M}^{\rm \sf}|^2 = \frac{1}{4}(1-P_-)(1+P_+)\,|{\cal M}^{\rm
\sf}|^2_L +\frac{1}{4}(1+P_-)(1-P_+)\,|{\cal M}^{\rm \sf}|^2_R\,,
\end{equation}
where
\begin{eqnarray}
|{\cal M}^{\rm \sf}|^2_{L,R} = N_C\,e^2 e_f^2\,2s
\left(T_{L,R}^{\g\g}+T_{L,R}^{\g
Z}+T_{L,R}^{ZZ}\right)\left(g^{\mu\nu}-\frac{2}{s}(p_1^{\mu}p_2^{\nu}
+p_1^{\nu}p_2^{\mu}) \right) T_{\mu\rho}T_{\nu}^{\rho} \,.
\end{eqnarray}
The tensor $T_{\mu\nu}$ is defined as
\begin{equation}
T_{\mu\nu} = \frac{1}{2k_1.k_3}(k_1-k_2+k_3)_{\mu}(2k_1+k_3)_{\nu}
- \frac{1}{2k_2.k_3}(k_1-k_2-k_3)_{\mu}(2k_2+k_3)_{\nu} -
2g_{\mu\nu} \,.
\end{equation}
The interference term of the squared hard photon amplitude is
\begin{eqnarray}
2\,\Re ({\cal M}^{\rm \sf}{\cal M}^{\rm e \dag}) &=&
\frac{1}{4}(1-P_-)(1+P_+)\,2\,\Re ({\cal M}^{\rm \sf}{\cal M}^{\rm
e \dag})_L +\frac{1}{4}(1+P_-)(1-P_+)\,2\,\Re ({\cal M}^{\rm
\sf}{\cal M}^{\rm e \dag})_R\,.
\end{eqnarray}
The chiral $L,R$ parts are
\begin{eqnarray}\non
&& 2\,\Re ({\cal M}^{\rm \sf}{\cal M}^{\rm e +})_{L,R} = N_C\,e^2
\left[(T_{\rm int}^{\g\g})_{L,R}+(T_{\rm int}^{\g Z})_{L,R}+
(T_{\rm int}^{ZZ})_{L,R}\right]\times
\\ \non
&& \left[{\cal R}_3 - {\cal R}_3(p_1\leftrightarrow p_2)- {\cal
R}_3(k_1\leftrightarrow k_2, m_{\sf_i} \leftrightarrow m_{\sf_j})
+ {\cal R}_3(p_1\leftrightarrow p_2, k_1\leftrightarrow k_2,
m_{\sf_i} \leftrightarrow m_{\sf_j})\right.
\\
&& \left. + {\cal R}_4 + {\cal R}_4(p_1\leftrightarrow p_2,
k_1\leftrightarrow k_2, m_{\sf_i} \leftrightarrow
m_{\sf_j})\right] \,,
\end{eqnarray}
where
\begin{eqnarray}
(T_{\rm int}^{\g\g})_{L,R}&=&\frac{e^4 e_f^2 (\d_{ij})^2}{s\,
s_{\rm red}}\, K_{L,R}^2\,,
\\
(T_{\rm int}^{\g Z})_{L,R}&=& -g_Z^2 e^2 e_f
a_{ij}^\sf\d_{ij}\left[\frac{1}{4 s_{\rm red} (s-m_Z^2)} +
\frac{1}{4 s (s_{\rm red}-m_Z^2)} \right]\, C_{L,R} K_{L,R}\,,
\\
(T_{\rm int}^{ZZ})_{L,R}
&=&\frac{g_Z^4(a_{ij}^\sf)^2}{16\,(s-m_Z^2)\,(s_{\rm
red}-m_Z^2)}\,C_{L,R}^2\,,
\end{eqnarray}
with $s_{\rm red} = (p_1+p_2-k_3)^2$.
\newline%
The functions ${\cal R}_3$ and ${\cal R}_4$ are given by
\begin{eqnarray}\non
{\cal R}_3 &=& -\frac{2}{(p_1.k_3)\,(k_1.k_3)}\,\left[
       -4\,(k_1.k_3)\,(k_1.p_1)\,(k_1.p_2) +
        4\,(k_1.p_1)^2(k_1.p_2) +
        2\,(k_1.p_1)\,(k_1.p_2)\,(k_2.k_3)
        \right. \\ \non && \hspace{-10pt} \left. +
        2\,(k_1.k_3)\,(k_1.p_2)\,(k_2.p_1) -
        4\,(k_1.p_1)\,(k_1.p_2)\,(k_2.p_1) +
        4\,(k_1.k_3)\,(k_1.p_1)\,(k_2.p_2) -
        4\,(k_1.p_1)^2(k_2.p_2)
        \right. \\ \non && \hspace{-10pt} \left. -
        2\,(k_1.p_1)\,(k_2.k_3)\,(k_2.p_2) -
        2\,(k_1.k_3)\,(k_2.p_1)\,(k_2.p_2) +
        4\,(k_1.p_1)\,(k_2.p_1)\,(k_2.p_2) +
        m_{\sf_i}^2(k_1.p_2)\,(k_3.p_1)
        \right. \\ \non && \hspace{-10pt} \left. -
        m_{\sf_j}^2(k_1.p_2)\,(k_3.p_1) -
        2\,(k_1.k_3)\,(k_1.p_2)\,(k_3.p_1) +
        4\,(k_1.p_1)\,(k_1.p_2)\,(k_3.p_1) +
        2\,(k_1.p_2)\,(k_2.k_3)\,(k_3.p_1)
        \right. \\ \non && \hspace{-10pt} \left. -
        2\,(k_1.p_2)\,(k_2.p_1)\,(k_3.p_1) -
        2\,m_{\sf_i}^2(k_2.p_2)\,(k_3.p_1) +
        2\,(k_1.k_2)\,(k_2.p_2)\,(k_3.p_1) -
        4\,(k_1.p_1)\,(k_2.p_2)\,(k_3.p_1)
        \right. \\ \non && \hspace{-10pt} \left. +
        2\,(k_2.p_1)\,(k_2.p_2)\,(k_3.p_1) +
            (k_1.p_2)\,(k_3.p_1)^2 -
            (k_2.p_2)\,(k_3.p_1)^2 +
        m_{\sf_i}^2(k_1.p_1)\,(k_3.p_2)
        \right. \\ \non && \hspace{-10pt} \left. +
        m_{\sf_j}^2(k_1.p_1)\,(k_3.p_2) -
        2\,(k_1.k_2)\,(k_1.p_1)\,(k_3.p_2) -
        2\,(k_1.k_3)\,(k_1.p_1)\,(k_3.p_2) +
        2\,(k_1.p_1)^2\,(k_3.p_2)
        \right. \\ \non && \hspace{-10pt} \left. +
        2\,(k_1.k_3)\,(k_2.p_1)\,(k_3.p_2) -
        2\,(k_1.p_1)\,(k_2.p_1)\,(k_3.p_2) +
        2\,m_{\sf_i}^2(k_3.p_1)\,(k_3.p_2) -
        2\,(k_1.k_2)\,(k_3.p_1)\,(k_3.p_2)
        \right. \\ \non && \hspace{-10pt} \left. +
            (k_1.p_1)\,(k_3.p_1)\,(k_3.p_2) -
            (k_2.p_1)\,(k_3.p_1)\,(k_3.p_2) +
        m_{\sf_i}^2(k_1.k_3)\,(p_1.p_2) +
        m_{\sf_j}^2(k_1.k_3)\,(p_1.p_2)
        \right. \\ \non && \hspace{-10pt} \left. -
        2\,(k_1.k_2)\,(k_1.k_3)\,(p_1.p_2) +
        2\,(k_1.k_3)^2(p_1.p_2) -
        2\,m_{\sf_i}^2(k_1.p_1)\,(p_1.p_2) -
        2\,m_{\sf_j}^2(k_1.p_1)\,(p_1.p_2)
        \right. \\ \non && \hspace{-10pt} \left. +
        4\,(k_1.k_2)\,(k_1.p_1)\,(p_1.p_2) -
        2\,(k_1.k_3)\,(k_1.p_1)\,(p_1.p_2) -
        2\,(k_1.k_3)\,(k_2.k_3)\,(p_1.p_2)
        \right. \\ \non && \hspace{-10pt} \left. +
        2\,(k_1.p_1)\,(k_2.k_3)\,(p_1.p_2) -
        m_{\sf_i}^2(k_3.p_1)\,(p_1.p_2) -
        m_{\sf_j}^2(k_3.p_1)\,(p_1.p_2) +
        2\,(k_1.k_2)\,(k_3.p_1)\,(p_1.p_2)
        \right. \\ && \hspace{-10pt} \left. -
            (k_1.k_3)\,(k_3.p_1)\,(p_1.p_2) +
            (k_2.k_3)\,(k_3.p_1)\,(p_1.p_2)\right] \,,
\\ \non
{\cal R}_4 &=& -\frac{8}{((p_1.k_3))}\left[
         -(k_1.p_2)\,(k_3.p_1) + (k_2.p_2)\,(k_3.p_1) +
           (k_1.p_1)\,(k_3.p_2) - (k_2.p_1)\,(k_3.p_2)
           \right. \\ && \left. -
           (k_1.k_3)\,(p_1.p_2) +
     (k_2.k_3)\,(p_1.p_2)\right]\,.
\end{eqnarray}

\end{document}